\documentclass[]{aa}
\usepackage{amsmath}
\usepackage{txfonts} 
\usepackage{graphicx}
\usepackage{subfigure}
\usepackage{color}
\usepackage{bm}
\usepackage{natbib}
\usepackage{url}

\usepackage{tabularx}
\usepackage{graphicx}
\usepackage{txfonts}
\usepackage{longtable}
\usepackage{hhline}
\usepackage{arydshln}
\usepackage{multirow}
\usepackage{lscape}
\usepackage{array}
\usepackage{rotating}

\newcommand{\lsun}{$L_\odot$}
\newcommand{\msun}{$M_\odot$}

\newcommand{\mic}{$\mu$m}

\newcolumntype{R}[1]{>{\raggedleft\arraybackslash }b{#1}}
\newcolumntype{L}[1]{>{\raggedright\arraybackslash }b{#1}}
\newcolumntype{C}[1]{>{\centering\arraybackslash }b{#1}}

\newlength{\pointwidth}
\begin{document}

\title{SMA observations of the polarized dust emission in solar-type Class 0 protostars: the magnetic field properties at envelope scales}

  \author{Maud Galametz 
  	       \inst{1},
	      Ana\"elle Maury
  	       \inst{1,5},
	      Josep M. Girart 
  	       \inst{2,3},
	      Ramprasad Rao 
  	       \inst{4},
	      Qizhou Zhang
  	       \inst{5},
	      Mathilde Gaudel
  	       \inst{1},
	      Valeska Valdivia
  	       \inst{1},
	      Eric Keto
  	       \inst{5},
	      Shih-Ping Lai
	       \inst{6}
}

\institute{
Astrophysics department, CEA/DRF/IRFU/DAp, Universit\'{e} Paris Saclay, UMR AIM, F-91191 Gif-sur-Yvette, France, \\
\email{maud.galametz@cea.fr} 
\and
Institut de Ci\`encies de l'Espai (ICE, CSIC), Can Magrans, S/N, E-08193 Cerdanyola del Vall\`es, Catalonia, Spain
\and
Institut d'Estudis Espacials de de Catalunya (IEEC), E-08034 Barcelona, Catalonia, Spain
\and
Institute of Astronomy and Astrophysics, Academia Sinica, 645 N. Aohoku Pl., Hilo, HI 96720, USA
\and
Harvard-Smithsonian Center for Astrophysics, 60 Garden street, Cambridge, MA 02138, USA 
\and
Institute of Astronomy and Department of Physics, National Tsing Hua University, Hsinchu 30013, Taiwan
}

\abstract
{}
{Although, from a theoretical point of view, magnetic fields 
are believed to have a significant role during the early stages of 
star formation, especially during the main accretion phase, 
the magnetic field strength and topology is poorly constrained in 
the youngest accreting Class~0 protostars that lead to the 
formation of solar-type stars. }
{We carried out observations of the polarized dust continuum 
emission with the SMA interferometer at 0.87mm, in order to 
probe the structure of the magnetic field in a sample of 12 
low-mass Class 0 envelopes, including both single 
protostars and multiple systems, in nearby clouds. 
Our SMA observations probe the envelope emission at scales 
$\sim 600-5000$ au with a spatial resolution ranging from 600 
to 1500 au depending on the source distance.}
{We report the detection of linearly polarized dust continuum emission 
in all of our targets, with average polarization fractions ranging from 
2\% to 10\% in these protostellar envelopes. 
The polarization fraction decreases with the continuum 
flux density, which translates into a decrease with the 
H$_2$ column density within an individual envelope.
Our analysis show that the envelope-scale magnetic field 
is preferentially observed either aligned or perpendicular to 
the outflow direction.
Interestingly, our results suggest for the first time
a relation between the orientation of the magnetic field and 
the rotational energy of envelopes, with a larger occurrence 
of misalignment in sources where strong 
rotational motions are detected at hundreds to thousands of au scales.
We also show that the best agreement between the magnetic field and 
outflow orientation is found in sources showing no small-scale 
multiplicity and no large disks at $\sim$100 au scales.  }
{}


\authorrunning{M. Galametz et al}
\titlerunning{Polarized dust emission in solar-type Class 0 protostars}
\maketitle


\section{Introduction} 

Understanding the physical processes at work during the earliest 
phase of star formation is key to characterize the typical outcome 
of the star formation process, put constraints on the efficiency 
of the accretion/ejection mechanisms and determine the pristine properties 
of the planet-forming material. 
Class 0 objects are the youngest accreting protostars \citep{Andre1993,Andre2000}: the bulk of 
their mass resides in a dense envelope that is 
being actively accreted onto the central protostellar embryo, during a short  
main accretion phase (t $< 10^5$ yr, \citealt{Maury2011,Evans2009}). 
The physics at work to reconcile the order-of-magnitude 
difference between the large angular momentum of prestellar cores \citep{Goodman93, Caselli02} 
and the rotation properties of the young main sequence stars 
\citep[also called the ``angular momentum problem";][]{Bodenheimer1995, Belloche2013}
is still not fully understood. It was proposed that the angular momentum 
could be strongly reduced by the fragmentation of the core in multiple 
systems, or the formation of a protostellar disk and/or launch 
of a jet carrying away angular momentum. None of these solutions, however, 
seems to be able to reduce the angular momentum of the core's material 
by the required 5 to 10 orders of magnitude. 

Magnetized models have shown that the outcome of the collapse 
phase can be significantly modified through `magnetic braking' \citep{Galli2006,Li2014}. 
The role of magnetic fields on Class 0 properties have also been investigated observationally. 
A few studies have been comparing the observations of Class 0 protostars to the outcome of models, 
either analytically \citep[see e.g. ][]{Frau2011} or from 
magneto-hydrodynamical simulations \citep[see e.g. ][]{Maury2010,Maury2018}.
They showed that magnetized models of protostellar formation better 
reproduce the observed small-scale properties of the youngest accreting protostars, 
for instance the lack of fragmentation and the paucity of large (100 $<$ r$_{disk}$ $<$ 500 au) 
rotationally-supported disks \citep{Maury2010,Enoch2011,Maury2014,SeguraCox2016} observed in Class 0 objects. \\

The polarization of the thermal dust continuum emission provides an indirect probe of the 
magnetic field topology since the long axis of non-spherical or irregular dust grains 
is suggested to align perpendicular to the magnetic field direction \citep{Lazarian2007}.
Recently, the {\it Planck Space Observatory} produced an all-sky map of the 
polarized dust emission at sub-millimeter (submm) wavelengths, revealing that 
the Galactic magnetic field shows regular patterns \citep{PlanckCollaboration2016_MF}. 
Observations at better angular resolution obtained from ground-based facilities seem 
to indicate that the polarization patterns are well ordered at the scale of H{\sc ii} 
regions or molecular cloud complexes \citep{Curran2007,Matthews2009,Poidevin2010}.
We refer the reader to \citet{Crutcher2012} and references therein for a 
review on magnetic fields in molecular clouds.

Few polarization observations have been performed to characterize the magnetic field 
topology at the protostellar envelope scales where the angular momentum problem is relevant. 
Submm interferometric observations of the polarized dust continuum emission with the SMA 
\citep[Submillimeter Array;][]{Ho2004} and CARMA \citep[Combined Array for Research in 
Millimeter-wave Astronomy;][]{Bock2006} interferometers have led to the detections of 
polarized dust continuum emission toward a dozen of low-mass Class 0 protostars 
\citep[][]{Rao2009,Girart2006,Hull2014}, in particular luminous and/or massive objects 
because of the strong sensitivity limitations to detect a few percent 
of the dust continuum emission emitted by low-mass objects. 
By comparing the large ($\sim$20\arcsec) and 
small ($\sim$2.5\arcsec) scale B fields of protostars at 1mm (TADPOL survey), 
\citet{Hull2014} find that sources in which the large and small magnetic field orientations 
are consistent tend to have higher fractional polarization, which 
could be a sign of the `regulating' role of magnetic fields during the infall of the protostellar core. 
No systematic relation, however, seems to exist between the core magnetic field direction 
and the outflow orientation \citep{Curran2007,Hull2013,Zhang2014}. An hourglass shape of the 
magnetic line segments was observed in several protostellar cores 
\citep{Girart1999,Lai2002,Girart2006,Rao2009,Stephens2013}.
Probably linked with the envelope contraction pulling the magnetic field lines toward 
the central potential well, this hourglass pattern could depend strongly on the alignment 
between the core rotation axis and the magnetic field direction \citep{Kataoka2012}. 
ALMA observations are now pushing the resolution and sensitivity enough to be probing down 
to 100 au scales in closeby low-mass Class 0 protostars. In the massive ($M_\textrm{env}$$\sim$20~\msun) 
binary protostar Serpens SMM1, they have revealed a chaotic magnetic field morphology 
affected by the outflow \citep{Hull2017}. In the solar-type Class 0 B335, they have, on the 
contrary, unveiled very ordered topologies, with a clear transition 
from a large-scale B-field parallel to the outflow direction to a 
strongly pinched B in the equatorial plane \citep{Maury2018}.\\

\begin{figure}
\begin{tabular}{m{8.5cm}}
\vspace{10pt}
\hspace{-10pt} \includegraphics[width=9cm]{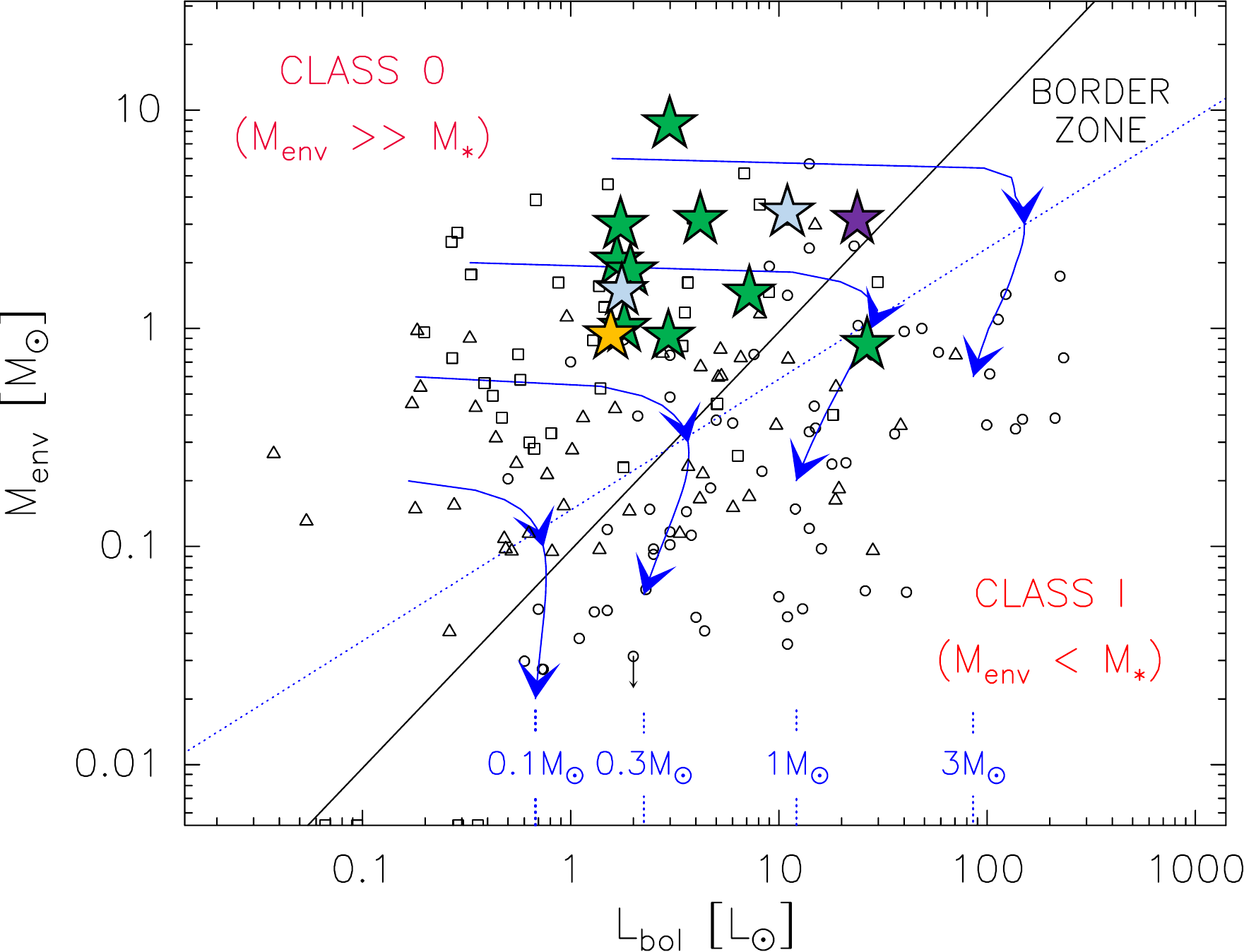} \\
\end{tabular}
\caption{Envelope mass versus bolometric luminosity 
diagram from \citet{Andre2000} and \citet{Maury2011}.
The small black squares and triangles are Class 0 
and I protostars detected in the Aquila rift, 
Ophiuchus, Perseus, and Orion regions \citep{Bontemps1996,Motte2001, Andre2000, Maury2011, Sadavoy2014}.
The lines mark the conceptual border zone between 
the Class 0 and Class I stage,
with $M_\textrm{env}$ $\propto$ $L_\textrm{bol}$ for the dashed 
line and $M_\textrm{env}$ $\propto$ $L_\textrm{bol}^{0.6}$
for the dotted line. Blue lines with arrows represent 
protostellar evolutionary tracks, computed 
for final stellar masses of 0.1, 0.3, 1 and 3\msun, 
with an evolution proceeding from top left
(objects with large envelope masses slowly accreting) to bottom  
(the envelope mass as well as the accretion luminosity are rapidly decreasing).
Our sample is overlaid with the positions of protostars in our sample: B335 in yellow, IRAS16293 in purple,
Perseus sources in green and Cepheus sources in light blue. 
It only contains low-mass Class 0 protostars, except for SVS13-A. 
The $M_\textrm{env}$ and $L_\textrm{bol}$ are from 
\citet{Massi2008} for CB230,
\citet{Launhardt2013} for B335, 
\citet{Crimier2010} for IRAS16293,
\citet{Sadavoy2014}, 
Ladjelate et al. (in prep.) 
and Maury et al. (in prep.) for the rest of the sources.}
\vspace{-5pt}
\label{Maury2011}
\end{figure}

Whether or not magnetic fields play a dominant role in regulating 
the collapse of protostellar envelopes needs to be further 
investigated from an observational perspective.
In this paper, we analyze observations of the 0.87 mm polarized dust 
continuum for a sample of 12 Class 0 (single 
and multiple systems) protostars with the SMA interferometer \citep{Ho2004}
to increase the current statistics on the polarized dust continuum
emission detected in protostellar envelopes on 750-2000 au scales. 
We provide details on the sample and data reduction steps in $\S$2. 
We study the continuum fluxes and visibility profiles and describe 
the polarization results in $\S$3. We analyze the polarization 
fraction dependencies with density and the potential causes of depolarization
in $\S$4.1. In $\S$4.2, we analyse the large-to-small-scale B orientation
and its variation with wavelength. We finally discuss the relation between 
the magnetic field orientation and both the rotation and fragmentation 
properties in the target protostellar envelopes. 
A summary of this work is presented in $\S$5.

\begin{table*}
\centering
\caption{Properties of the sample}
\begin{tabular}{cccccc}\hline
Name 			& Cloud 		& Distance $^a$	& Class		
& Multiplicity $^b$  	 	& Observation Phase Reference Center  	\\ 
\hline
B335    		& isolated			& 100 pc 			& Class 0		
& Single					& 19h37m00.9s~~+07$\degr$34$\arcmin$09\farcs6	\\
SVS13 		 	& Perseus	& 235 pc			& Class 0/I	
& Multiple (wide)			& 03h29m03.1s~~+31$\degr$15$\arcmin$52\farcs0 	\\
HH797	 		& Perseus	& 235 pc			& Class 0		
& Multiple					& 03h43m57.1s~~+32$\degr$03$\arcmin$05\farcs6	\\
L1448C		 	& Perseus	& 235 pc			& Class 0 		
& Multiple (wide)			& 03h25m38.9s~~+30$\degr$45$\arcmin$14\farcs9 	\\
L1448N		 	& Perseus	& 235 pc			& Class 0 		
& Multiple (close and wide)	& 03h25m36.3s~~+30$\degr$44$\arcmin$05\farcs4	\\
L1448-2A		& Perseus	& 235 pc			& Class 0		
& Multiple (close and wide)	& 03h25m22.4s~~+30$\degr$45$\arcmin$12\farcs2	\\
IRAS03282		& Perseus	& 235 pc			& Class 0		
& Multiple (close)			& 03h31m20.4s~~+30$\degr$45$\arcmin$24\farcs7 	\\
NGC~1333 IRAS4A 			& Perseus	& 235 pc				& Class 0
& Multiple (close)			& 03h29m10.5s~~+31$\degr$13$\arcmin$31\farcs0	\\
NGC~1333 IRAS4B 			& Perseus	& 235 pc				& Class 0
& Single					& 03h29m12.0s~~+31$\degr$13$\arcmin$08\farcs0	\\
IRAS16293 		& Ophiuchus	& 120 pc			& Class 0
& Multiple (wide)			& 16h32m22.9s~~-24$\degr$28$\arcmin$36\farcs0	\\
L1157		 	& Cepheus	& 250 pc			& Class 0 		
& Single					& 20h39m06.3s~~+68$\degr$02$\arcmin$15\farcs8	\\
CB230			& Cepheus	& 325 pc			& Class 0 		
& Binary (close)	 		& 21h17m40.0s~~+68$\degr$17$\arcmin$32\farcs0	\\
\hline
\end{tabular}
\begin{list}{}{}
\vspace{5pt}
\item[$^a$] Distance references are \citet{Knude_Hog_1998}, \citet{Stutz2008}, \citet{Hirota2008}, \citet{Hirota2011}, \citet{Looney2007} and \citet{Straizys1992}.
\item[$^b$] Multiplicity derived from 
\citet{Launhardt2004} for IRAS03282 and CB230,
\citet{Rao2009} for IRAS16293,
\citet{Palau2014} for HH797, 
\citet{Evans2015} for B335 
and from the CALYPSO survey dust continuum maps 
at 220 GHz (PdBI, 
see {\url{http://irfu.cea.fr/Projets/Calypso/}}), \citet{Maury2014}) for the rest of the sources. 
\end{list}
\label{SourceCharacteristics}
\end{table*}


\section{Observations and data reduction}

\subsection{Sample description}

Our sample contains 12 protostars that include single objects as well 
as common-envelope multiple systems and separate envelopes multiple 
system \citep[][following the classification proposed by]{Looney2000}. 
Figure~\ref{Maury2011} shows the location of the protostars 
in an envelope mass ($M_\textrm{env}$) versus bolometric luminosity 
($L_\mathrm{bol}$) diagram \citep{Bontemps1996,Maury2011}. 
With most of the object mass contained in the envelope, our selection 
is a robust sample of low-mass Class 0 protostars. Only SVS13-A can 
be classified as a Class I protostar. Details on each object are provided in 
Table~\ref{SourceCharacteristics} and in appendix.

\subsection{The 0.87mm SMA observations}

Observations of the polarized dust emission of 9 low-mass 
protostars at 0.87mm were obtained using the SMA 
(Projects 2013A-S034 and 2013B-S027, 
PI: A. Maury) in the compact and subcompact configuration.  
To increase our statistics, we also include SMA observations from three
additional sources from Perseus (NGC~1333 IRAS4A and IRAS4B) and Ophiuchus (IRAS16293)
observed in 2004 and 2006 (Projects 2004-142 and 2006-09-A026, PI: R. Rao; 
Project 2005-09-S061, PI: D.P. Marrone). The observations of NGC~1333 IRAS4A and IRAS16293 
are presented in \citet{Girart2006} and \citet{Rao2009} respectively.
We refer the reader to \citet{Marrone2006} and \citet{Marrone2008} for 
a detailed description of the SMA polarimeter system, 
but provide a few details on the SMA and the polarization design here. 
The SMA has eight antennas. Each optical path is equipped 
with a quarter-wave plate (QWP), an optical element 
that adds a 90$\degr$ phase delay between orthogonal linear 
polarizations and is used to convert the linear into circular polarization. 
The antennas are switched between polarizations 
(QWP are rotated at various angles) in a coordinated temporal sequence to 
sample the various combinations of circular polarizations on each baseline. 
The 230, 345 and 400 band receivers are installed in all 
eight SMA antennas. Polarization can be measured in single-receiver 
polarization mode as well as in dual-receiver mode when two 
receivers with orthogonal linear polarizations are tuned 
simultaneously. In this dual-receiver mode, all correlations (the 
parallel-polarized RR and LL and the cross-polarized RL and LR; 
with R and L for right circular and left circular respectively) 
are measured at the same time. Both polarization modes were used in 
our observations. This campaign was used to partly commission the dual-receiver full 
polarization mode for the SMA. A fraction of the data was 
lost during this period due to issues with the correlator software. 
Frequent observations of various calibrators were interspersed to ensure 
that such issues were detected as early as possible to minimize data loss.

\begin{table*}
\caption{Details on the observations}
\begin{tabular}{ccccccc}
\hline
\hline
\vspace{-5pt}
&\\
Date & Mode $^a$ & Flux 	& Bandpass 	& Gain 	 & Polarization & Antenna \\
	 & 			 & Calib.   & Calib. 	& Calib. & Calib. 		& used \\

\vspace{-5pt}
&\\
\hline
\vspace{-5pt}
&\\
Dec 05 2004		& Single Rx - Single BW & Ganymede 		& 3c279	
& 3c84 									& 3c279 		& 1, 2, 3, 5, 6, 8   \\
Dec 06 2004		& Single Rx - Single BW & Ganymede 		& 3c279
& 3c84 									& 3c279 		& 1, 2, 3, 5, 6, 8   \\
April 08 2006	& Single Rx - Single BW & Callisto		& 3c273
& 1517-243, 1622-297					& 3c273			& 1, 2, 3, 5, 6, 7, 8 \\
Dec 23 2006		& Single Rx - Single BW & Titan			& 3c279
& 3c84									& 3c279			& 1, 2, 3, 4, 5, 6, 7 \\
Aug 26 2013    	& Single Rx - Double BW	& Neptune 		& 3c84	    	
& 1927+739, 0102+584					& 3c84  		& 2, 4, 5, 6, 7, 8 \\
Aug 31 2013     & Dual Rx - Autocorrel	& Callisto		& 3c84			
& 1751+096, 1927+739, 0102+584, 3c84	& 3c84 			& 2, 4, 5, 6, 7, 8 \\ 
Sept 1 2013      & Dual Rx - Autocorrel	& Callisto		& 3c84			
& 1927+739, 0102+584					& 3c84 			& 2, 4, 5, 6, 7, 8 \\
Sept 2 2013      & Dual Rx - Autocorrel	& Callisto		& 3c454.3		
& 1927+739, 0102+584,3c84				& 3c84 			& 2, 4, 5, 6, 7, 8 \\
Sept 7 2013	& Dual Rx - Full pol. 		& Callisto   	& 3c454.3 		
& 3c84, 3c454.3	 						& 3c454.3 		& 2, 4, 5, 6, 7, 8 \\ 
Dec 7 2013	& Dual Rx - Full pol.		& Callisto		& 3c84			
& 3c84									& 3c84 			& 2, 4, 5, 6, 7, 8 \\ 
Feb 24 2014	& Dual Rx - Full pol. 		& Callisto   	& 3c279   		
& 3c84, 3c279, 0927+390					& 3c279 		& 1, 2, 4, 5, 6, 7 \\ 
Feb 25 2014	& Dual Rx - Full pol. 		& Callisto   	& 3c84	  		
& 3c84									& 3c84 			& 1, 2, 4, 5, 6, 7 \\
\vspace{-5pt}
&\\
\hline
\end{tabular}
\begin{list}{}{}
\item[$^a$] Rx = Receiver; BW = bandwidth.
\end{list}
\label{Observations}
\end{table*}

\begin{table}
\centering
\caption{Characteristics of the SMA maps}
\begin{tabular}{ccccccc}
\hline
\hline
\vspace{-5pt}
&\\
Name 	 	& 	\multicolumn{2}{c}{Synthesized beam} 	&& \multicolumn{3}{c}{rms (mJy/beam)} \\
\vspace{-5pt}
&\\
\cline{5-7} 
\vspace{-5pt}
&\\
&&&& I $^{a}$  	&	Q 	& U			\\ 	
\vspace{-5pt}
&\\
\hline
\vspace{-5pt}
&\\
B335    	& 	\hspace{10pt}4\farcs8$\times$4\farcs1 	& \hspace{-10pt}(84$\degr$)		
&& 4.6 			&	4.0			& 4.3	\\
SVS13 		& 	\hspace{10pt}4\farcs9$\times$4\farcs3 		& \hspace{-10pt}(-88$\degr$)		
&& 7.0 			&	3.0			& 3.1	\\
HH797		& 	\hspace{10pt}4\farcs7$\times$4\farcs3 		& \hspace{-10pt}(89$\degr$)		
&& 4.2 			&	1.9			& 1.8	\\
L1448C		& 	\hspace{10pt}4\farcs7$\times$4\farcs4 		& \hspace{-10pt}(87$\degr$)		
&& 4.7 			&	3.6			& 3.7	\\ 
L1448N		& 	\hspace{10pt}2\farcs6$\times$2\farcs4 		& \hspace{-10pt}(55$\degr$)		
&& 5.5 			&	3.3			& 3.0	\\
L1448-2A	& 	\hspace{10pt}2\farcs5$\times$2\farcs2 	& \hspace{-10pt}(49$\degr$)		
&& 1.8 			&	1.6			& 1.6	\\ 
IRAS03282	& 	\hspace{10pt}4\farcs8$\times$4\farcs3 		& \hspace{-10pt}(88$\degr$)		
&& 5.3 			&	3.3			& 3.5	\\ 
NGC~1333 IRAS4A		& 	\hspace{10pt}2\farcs0$\times$1\farcs4 		& \hspace{-10pt}(-46$\degr$) 
&& 18.7 			& 7.5 			& 8.0 	\\ 
NGC~1333 IRAS4B		& 	\hspace{10pt}2\farcs2$\times$1\farcs7 		& \hspace{-10pt}(10$\degr$) 
&& 9.0 			& 2.2 			& 2.3	\\ 
IRAS16293	& 	\hspace{10pt}3\farcs0$\times$1\farcs8 		& \hspace{-10pt}(-6$\degr$) 
&& 10.4 			& 5.6 			& 6.0	\\ 
L1157		& 	\hspace{10pt}5\farcs9$\times$4\farcs6 		& \hspace{-10pt}(1$\degr$)		
&& 8.8 			& 6.4			& 7.6	\\
CB230		& 	\hspace{10pt}6\farcs1$\times$4\farcs2 		& \hspace{-10pt}(7$\degr$)		
&& 5.5 			& 4.11			& 4.0	\\
\vspace{-5pt}
&&&&\\
\hline
\end{tabular}
\begin{list}{}{}
\item[$^a$] Stokes I rms noise values reported here come from the maps obtained after self-calibration.
\vspace{-5pt}
\end{list}
\label{SMAmaps}
\vspace{10pt}
\end{table}

\subsection{Data calibration and self-calibration}

We perform the data calibration in the IDL-based software \texttt{MIR} 
(Millimeter Interferometer Reduction) and the data reduction package 
\texttt{MIRIAD}\footnote{https://www.cfa.harvard.edu/sma/miriad/}. The calibration includes: an initial flagging of high 
system temperatures $T_\mathrm{sys}$ and other wrong visibilities, a bandpass 
calibration, a correction of the cross-receiver delays, 
a gain and a flux calibration. The various calibrators observed for each of 
these steps and the list of antennae used for the observations 
are summarized in Table~\ref{Observations}.  
The polarization calibration was performed in \texttt{MIRIAD}.
Quasars were observed to calculate the leakage terms:
the leakage amplitudes (accuracy: $\sim$0.5\%) are $<$2\% in the two sidebands 
for all antenna except for 7 than can reach a few percent. 
Before the final imaging, we use an iterative procedure to self-calibrate the 
Stokes I visibility data.

\subsection{Deriving the continuum and polarization maps}

The Stokes parameters describing the polarization state are defined as
\begin{equation}
   \overrightarrow{S} = \begin{bmatrix}
          	 I \\
          	 Q \\
          	 U \\
	 	 	 V \\
         \end{bmatrix}
\end{equation}

\noindent with Q and U the linear polarization 
and V the circular polarization.
We use a robust weighting of 0.5 to transform the visibility data into a dirty map. 
The visibilities range from 5 to 30k$\lambda$ in 7 sources
(B335, SVS13, HH797, L1448C, IRAS03282, L1157 and CB230)
and 10 to 80k$\lambda$ for the other 5 sources. Visibilities
beyond 80k$\lambda$ are available for NGC~1333 IRAS4A but not used
to allow an analysis of comparable scales for all our sources.
The Stokes I dust continuum emission maps are shown in 
Fig.~\ref{StokesI_Borientation}. The Stokes Q and U maps are shown
in appendix (Fig.~\ref{PolaMaps}). Their combination probes the polarized component of the dust emission.
The synthesized beams and rms of the cleaned maps are 
provided in Table~\ref{SMAmaps}.
We note that because of unavoidable missing flux and 
the dynamic-range limitation, the rms of the Stokes I 
maps are systematically higher than those of Stokes Q 
and U. Following the self-calibration procedure, the 
rms of the continuum maps have decreased by 10 to 45\%. \\

The polarization intensity (debiased), fraction and angle are derived 
from the Stokes Q and U as follow:

\vspace{-2pt}
\begin{equation}
P_{i} = \sqrt[]{Q^2+U^2 - \sigma_{Q,U}^2},
\vspace{-2pt}
\end{equation}

\begin{equation}
p_{frac} = P_{i}~/~I,
\vspace{-2pt}
\end{equation}

\begin{equation}
PA = 0.5 \times\ arctan(U/Q).
\end{equation}

\noindent with $\sigma$$_{Q,U}$ the average rms of the Q and U maps.
We apply a 5-$\sigma$ cutoff on Stokes I and 3-$\sigma$ cutoff on 
Stokes Q and U to only select locations where polarized emission is 
robustly detected. The polarization intensity and fraction maps are 
provided in appendix (Fig.~\ref{PolaMaps}).

\begin{figure*}
\vspace{20pt}
\hspace{-10pt}\includegraphics[width=19cm]{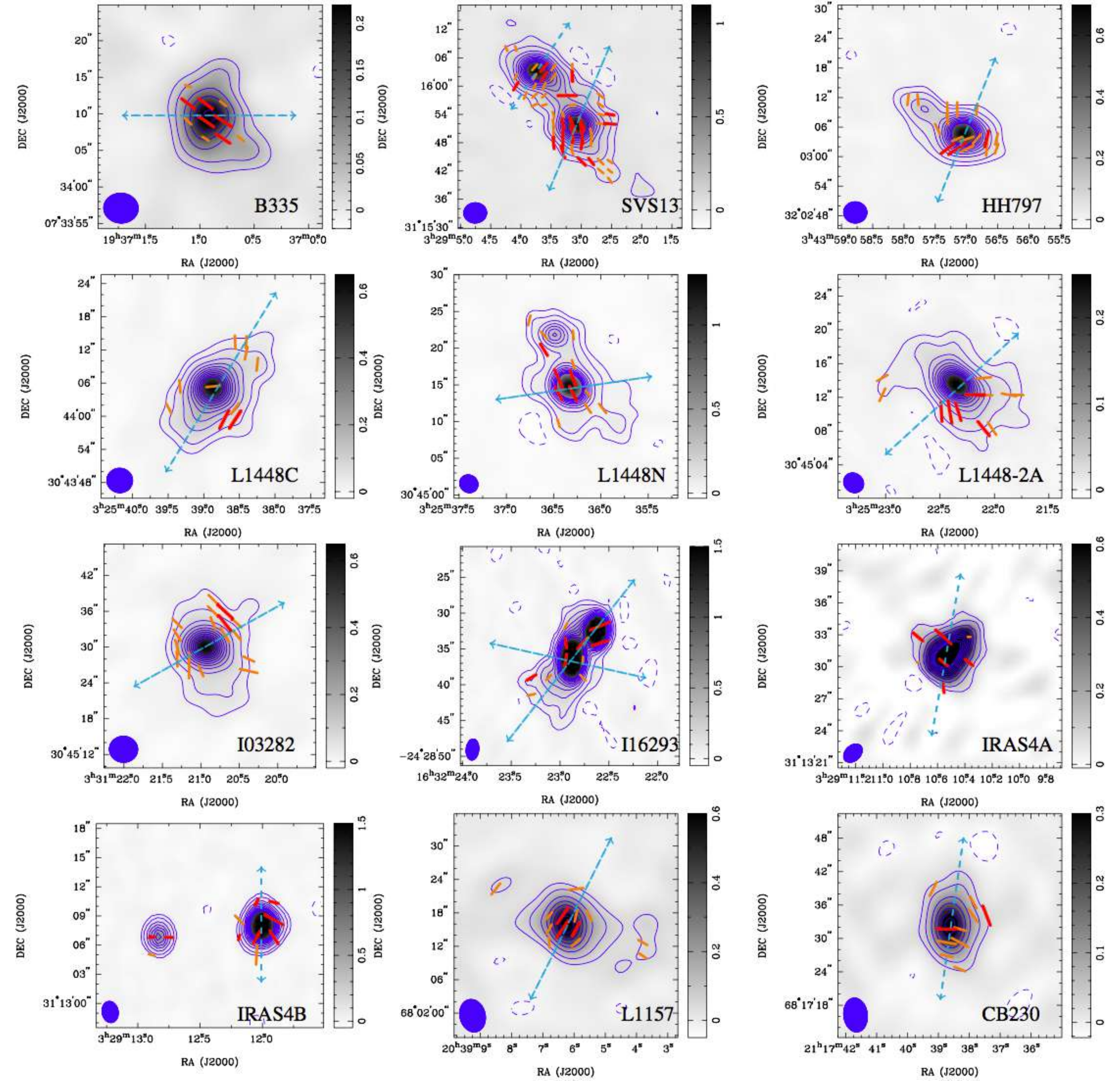} 
\caption{The SMA 850 \mic\ Stokes I continuum maps. 
Color scales are in mJy/beam. 
Contours at [-3,5,10,20,30,40,50,60,70,80,90,100] $\sigma$ appear in blue. 
The filled ellipses on the lower left corner indicate the 
synthesized beam of the SMA maps. Their sizes are reported in Table~\ref{SMAmaps}.
The blue arrows indicate the outflow direction. B orientation 
(derived from the polarization angles assuming a 90$\degr$ 
rotation) are overlaid on the Stokes I map. 
Red bars show $>$3-$\sigma$ detections. 
We also indicate the $>$2-$\sigma$ detections with orange 
bars but these detections are not used for this analysis. }
\label{StokesI_Borientation}
\vspace{10pt} 
\end{figure*}

\begin{table*}
\centering
\caption{0.87mm integrated flux densities, source sizes, polarization intensities and fractions.}
\begin{tabular}{cccccccc}
\hline
\hline
\vspace{-5pt}
&\\

Name 	 	& 	\multicolumn{5}{c}{On the 0.87mm reconstructed maps} 
&&  From the visibility amplitude curve \\ 
\vspace{-5pt}
&\\
\cline{2-6}
\cline{8-8}
\vspace{-5pt}
&\\
&	Flux density$^a$ 	&	Mass	&	Peak intensity		
& Peak P$_i$ $^b$	&	p$_{frac}$ $^c$  &&	0.87mm Flux density	 \\
& (Jy) & (\msun) & (Jy/beam) & (mJy/beam) & (\%) && (Jy) 		 \\ 
\vspace{-5pt}
&\\
\hline
\vspace{-5pt}
&\\
B335    		&   0.21$\pm$0.04 	& 0.02	& 0.24		
& 25.3			&	10.5			&&     0.51$\pm$0.05	\\
SVS13 			&	1.40$\pm$0.28	& 0.29	& 1.06		
& 14.8			&   2.5				&&	 1.54$\pm$0.15$^d$	\\
HH797			&	0.47$\pm$0.09	& 0.10	& 0.71		
& 5.7			&	3.0				&&	 0.94$\pm$0.09		\\
L1448C			& 	0.57$\pm$0.11	& 0.11	& 0.65		
& 12.5			&	-				&&	 1.00$\pm$0.01		\\ 
L1448N			& 	1.50$\pm$0.30	& 0.31	& 1.34		
& 20.3			&	2.7				&&	 1.93$\pm$0.19$^d$	\\ 
L1448-2A		& 	0.34$\pm$0.07	& 0.07	& 0.27		
& 5.7			&	-				&&	 0.48$\pm$0.05		\\ 
IRAS03282		& 	0.44$\pm$0.09	& 0.09	& 0.65		
& 9.9			&	-				&&	 1.17$\pm$0.12		\\ 
NGC~1333 IRAS4A 			&	2.38$\pm$0.48	& 0.50	& 2.57 
& 77.3 			& 	8.8 			&&	 6.79$\pm$0.68		\\
NGC~1333 IRAS4B 			&	1.42$\pm$0.28	& 0.30	& 2.89 
& 14.9 			&	4.7				&&	 4.10$\pm$0.41		\\
IRAS16293 		&	4.81$\pm$0.96	& 0.26	& 3.83 
& 39.2 			&	1.7				&&	 5.96$\pm$0.60$^d$ \\
L1157			& 	0.43$\pm$0.09	& 0.10	& 0.71		
& 31.4			&	4.5				&&	 1.07$\pm$0.11		\\
CB230			& 	0.22$\pm$0.04	& 0.09	& 0.37		
& 14.3			&	6.4				&&	 0.58$\pm$0.06		\\
\vspace{-5pt}
&&&\\
\hline
\end{tabular}
\begin{list}{}{}
\item[$^a$] Flux density estimated within the central 6000 au for SVS13, 
3000 au for other sources.
\item[$^b$] We remind the reader that not all the sources 
have an intensity peak co-spatial with the polarized intensity peak.
\item[$^c$] Polarization fraction defined as the unweighted 
ratio between the median polarization over total flux.
\item[$^d$] The visibility data allows us to separate the two components
of wide binaries. The fluxes reported here are those of the object located
at the phase center, e.g. L1448N-B, SVS13-B and IRAS16293-A.
\end{list}
\label{Results}
\vspace{-10pt}
\end{table*}

\begin{table*}
\centering
\caption{Comparison with SCUBA observations from \citet{DiFrancesco2008}}
\begin{tabular}{cccccccc}
\hline
\hline
\vspace{-5pt}
&\\
Name & \multicolumn{3}{c}{SCUBA 850 \mic} & & \multicolumn{3}{c}{Comparison SMA / SCUBA}\\
\cline{2-4}
\cline{6-8} \\
& Peak Intensity & Effective radius & Flux density && 
Rescaled Peak Intensity$^a$ 	& 
Peak intensity Ratio  & Total Flux Ratio 	\\
&(Jy/beam) & (\arcsec) & (Jy) && (Jy/SMA beam) 
& SMA/SCUBA & SMA/SCUBA \\
\vspace{-5pt}
&\\
\hline
\vspace{-5pt}
&\\
B335		& 1.45$\pm$0.14 & 34.5 	& 2.38$\pm$0.36 && 
0.46$\pm$0.07 	& 0.52$\pm$0.09 & 0.09$\pm$0.02 \\
HH797		& 1.76$\pm$0.18 & 43.2 	& 4.67$\pm$0.70	&& 
0.57$\pm$0.09 	& 1.24$\pm$0.22 & 0.12$\pm$0.03 \\
L1448C  	& 2.37$\pm$0.24 & 43.5	& 4.15$\pm$0.62	&& 
0.77$\pm$0.12 	& 0.84$\pm$0.15 & 0.14$\pm$0.03 \\
L1448N 		& 5.46$\pm$0.55 & 49.3	& 10.18$\pm$1.53&& 
0.96$\pm$0.14 	& 1.27$\pm$0.23 & 0.15$\pm$0.04 \\
L1448-2A	& 1.41$\pm$0.14 & 36.8	& 2.24$\pm$0.34	&& 
0.24$\pm$0.04 	& 0.97$\pm$0.17 & 0.15$\pm$0.04 \\
IRAS03282 	& 1.30$\pm$0.13 & 45.2	& 2.28$\pm$0.34	&& 
0.42$\pm$0.06 	& 1.54$\pm$0.28 & 0.18$\pm$0.04 \\
NGC~1333 IRAS4A 		& 11.4$\pm$0.11 & 38.5 	& 14.4$\pm$2.16	&&
1.40$\pm$0.21	& 1.83$\pm$0.33 & 0.17$\pm$0.04\\
NGC~1333 IRAS4B 		& 5.46$\pm$0.55 & 40.7 	& 8.90$\pm$1.34	&&
0.75$\pm$0.11	& 3.83$\pm$0.69$^b$	& 0.16$\pm$0.04 \\
IRAS16293 	& 20.2$\pm$0.20	& 56.8 	& 29.5$\pm$4.43	&&
3.43$\pm$0.51	& 1.12$\pm$0.20	& 0.16$\pm$0.04 \\
L1157		& 1.58$\pm$0.16 & 42.4	& 2.46$\pm$0.37 && 
0.59$\pm$0.09 	& 1.20$\pm$0.22 & 0.17$\pm$0.04 \\
CB230 		& 1.22$\pm$0.12 & 43.2	& 2.35$\pm$0.35	&& 
0.45$\pm$0.07 	& 0.82$\pm$0.15 & 0.09$\pm$0.02 \\
\vspace{-5pt}
&\\
\hline
\end{tabular}
\begin{list}{}{}
\item[$^a$] In order to estimate what the peak intensity of the SCUBA maps 
would correspond to in a SMA beam, we considered that the envelope follows a 
r$^{-2}$ density profile, thus that the intensity would scale in 1/r. The SMA 
beam for each source is provided in Table~\ref{SMAmaps}. The SCUBA FWHM at 
850 \mic\ is 14\arcsec. SVS13 is not included in the table because SVS13-A 
and B are not resolved by SCUBA.
\item[$^b$] This high value can be explained by the very flat intensity profile
of the source that is barely resolved in our SMA observations (see Fig.~3).
\end{list}
\label{SCUBA}
\end{table*}

\begin{figure*}
\begin{tabular}{m{5.5cm}m{5.5cm}m{5.5cm}}
 \includegraphics[width=6cm]{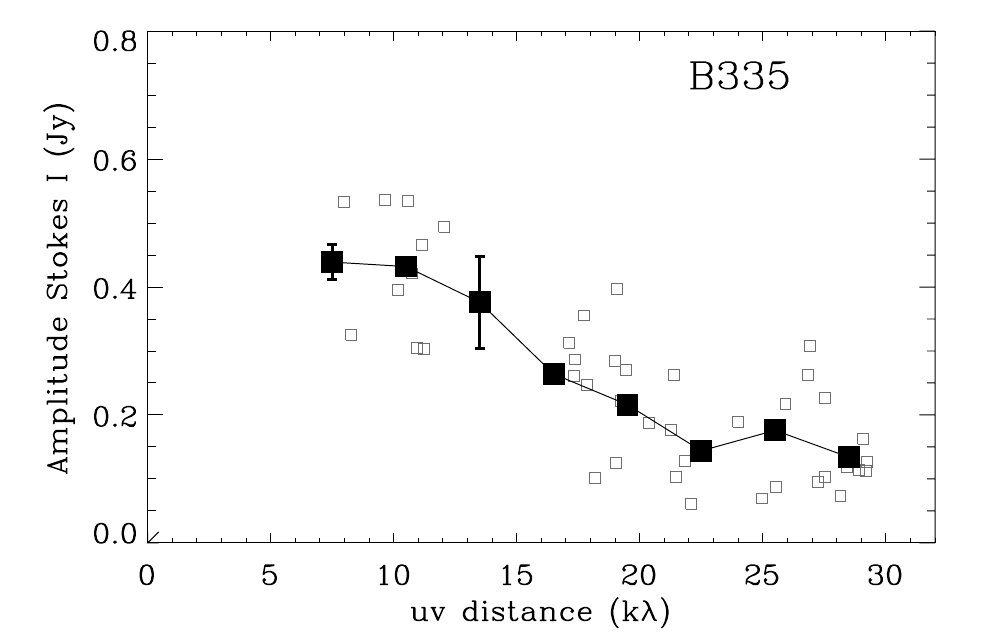} &
 \includegraphics[width=6cm]{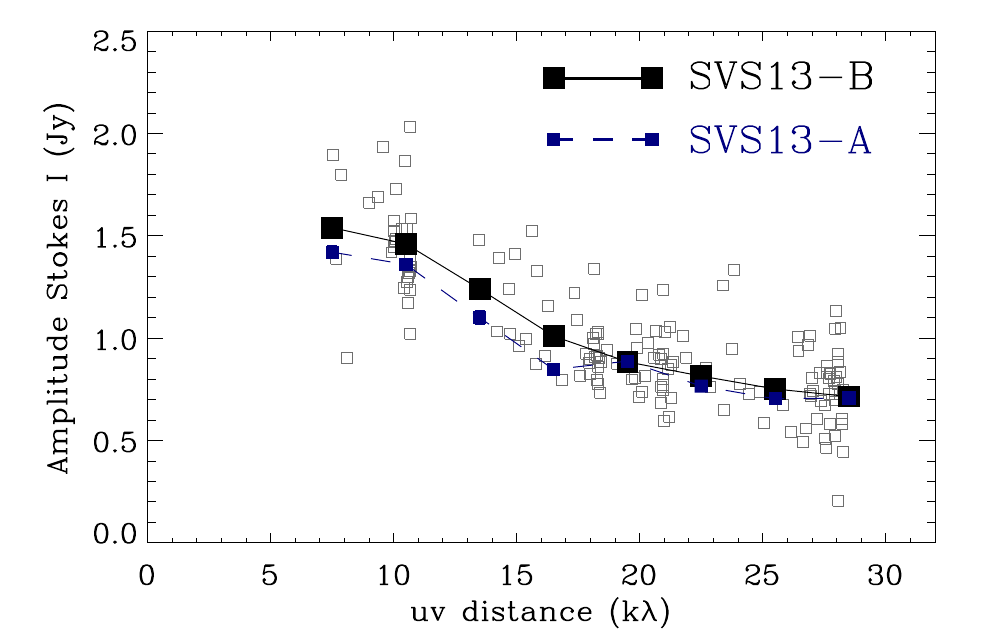} &
 \includegraphics[width=6cm]{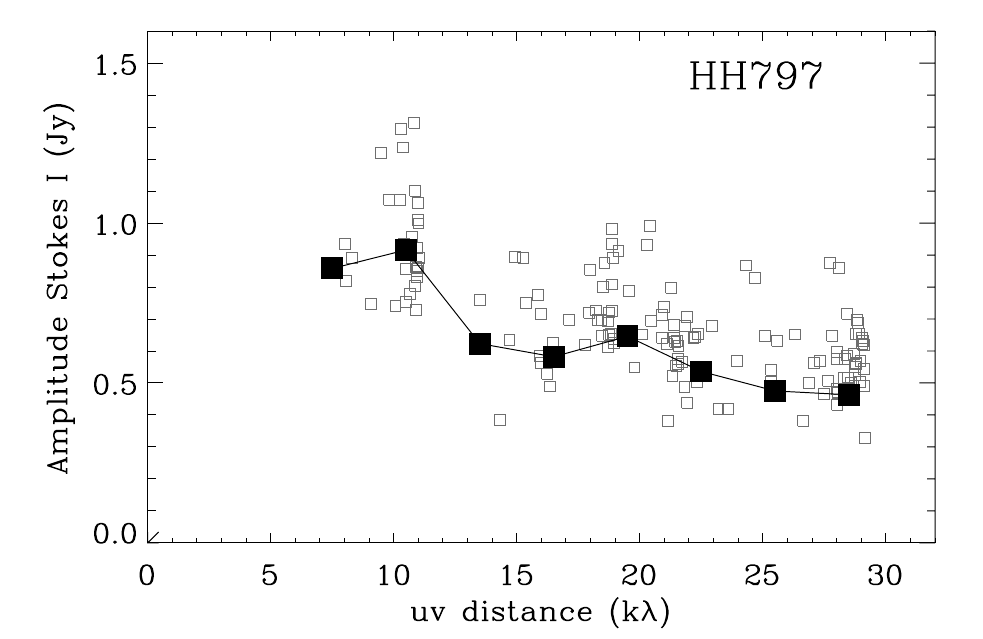} \\
 \vspace{-5pt} \includegraphics[width=6cm]{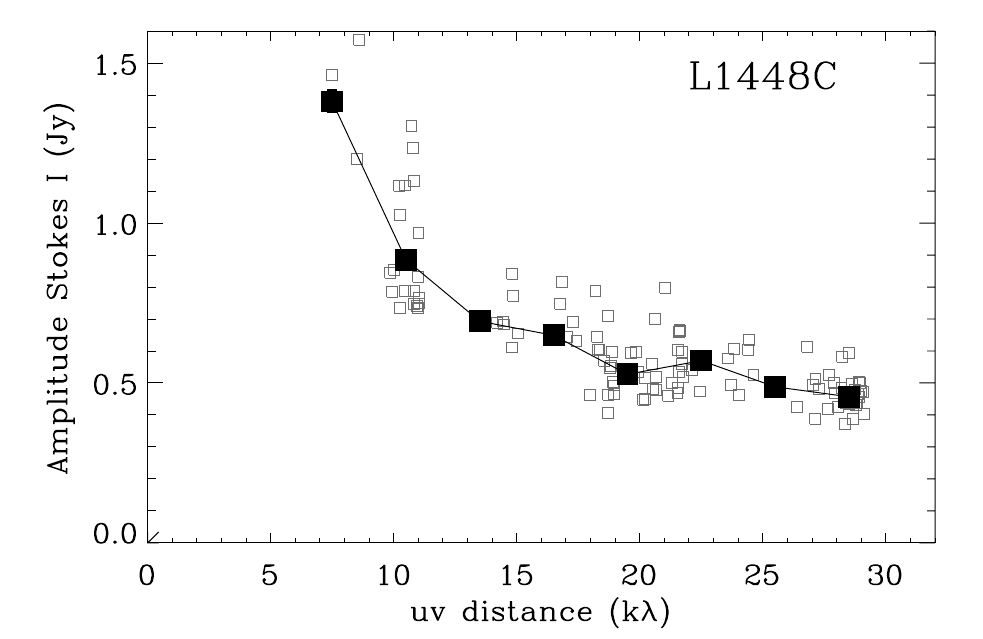} &
 \vspace{-5pt} \includegraphics[width=6cm]{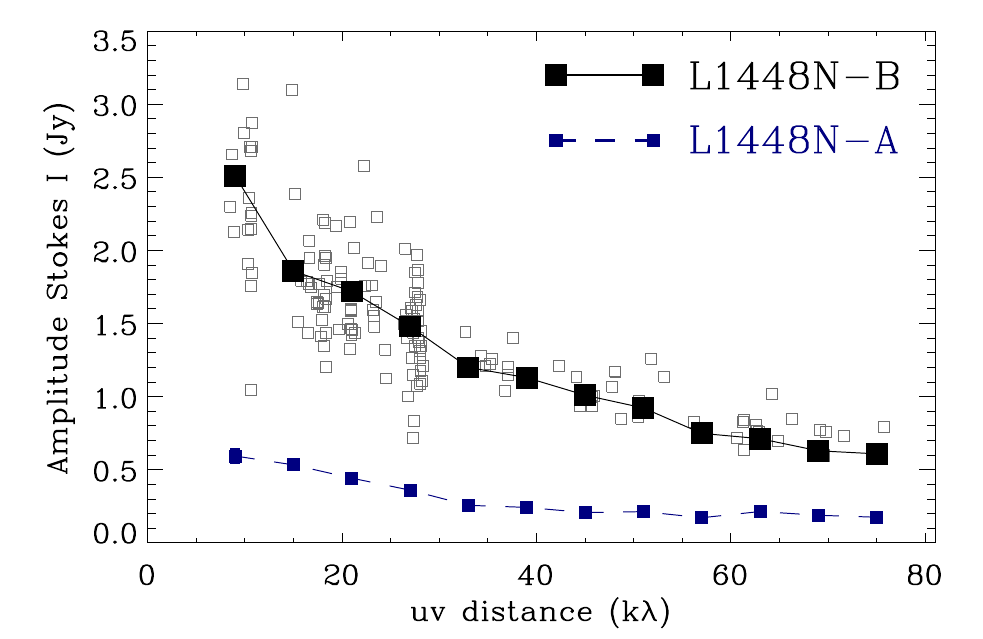} &
 \vspace{-5pt} \includegraphics[width=6cm]{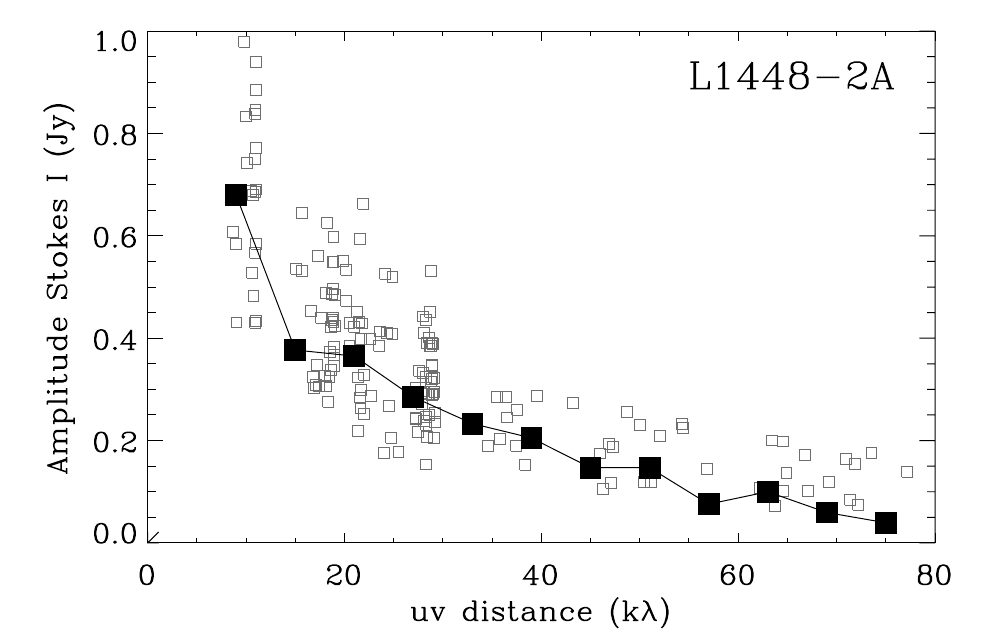} \\
 \vspace{-5pt} \includegraphics[width=6cm]{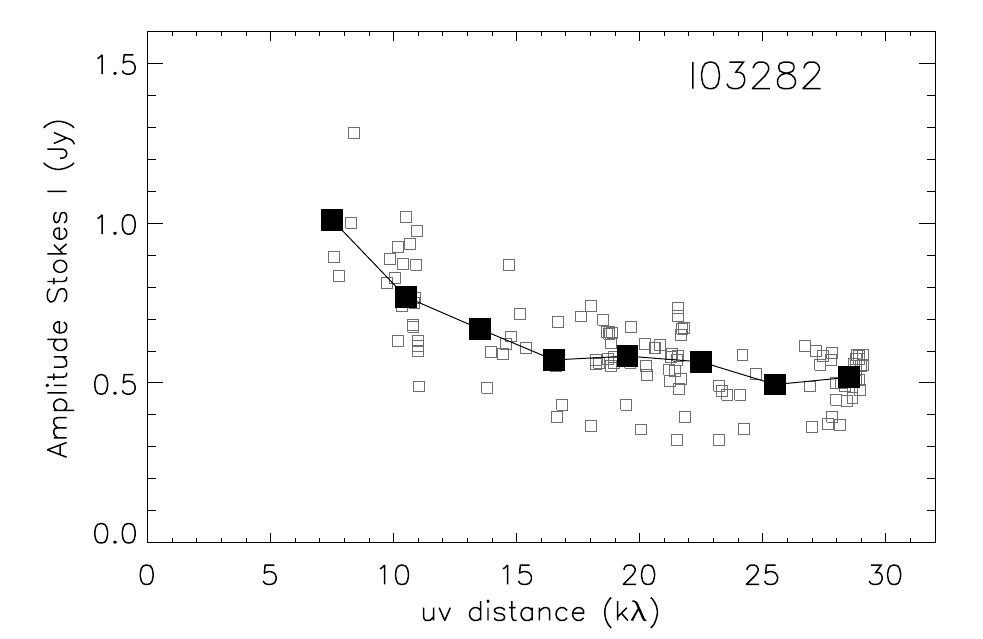} &
  \vspace{-5pt} \includegraphics[width=6cm]{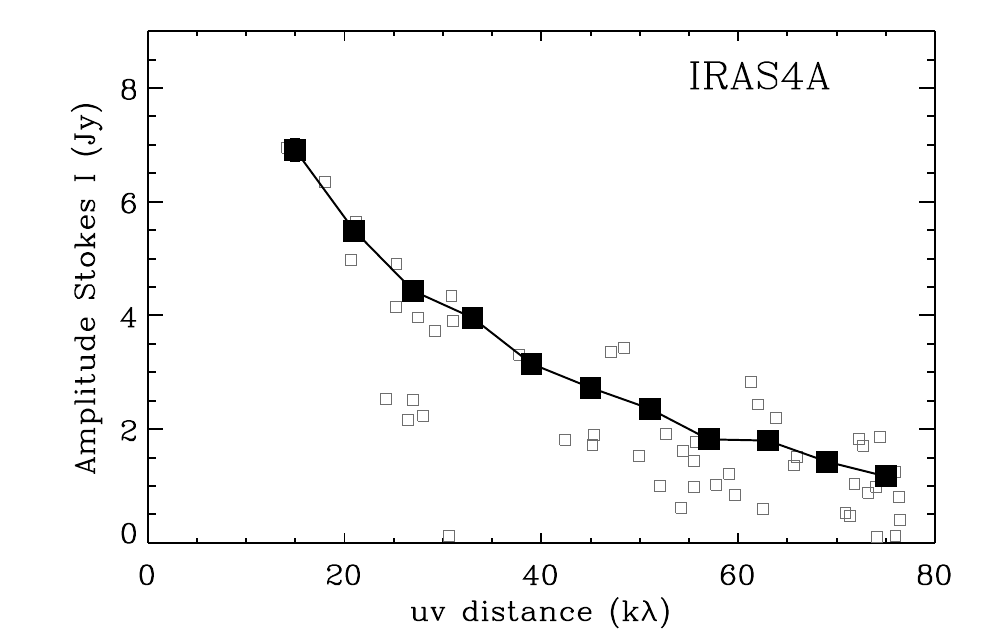} &
  \vspace{-5pt} \includegraphics[width=6cm]{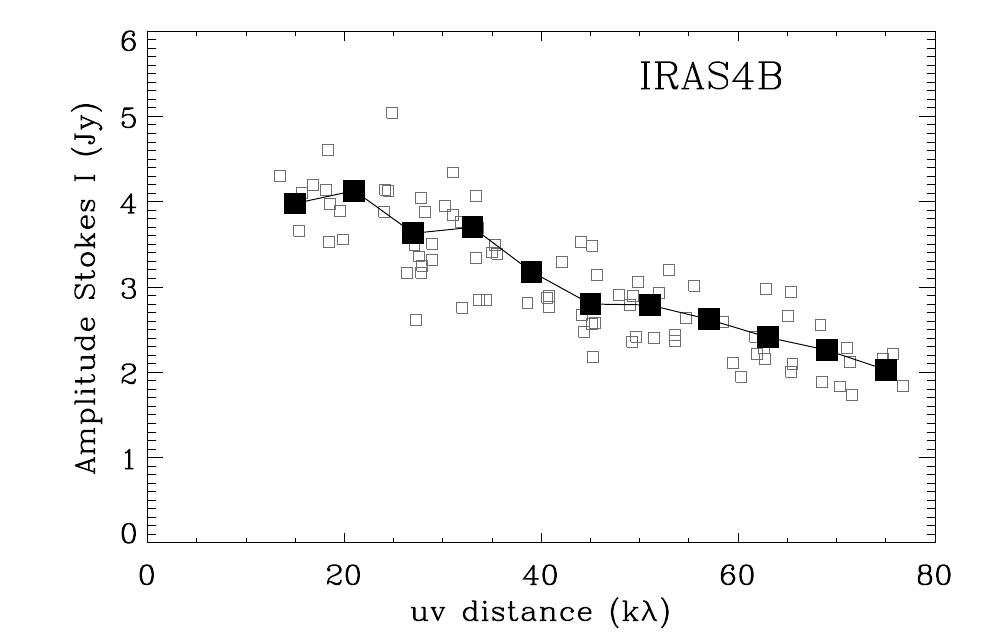} \\
 \vspace{-5pt} \includegraphics[width=6cm]{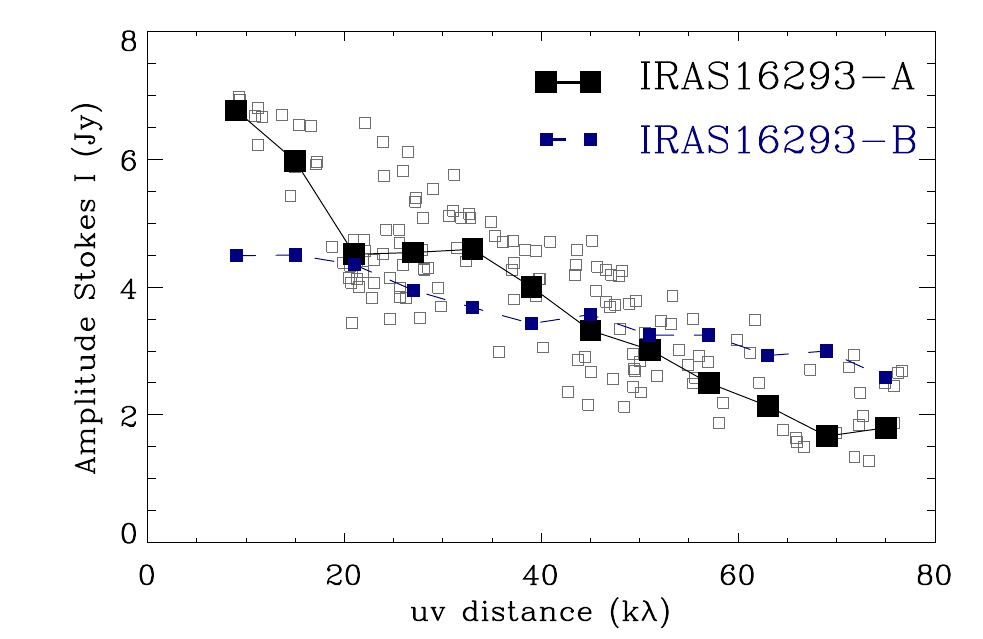} &
 \vspace{-5pt} \includegraphics[width=6cm]{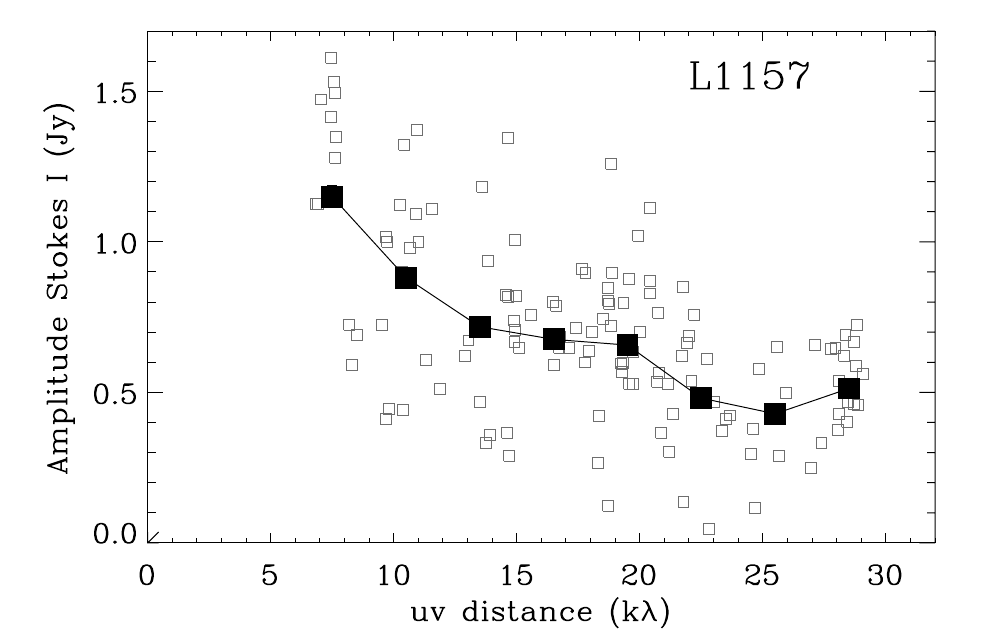} &
 \vspace{-5pt} \includegraphics[width=6cm]{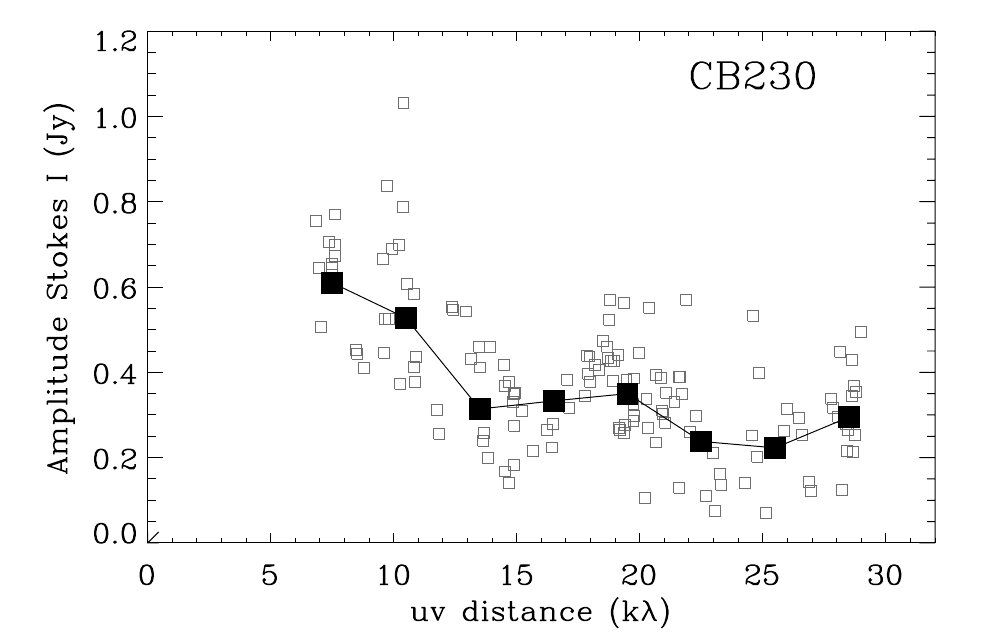} \\
\end{tabular}
\caption{Stokes I visibility amplitude as a function of uv distance 
(in units of k$\lambda$). The open squares are the amplitudes of the 
visibilities channel-averaged and time-averaged every 120s. 
The black squares are derived averaging these amplitudes of the 
visibilities in 6 k$\lambda$ bins for L1448N, L1448-2A, NGC~1333 IRAS4A, IRAS4B and IRAS16293 and 3 k$\lambda$ bins for the other sources. 
For the wide-binary L1448N, SVS13 and IRAS16293, we model and isolate 
the two sources separately. For IRAS4B, we also isolate and
remove NGC~1333 IRAS4B2 from the visibility data.}
\label{Visibilities_StokesI}
\end{figure*}


\section{Results}

\subsection{0.87mm continuum fluxes}

The 0.87mm dust continuum maps are presented in Fig.~\ref{StokesI_Borientation} 
(the Stokes I dust continuum emission is described in appendix A).
We overlay the outflow direction (references from the literature can be 
found in Table~\ref{Angles}). We provide the peak intensities and the integrated 0.87mm 
flux densities in Table~\ref{Results}. 
We also provide the associated masses calculated using the relation from 
\citet{Hildebrand1983}:

\begin{equation}
M=\frac{S_{\nu}~\mathcal{D}~d^2}{\kappa_{\nu}~B_{\nu}(T)}
\end{equation}

\noindent with 
S$_{\nu}$ the flux density, 
$\cal{D}$ the dust-to-gas mass ratio assumed to be 0.01,
d the distance to the protostar,
$\kappa_{\nu}$ the dust opacity (tabulated in \citet{Ossenkopf1994}, 1.85 cm$^2$~g$^{-1}$ 
at 0.87mm) and B$_{\nu}$(T) the Planck function. We choose a dust temperature of 25K,
coherent with that observed at 1000 au in IRAS16293 by \citet{Crimier2010}. 
A dust temperature of 50K would only decrease the gas mass reported in 
Table~\ref{Results} by a factor of $\sim$2. \\

The sensitivity of SMA observations decreases outside the 
primary beam (i.e. 34\arcsec) and the coverage of the shortest 
baselines is not complete:
we thus expect that only part of the total envelope flux will 
be recovered by our SMA observations, especially for the closest sources. 
To estimate how much of the extended flux is missing, we 
compare the SMA fluxes with those obtained with the 
SCUBA single-dish instrument. 
\citet{DiFrancesco2008} provide a catalogue of 0.87mm 
continuum fluxes and peak intensities for a large range of 
objects observed with SCUBA at 450 and 850 \mic, including our sources. 
The SCUBA integrated flux uncertainties are dominated by the 15\% calibration 
uncertainties while the absolute flux uncertainty for the SMA 
is $<$10\%. Results are summarized in Table~\ref{SCUBA}. 
We find that the SMA total flux, integrated in the reconstructed cleaned
maps of the sources, accounts for 9 -- 18\% of the SCUBA 
total fluxes. This is consistent with the results 
of the PROSAC low-mass protostar survey for which only 
10 -- 20\% of the SCUBA flux is recovered with the SMA 
\citep{Jorgensen2007}.
As far as the dust continuum peak flux densities are concerned, 
if we rescale the SCUBA peak flux densities to the one expected 
in the SMA beam, assuming that the intensity scales with radius 
(i.e. a density dependence $\rho~\propto~$r$^{-2}$), we find 
that SCUBA and SMA peak flux densities are consistent with each 
other. This means that most of the envelope flux is recovered at SMA beam scales. 
The largest discrepancies appear for B335, NGC~1333 IRAS4A and NGC~1333 IRAS4B.
For B335, the SCUBA value is twice that derived with SMA. This 
indicates that for this source, one of the closest of our sample, 
part of the SMA continuum flux might be missing even at the peak of 
continuum emission, and the polarization fraction could be lower 
than that determined from the SMA observations ($\S$3). 
For NGC~1333 IRAS4A and IRAS4B, the inverse is observed: the SMA peak value is
2 and 4 times larger than the rescaled SCUBA. The two sources are
the most compact envelopes of the sample \citep[][]{Looney2003,Santangelo2015}. The 
difference is thus
probably due to beam dilution of the compact protostellar
cores in the large SCUBA beam as well as possible contamination
of the SCUBA fluxes due to the fact that NGC~1333 IRAS4A and B are both located within
a long (1\arcmin$\times$40\arcsec) filamentary cloud
extending southeast-northwest \citep{Lefloch1998}.

\begin{figure}
\vspace{10pt}
\includegraphics[width=9cm]{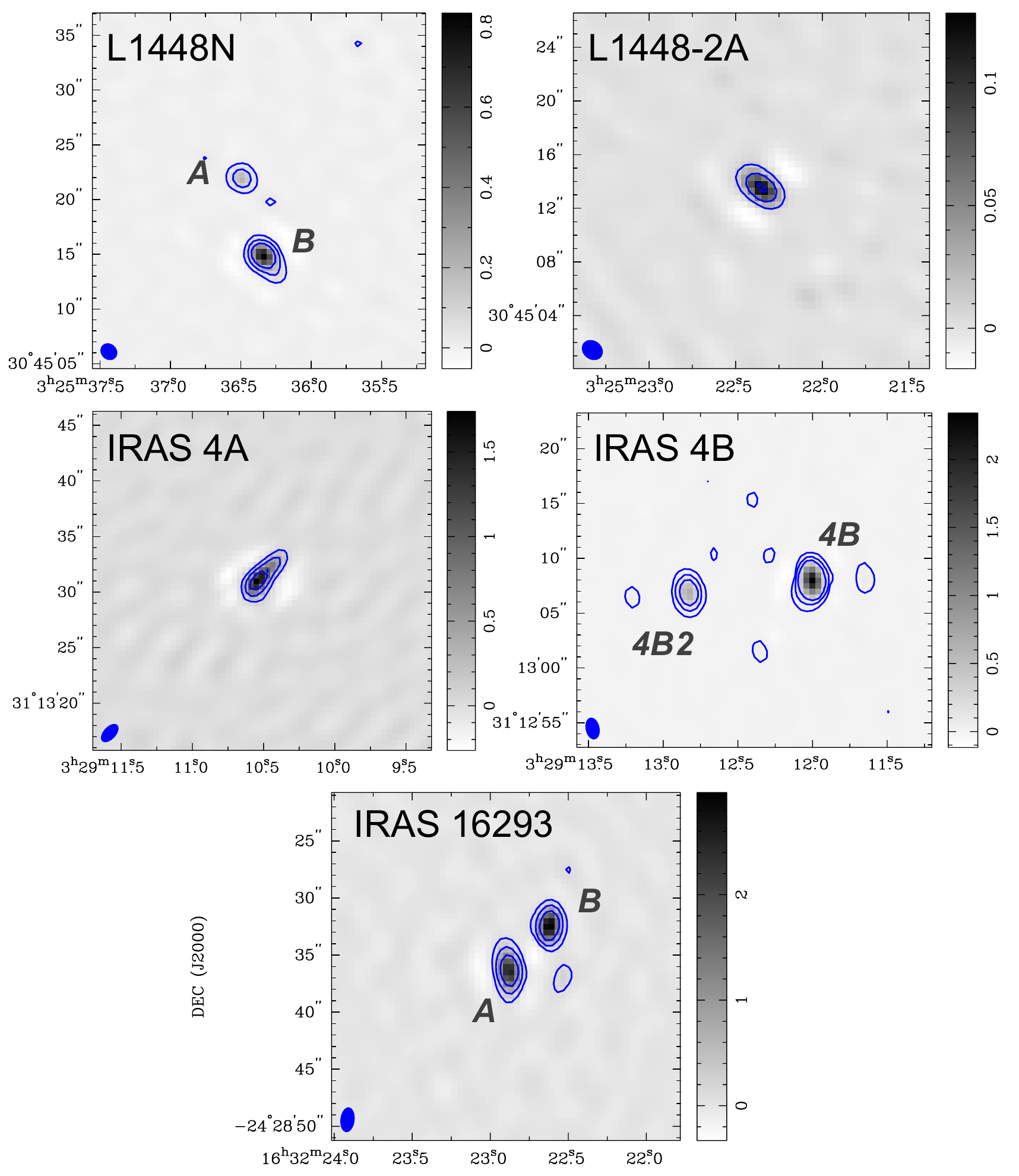}
\caption{345GHz dust continuum emission in 
L1448N, L1448-2A, NGC~1333 IRAS4A, NGC~1333 IRAS4B and IRAS16293 
when only visibilities above 40 k$\lambda$ are kept. 
We apply a natural weighting to produce these maps.
Contours at [5,20,50] $\sigma$ appear in blue. 
The beam sizes are $\sim$1\farcs9$\times$1\farcs2.}
\label{only40klambda}
\end{figure}

\subsection{Analysis of the Stokes I visibilities}

Part of the extended dust continuum emission is lost during the reconstruction
of the final image itself, since an interferometer can 
only sparsely sample Fourier components of different spatial frequencies
of the incoming signal. To analyze the impact of the map reconstruction on the 
flux values derived, we can compare the flux densities derived from the
reconstructed map with those deduced directly from the visibility amplitude curves.
The Stokes I amplitudes of the visibilities, averaged every 120s, 
are presented in Fig.~\ref{Visibilities_StokesI}.
The sample includes three wide binary objects resolved by 
our SMA observations: L1448N, SVS13 and IRAS16293. We use the visibility data
to separate the binary sources. As the sources cannot be modeled using a simple 
Gaussian fitting, we use the \texttt{MIRIAD}~/~{\it imsub} function 
to produce a sub-image containing L1448N-A, SVS13-A 
and IRAS16293-B from the cleaned image. We then use
\texttt{MIRIAD}~/~{\it uvmodel} to subtract these sources 
from the visibility data and make an image of 
L1448N-B, SVS13-B and IRAS16293-A from the modified visibility datasets. 
We proceed the same way to independently map L1448N-A, SVS13-A and IRAS16293-B. 
The separated visibility profiles are shown in Fig.~\ref{Visibilities_StokesI}. 
The region mapped around NGC~1333 IRAS4B contained IRAS4A (outside the primary beam)
and IRAS4B2 located about 10\arcsec\ east of IRAS4B. These two sources are also 
isolated and removed from the visibility data presented in Fig.~\ref{Visibilities_StokesI}.

We use a Gaussian fitting method to fit the visibilities and extrapolate these models at 
0~k$\lambda$ to derive the integrated 0.87mm continuum fluxes. 
The fluxes derived from the visibility amplitude curves are reported in 
Table~\ref{Results}. The fluxes derived from the visibility amplitude curves
are close but, as expected, systematically larger (on average 2.2 times) than 
those derived on the reconstructed maps. 
We also indicate the 0.87mm continuum fluxes of L1448N-B and 
SVS13-B (the sources located at the phase center) in Table~\ref{Results}.
We find that SVS13-A and B have similar fluxes while L1448N-B is 4 
times brighter than L1448N-A. 

For five sources, observations sampled baselines above 40k$\lambda$, 
i.e. sampling smaller spatial scales and most compact components of the envelope. 
We produce new maps of these sources using only visibilities that have 
a radius in the u-v plane larger than 40 k$\lambda$. 
We use a natural weighting scheme to produce the maps in order to achieve 
the highest point-source sensitivity. These maps are shown in Fig.~\ref{only40klambda}. 
The flux densities of the compact components are 
$\sim$150, 25, 580, 530 and 1120mJy for L1448N, 
L1448-2A, NGC~1333 IRAS4A, NGC~1333 IRAS4B and IRAS16293 
respectively: for L1448N and L1448-2A, 
the compact component only accounts for less than 1/10 
of the total flux and for NGC~1333 IRAS4A and IRAS16293 for 1/4. 
This means that for these objects, most of the mass is still in 
the large-scale envelope rather than in a massive disk and 
that our SMA observations will allow to trace the polarized 
emission at the envelope scales where most of the mass resides,
with only little contamination from a possible central protostellar disk. 
The visibility profile of NGC~1333 IRAS4B is the flattest in our sample 
(Fig.~\ref{Visibilities_StokesI}). 1/3 of the total 0.87mm flux is contained 
in the compact component. 
This flat visibility profile is consistent with those analyzed 
in \citet{Looney2003} and the ones obtained at both 1mm and 3mm 
by the IRAM-PdBI CALYPSO\footnote{http://irfu.cea.fr/Projets/Calypso} 
survey. The profile drops steeply at u$-$v distances larger than 
60k$\lambda$, which suggests that NGC~1333 IRAS4B either has a compact 
(FWHM$<$6\arcsec) envelope or that its submm emission at scales 
3 -- 6\arcsec\ is dominated by the emission from a large Gaussian 
disk-like structure. Note that both NGC~1333 IRAS4A and IRAS4B have
very compact envelopes (envelope outer radii $<$ 6\arcsec) as seen 
with the PdBI observations at 1mm and 3mm (Maury et al. in prep).


\subsection{Polarization results}

Figure~\ref{PolaMaps} shows the maps of the Stokes Q and U parameters 
and the polarization intensity and polarization fraction maps obtained when combining them. 
The peak polarization intensities and the median polarization fractions (defined as the 
unweighted ratio between the median polarization intensity over the flux intensity) 
are listed in Table~\ref{Results}. In spite of their low luminosity at SMA scales, we 
detect linearly polarized continuum emission in all the low-mass protostars, even 
using a conservative 3-$\sigma$ threshold. Median polarization fractions range 
from a few \% for IRAS16293 and L1448N to 10\% for B335. 
We only obtain a weak 2-$\sigma$ detection toward the northeastern companion of SMME/HH797 
(around 03:43:57.8; +32:03:11.3). This object was classified as a Class 0 proto-brown 
dwarf candidate by \citet{Palau2014}. When polarization is detected in the center, 
we observe that the polarization fraction drops toward this center. 
For HH797, L1448C, L1448-2A and IRAS03282, we detect polarization in the envelope at 
distances between 700 and 1600 au from the Stokes I emission peak, but not in the center itself. 
The absence of polarization could be linked with an averaging effect at high column 
density in the beam and/or along the line-of-sight in those objects. 
We analyze potential causes for depolarization in $\S${4.1}.

\section{Analysis and discussion} 

\subsection{Distribution of the polarized emission}

Polarization is detected in all of our maps, but not 
always towards the peak of dust continuum emission 
(Stokes I). A constant polarization fraction would 
predict on the contrary that the polarized intensity 
scales as Stokes I. Yet, polarization `holes' or 
depolarization are often observed in the high-density 
parts of star-forming molecular cores and protostellar systems 
\citep{Dotson1996,Rao1998,Wolf2003,Girart2006,Tang2013,Hull2014}. 
We analyse the distribution of both polarized 
fraction and intensity in this section 
in order to probe if and where depolarization is observed.

\subsubsection{Variation of the polarized fraction with environment}

\begin{figure}
\begin{tabular}{c}
\vspace{-10pt}
\hspace{-5pt} 
\includegraphics[width=8.5cm]{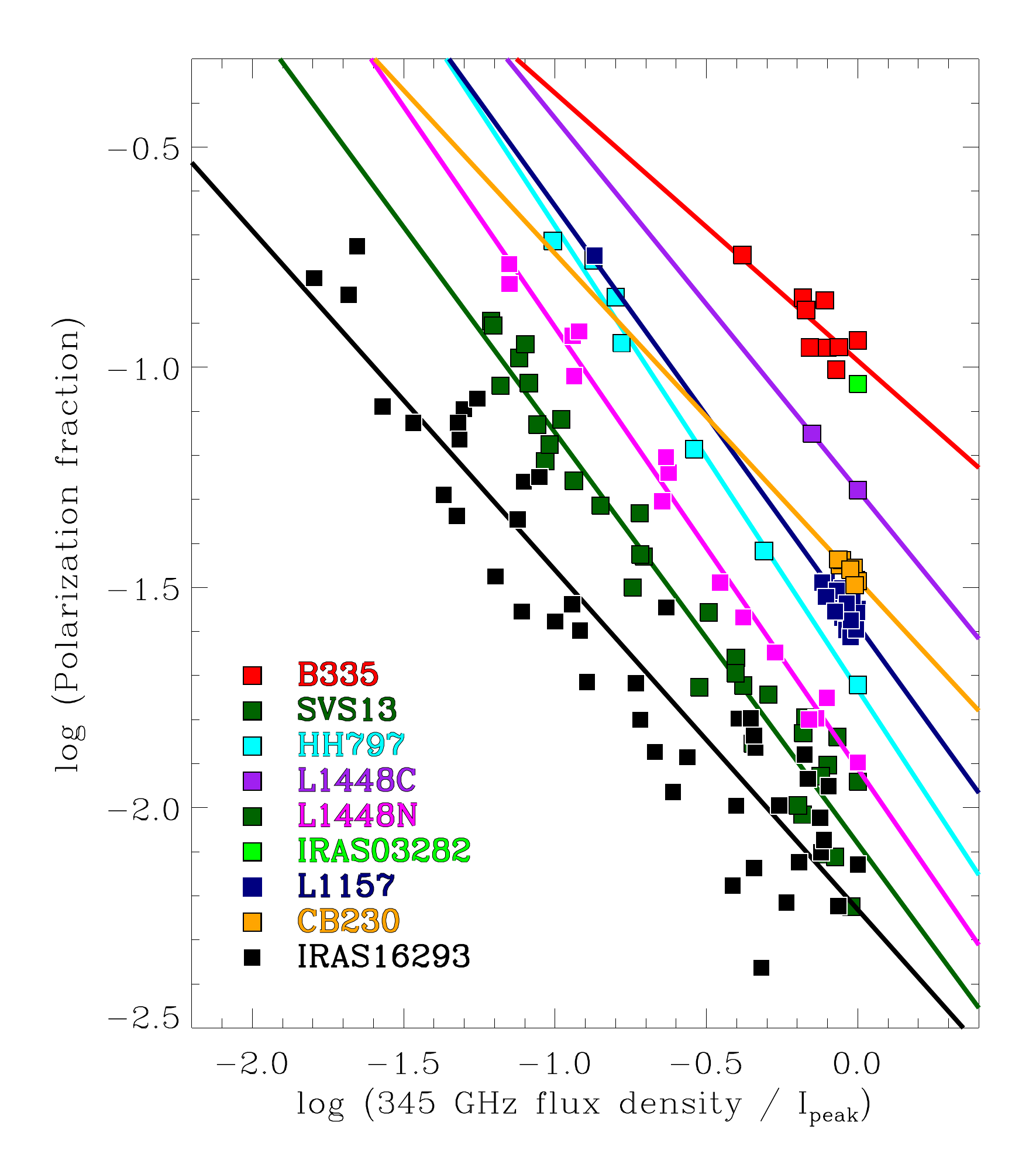} \\
\vspace{-5pt}
\hspace{-5pt} 
\includegraphics[width=8.5cm]{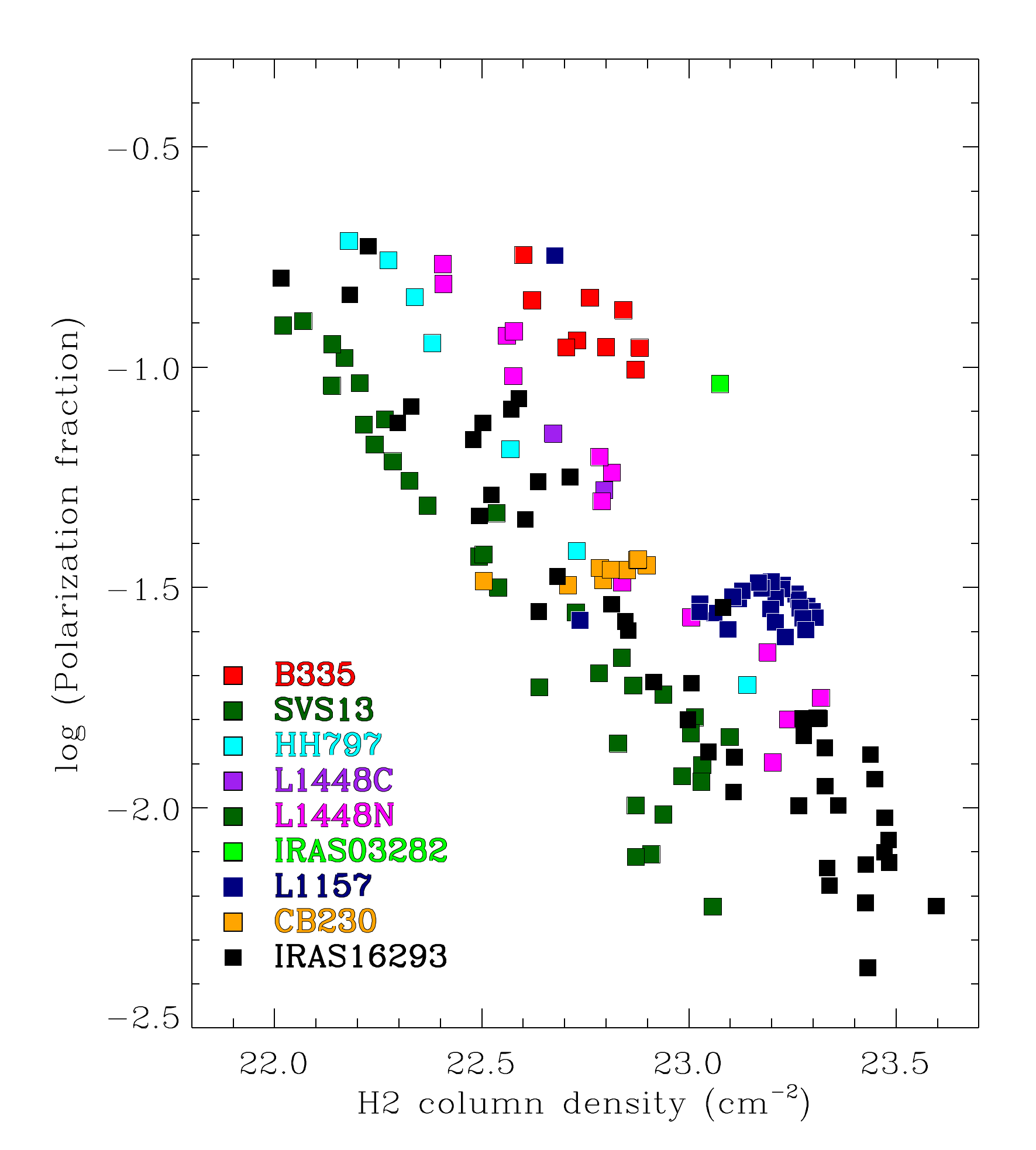} \\
\end{tabular}
\caption{Polarization fractions as a function of Stokes I continuum intensities at 
0.87mm (top) or as a function of the H$_2$ column densities (bottom). Maps have 
been regenerated to share a common synthesized beam of about 5.5\arcsec\ and 
rebinned to have a grid with a common pixel size of 1.8\arcsec. The continuum 
flux densities plotted in the top panel are in units of Jy/beam but are normalized 
to the peak intensity for each object. The solid lines are the best fit to the function 
p$_{frac}\propto$ (I/I$_{max}$)$^{a}$. The H$_2$ column densities plotted on the 
bottom panel are in units of cm$^{-2}$ and are derived using Eq. 5 and 6.}
\label{PolaFvsI}
\end{figure}

For sources where polarization is detected within the central 5\arcsec, we 
observe a drop in the polarization fraction (Fig.~\ref{PolaMaps}). To 
quantify this decrease, we analyze how the polarization fraction 
varies as a function of the 0.87mm flux density. 
We first re-generate our polarization maps with an homogeneous synthesized 
beam of 5.5\arcsec\ in all our sources, using independent pixels 1/3 the 
synthesized beam (1.8\arcsec). IRAS 4A and B are not included 
in this analysis because their dust continuum emission is barely resolved 
at 5.5\arcsec. L1448-2A is not either since polarization is not robustly 
detected anymore at this new resolution.
The relation linking p$_{\rm{frac}}$ versus the Stokes I flux density 
in each individual protostellar envelope is shown in 
Fig.~\ref{PolaFvsI} (top). The continuum flux densities 
is normalized to the peak intensity for each object to allow 
a direct comparison. 
We observe a clear depolarization toward inner envelopes with higher 
Stokes I fluxes, with a polarization fraction $\propto$ I$^{-0.6}$ for B335 down to 
$\propto$ I$^{-1.0}$ for L1448N or HH797\footnote{A constant polarized flux 
gives a polarization fraction $\propto$ I$^{-1}$.}.
Those coefficient are consistent with results from the literature: 
the polarization fraction has, for instance, been found $\propto I^{-0.6}$ 
in Bok globules \citep{Henning2001}, $\propto I^{-0.7}$-I$^{-0.8}$ in dense cores 
\citep{Matthews_Wilson_2000} and down to $\propto I^{-0.97}$ in the main 
core of NGC 2024 FIR 5 \citep{Lai2002}.\\

Since polarization `holes' are usually associated with a high-density medium, 
we also analyze how the polarization fraction varies as a function of the gas 
column density N(H$_2$). We assume that the emission at 0.87\mic\ is optically thin. 
The column density is derived using the formula from \citet{Schuller2009}:

\begin{equation}
N(\rm H_2) = \frac{S_{\nu}}{\mathcal{D}~\mu_{\rm H_2}~m_{\rm H}~\Omega~\kappa_{\nu}~B_{\nu}(T)}
\end{equation}

\noindent with S$_{\nu}$ the flux density, $\mu_{H_2}$ the molecular weight of the ISM
($\mu_{H_2}$=2.8; from \citealt{Kauffmann2008}), m$_{\rm H}$ the mass of an hydrogen 
atom, $\Omega$ the solid angle covered by the beam, $\kappa_{\nu}$ the dust opacity 
(tabulated in \citet{Ossenkopf1994}, 1.85 cm$^2$~g$^{-1}$ at 0.87mm) and B$_{\nu}$(T) 
the Planck function at a dust temperature T. The dust-to-gas mass ratio $\cal{D}$ 
is assumed to be 0.01.

The temperature is not constant throughout the object envelopes: we assume that the 
temperature profile in optically thin outer envelopes follows T(r) $\propto$ r$^{-0.4}$ 
and scale the profiles using the protostellar internal luminosities \citep{Terebey1993}.

\begin{equation}
T = 38~L_{int}^{0.2}~\left(\frac{\rm r}{100 au}\right)^{-0.4}
\end{equation}

\vspace{5pt}
Internal luminosities L$_{int}$ were calculated from {\it Herschel}
Gould Belt Survey in \citet{Sadavoy2014} 
for the Perseus sources and from Maury et al. (in prep) for the other 
sources. L$_{int}$ was approximated to the bolometric luminosity 
if no internal luminosity could be found in the literature. 
Figure~\ref{PolaFvsI} (bottom) presents the dependency 
of p$_{\rm{frac}}$ with local column density. Even if the relation 
observed presents a large scatter, our results
confirm that the polarization fraction decreases with increasing 
column density in the envelopes of our sample. \\

Along with the polarization fraction variations, the 
distribution of polarized emission intensities can also 
help us determine the cause of depolarization. 
SCUBAPOL observations have shown that the polarization 
intensity can decrease at high densities \citep{Crutcher2004}. 
We do not observe such trend in our sample. In Fig.~\ref{PolaFvsI}, 
the slope of the polarization fraction relation with the dust 
continuum intensity is close to -1 or flatter for most sources.
This means that the polarization intensity is constant and even 
increases (i.e. B335, SVS13 or IRAS16293) toward the central regions 
(see also the polarization intensity and fraction maps provided 
in Fig~\ref{PolaMaps}). These results suggest that for these sources, 
the decrease of the polarization fraction in the center is not linked with 
a full depolarization but could indicate a variation in the polarization 
efficiency itself (see next section).

\begin{figure}
\centering
\vspace{10pt}
\hspace{20pt}{\large SCUBA}\hspace{95pt}{\large SMA}
\includegraphics[width=8.7cm]{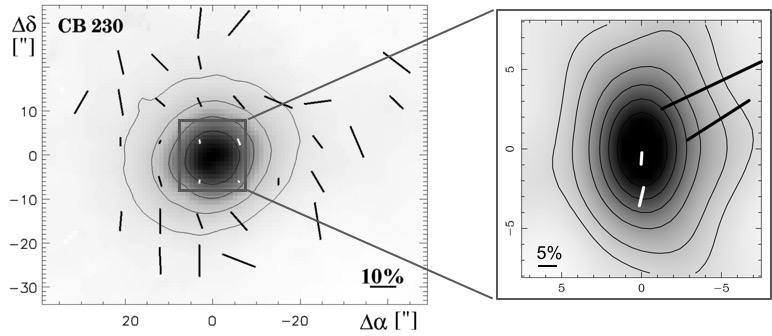}
\caption{{\it Left:} SCUBA 0.85mm continuum maps with polarization 
orientation overlaid from \citet{Wolf2003}. The contour lines mark 
20\%, 40\%, 60\%, and 80\% of the maximum intensity. 
{\it Right:} Zoom-in into the scales covered with the SMA 
at 0.87mm. The contours are the same than in Fig.~\ref{StokesI_Borientation}.
For both, the bar length is proportional to the polarization fraction. }
\label{SCUBA_SMA_CB230}
\end{figure}

\subsubsection{Potential causes for depolarization}

Observations as well as models or simulations \citep{Padoan2001,Bethell2007,FalcetaGoncalves2008,Kataoka2012}
have shown that depolarization can be linked with both geometrical 
effects (e.g. averaging effects linked with the complex structure of 
the magnetic field along the line of sight) as well as physical 
effects (e.g. collisional/mechanical disalignment, lower grain 
alignment efficiency, variation of the grain population). 
We discuss here some of these `depolarization' effects.
We note that our sample is sensitivity limited, which means that there is a 
potential bias toward strong polarization intensities.
We also remind the reader that the polarized emission presented in this paper
comes from the envelope and that we are not sampling the polarized (or unpolarized) 
emission from the inner ($<500$ au) envelope here.\\

\noindent{\it Geometrical effects -} Our SMA observations recover polarized emission from 
previously reported ``polarization holes" observed with single-dish observations. 
Observations of B335 and CB230 at a 2200 au resolution using SCUBAPOL \citep{Holland1999} 
performed by \citet{Wolf2003} show that the polarization fraction decreases from 6 -- 15\% 
in the outer parts of the cores to a few \%, thus at scales where 
3 -- 10\% polarization is detected with the SMA (see Fig.~\ref{SCUBA_SMA_CB230} 
for an illustration in CB230). 
The central depolarization observed by \citet{Wolf2003} in B335 and CB230 is thus 
partially due to beam averaging effects or mixing of the polarized signal along the 
line-of-sight, an envelope pattern that our SMA resolution now allows to recover. 
Recent polarization observations of B335 with ALMA at even higher resolution by 
\citet{Maury2018} confirm that polarization is still detected towards the center of the B335 
envelope at 0.5 -- 5\arcsec\ scales, a sign that the dust emission stays polarized 
(p $\sim$ 3 -- 7\%) at small scales in the high density regions. \\

\noindent{\it Optical depth effects -} As already reported in \citet{Girart2006}, 
a two-lobe distribution of the polarization intensity is detected toward 
NGC~1333 IRAS4A. We report a similar two-lobe structure for the first time 
in NGC~1333 IRAS4B, with peaks of the polarization intensity on each side of the 
north-south outflow. \citet{Liu2016} find that the polarization intensity of
NGC~1333 IRAS4A observed at 6.9mm has a more typical distribution, with the 
polarization intensity peaking toward the source center and attributed 
this variation of the polarization distribution with wavelength to optical 
depth effects. Our temperature brightnesses rather suggest that the 0.87mm 
emission is optically thin at the envelope scales probed in this paper, for all our objects. We 
note that modeling the magnetic field in NGC~1333 IRAS4A, \citet{Goncalves2008} 
show that strong central concentrations of magnetic field lines could 
reproduce a two-lobe structure in the polarized intensity distribution. \\

\noindent{\it Grain alignment, grain growth -} 
Several studies of dense molecular clouds or starless cores have shown that the slope 
of the polarization degree can reach values of `-1' (as in some of our objects) 
or lower with respect to A$_{\rm V}$ or I/I$_{max}$ and interpret 
this slope as a sign of a decrease or absence of 
grain alignment at higher densities \citep{Alves2014,Jones2015,Jones2016}. Even if we 
are not in such case (our objects all have a central heating source), 
the degree of polarization we detect is still extremely sensitive to 
the efficiency of grain alignment, which is itself very sensitive to the grain 
size distribution and potential grain growth \citep{Bethell2007,Pelkonen2009,Brauer2016}. This grain growth has been already suggested
in Class 0 protostars like L1448-2A or L1157 \citep{Kwon2009}, with a faster 
grain growth toward the central regions. More observations combining 
different sub-mm wavelengths at various scales will help us probe the dust spectral 
index throughout the envelope, analyse the scales at which grain growth is 
expected \citep{Chacon-Tanarro2017}, probe the various environmental dependences 
of the grain alignment efficiency \citep{Whittet2008} and observationally 
constrain the dependence on the polarization degree and the grain size as 
predicted for instance by radiative torque models \citep[see][]{Bethell2007,Hoang2009,Andersson2015,Jones2015}.  \\

\subsection{Orientation of the magnetic field}

The B-field lines are inferred from the polarization angles 
by applying a 90$\degr$ rotation. The mean magnetic 
field orientations in the central 1000 au are provided 
in Table~\ref{Angles}. The B orientation is overlaid with 
red bars in Figure~\ref{StokesI_Borientation}.
We also show the $>$2-$\sigma$ detections (orange bars): although these 
are subject to higher uncertainties we stress 
that they are mostly consistent with the neighboring 3-$\sigma$ detection 
line segments. For robustness, however, these lower significance 
detections are not used in this analysis.
Under flux-freezing conditions, the pull of field lines in 
strong gravitational potentials is expected to create an hourglass 
morphology of the magnetic field lines, centered around the dominant 
infall direction, during protostellar collapse. 
Using CARMA observations, \cite{Stephens2013} show that in L1157, 
the full hourglass morphology becomes apparent around 550 au. 
Using SMA observations, \cite{Girart2006} also observed this 
hourglass shape in NGC~1333 IRAS4A. The resolution we select for this analysis 
of B at envelope scale is not sufficient to reveal the hourglass 
morphology in most of our objects. 
Only one of our sources, L1448-2A, shows hints of an hourglass shape 
at envelope scales (see Fig.~2) but the non detection of polarization 
in the northern quadrant does not allow to detect a full hourglass 
pattern in this source: better sensitivity observations are needed 
to confirm this partial detection.

\subsubsection{Large scales versus small scales}

Our B-field orientations can be compared to SHARP 350\mic\ and SCUBAPOL 
850\mic\ polarization observations that provide the orientation of the 
magnetic field at the surrounding cloud/filament scales for half of our sample.
We use the SHARP results presented in \citet{Attard2009}, \citet{Davidson2011} and 
\citet{Chapman2013} and the SCUBAPOL of \citet{Wolf2003} and \citet{Matthews2009}. 
The large scale and smaller scale B orientations match for NGC~1333 IRAS4A, 
L1448N, L1157, CB230 and HH797. The SMA B orientation in L1448-2A is quite 
different from that detected with SHARP. In B335, the SCUBA observations 
seem also inconsistent (north-south direction) with the SMA B orientation.
However, only two significant line segments are robustly detected with SCUBA and
\citet{Bertrang2014} detect a mostly poloidal field in the B335 core using near-infrared 
polarization observations at 10000 au scales.
Recent observations of B335 with ALMA have revealed a very ordered topology of 
the magnetic field structure at 50 au, with a combination of a large-scale 
poloidal magnetic field (outflow direction) and a strongly pinched magnetic 
field in the equatorial direction \citep{Maury2018}. Our SMA observations 
suggest that above $>$1000 au scales, the poloidal component dominates in B335. 
All these results show that an ordered B morphology from the cloud to the 
envelope is observed for most of our objects. Our results also seem to confirm
that sources with consistent large-to-small-scale fields (e.g. B335, 
NGC~1333 IRAS4A, L1157 and CB230) tend to have a high polarization 
fraction , similar to what is found in \citet{Hull2013}.

\subsubsection{Variation of the B orientation with wavelength}

\begin{figure*}
\includegraphics[width=17cm]{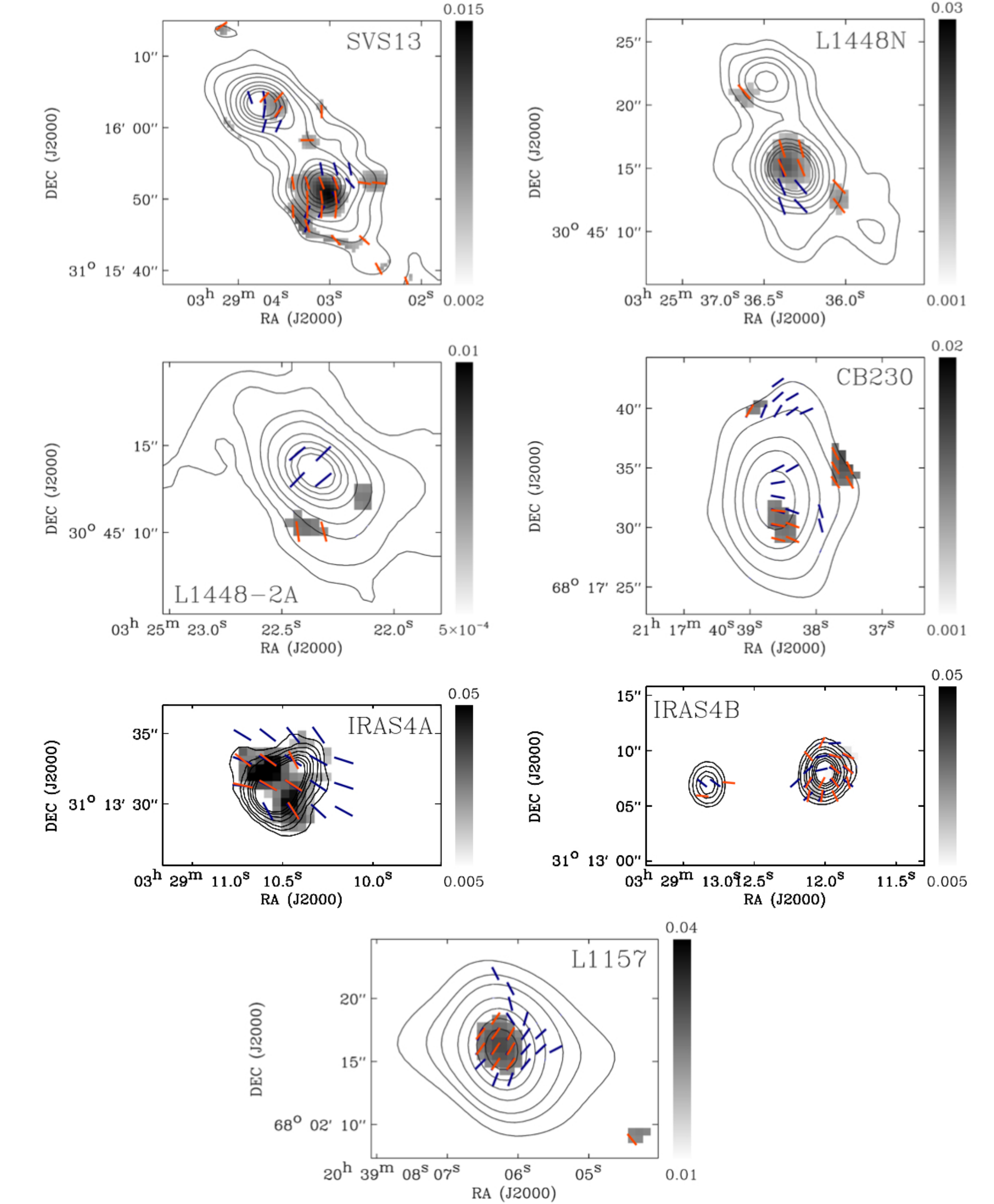} 
\caption{B-field orientation derived from SMA 0.87mm observations 
in red (3-$\sigma$ detections) and from CARMA 1.3mm observations 
in blue \citep[3.5-$\sigma$ detections from][]{Hull2014}. 
The underlying map is the polarization intensity in mJy/beam 
and contours are the Stokes I continuum emission 
(same levels as in Fig.~\ref{StokesI_Borientation}).}
\label{CARMA}
\end{figure*}

Only few studies have investigated the relationship 
between the 
magnetic field orientation and wavelength on similar scales. 
Using single-dish observations, \citet{Poidevin2010} found 
that in star-forming molecular clouds, most of the 0.35 and 
0.85mm polarization data had a similar polarization pattern.
In Fig.~\ref{CARMA}, we compare the 0.87mm SMA B-field orientation 
to those observed at 1.3mm by  \citet[][]{Stephens2013} and \citet{Hull2014}. 
CARMA maps are rebinned to match the grid of our SMA maps. 
7 objects are common to the two samples and have co-spatial detections. 
We do not compare the SMA and CARMA results for B335 and L1448C, as 
polarization is only marginally ($<$3.5-$\sigma$) detected at 1.3 mm \citep{Hull2014}.
The orientations at 0.87 and 1.3mm match for L1157, SVS13, L1448N-B, 
NGC~1333 IRAS4A and IRAS4B. 
For SVS13-B, the orientations only slightly deviate in the southeast part, 
but the position angles remain consistent within the 
uncertainties added in quadrature (8$\degr$ uncertainties for 
SMA, 14$\degr$ for CARMA). 
In L1448N-B, the orientation of the B-field, perpendicular to the 
outflow, is also consistent with BIMA 
\citep[Berkeley Illinois Maryland Array;][]{Welch1996} 
1.3mm observations presented in \citet{Kwon2006}.
The orientations at 0.87 and 1.3mm deviate in the outer parts of the 
envelope in NGC~1333 IRAS4B, in regions with low detected polarized intensities.
In CB230, the 3-$\sigma$ detection at 0.87 and 1.3mm do not 
exactly overlap physically but the east-west orientation of our central 
detections is consistent with the orientation found 
with CARMA by \citet{Hull2014}.
For L1448-2A finally, the SMA 0.87 and CARMA 1.3mm detections do not overlap, 
making a direct comparison between wavelength difficult for this object. 
We note that the flat structure of the dust continuum 
emission oriented northeast-southwest observed at 0.87mm is consistent with that 
observed at 1.3 mm by \citet{Yen2015}. The 1.3mm observations from TADPOL 
are not in agreement: a northwest-southeast extension is observed along the outflow,  
the polarization detections (blue bars in Fig.~\ref{CARMA}) from TADPOL in the central 
part of L1448-2A might be suffering from a contaminating by CO polarized 
emission from the outflow itself. Their 2.5-$\sigma$ detections along 
the equatorial plane are, however, consistent with our 2-$\sigma$ detections. 
Our conclusion is that the orientations observed at 0.87 and 1.3mm 
are consistent. Those two wavelengths being close to each other, they 
probably trace optical depths on which the magnetic field orientations stay ordered.

\subsubsection{Misalignment with the outflow orientation}

Whether or not the rotation axis of protostellar cores 
are aligned with the main magnetic field direction is still an open question.
From an observational point of view, it remains difficult to precisely 
pinpoint the rotation axis of protostellar cores. The outflow axis 
is often used as a proxy for this key parameter, as protostellar bipolar 
outflows are believed to be driven by hydromagnetic winds in the circumstellar 
disk \citep{Pudritz1983,Shu2000,Bally2016}. The core rotation and outflow axes are 
thus usually considered as aligned\footnote{We note that MHD simulations 
have shown that the outflowing gas tends to follow the magnetic field 
lines when reaching scales of a few thousands au even when the rotation axis 
is not aligned with the large-scale magnetic field: hence one should always 
be careful when assuming that the orientation of the outflow traces the 
rotation axis.}.  
Synthetic observations of magnetic fields in protostellar cores have shown 
that magnetized cores have strong alignments of the outflow axis with the B orientation 
while less magnetized cores present more random alignment \citep{Lee2017}.
MHD collapse models predict in particular that magnetic braking should be 
less effective if the envelope rotation axis and the magnetic field are 
not aligned \citep{HennebelleCiardi2009,Joos2012,Krumholz2013,Li2013}.
The comparison between the rotation axis and the B-field orientation 
is thus a key to understand the role of B in regulating the collapse. 
Previous observational works found that outflows do not seem to show a
preferential direction with respect to the magnetic field direction
both on large scales \citep{Curran2007} and at 1000 au scales \citep{Hull2013}.
A similar absence of correlation is also reported in a sample of 
high-mass star-forming regions by \citet{Zhang2014}.
Our detections of the main B-field component at envelope scales for all 
the protostars of our sample allow us to push the analysis further for low-mass Class 0
objects.

\begin{figure}
\begin{tabular}{m{8cm}}
\hspace{-10pt} \includegraphics[width=9cm]{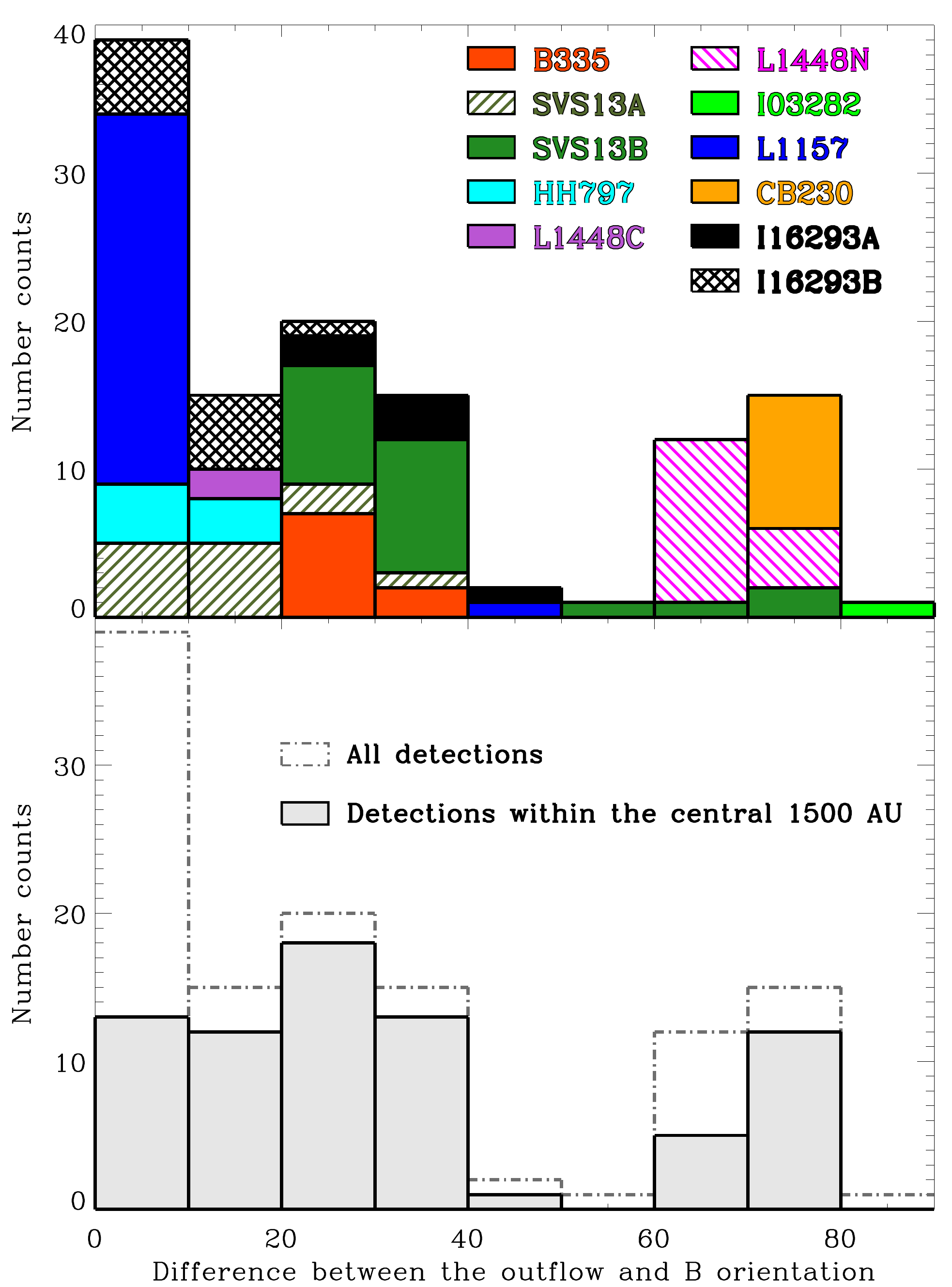} \\
\vspace{4pt}
\end{tabular}
\vspace{-8pt}
\caption{Histogram of the projected angles between the magnetic field 
and the outflow direction. The data share a common synthesized beam size 
of about 5.5\arcsec\ and are binned with pixels of 1.8\arcsec\ for this analysis.
The top panel shows the histogram color-coded 
per source while the bottom panel shows the global
histogram (in white) with the histogram restricted 
to detections within the central 1500 au overlaid in gray.}
\label{Misalignement}
\end{figure}

The misalignments between the B orientation and the outflow axis driven by the 
protostars can be observed directly from Figure~\ref{StokesI_Borientation}. 
Table~\ref{Angles} provides the projected position angle of the outflows 
as well as the angle difference between this 
outflow axis and the main envelope magnetic field orientation when B is
detected in the central region.
For half of the objects, the magnetic field lines are oriented within 
40$\degr$ of the outflow axis but some sources show rather large 
($>$60$\degr$) difference of angles (e.g. L1448N-B and CB230). 
Using the maps generated with a common 5.5\arcsec\ synthesized beam 
(thus excluding NGC~1333 IRAS 4A, 4B and L1448-2A, see $\S$4.1), we build a 
histogram of the projected angles between the magnetic field orientation 
and the outflow direction (hereafter $\cal{A}$$_{B-O}$) shown in Fig.~\ref{Misalignement}. 
We also compare the full histogram with that restricted to detections within 
the central 1500 au (Fig.~\ref{Misalignement}, bottom panel).   
The distribution of $\cal{A}$$_{B-O}$ looks roughly bimodal, suggesting that at 
the scales traced with the SMA, the B-field lines are either aligned or perpendicular 
to the outflow direction. \citet{Hull2014} find hints that for objects with low 
polarization fractions, the B-field orientations tend to be preferentially perpendicular 
to the outflow. In our sample focusing on solar-type Class 0 protostars, we do not observe a relation 
between $\cal{A}$$_{B-O}$ and the polarization fraction nor intensity. 
For instance, L1448N-B and SVS13-B have very similar polarization fractions 
but very different $\cal{A}$$_{B-O}$. In the same way, the two Bok globules 
B335 and CB230 both have a high polarization fraction but exhibit very different 
$\cal{A}$$_{B-O}$.

\begin{table}
\centering
\caption{Misalignment between the outflow angles $\chi$$_o$ and the mean 
magnetic field orientation $\chi$$_B$$_{1000}$ in the central 1000 au.}
\begin{tabular}{cccc}
\hline
\hline
Name 	 	& $\chi$$_o$$^a$ & 	$\chi$$_B$$_{1000}$ 	& 
$\lvert$~$\chi$$_o$~-~$\chi$$_B$$_{1000}$~$\rvert$	\\
    		& ($\degr$)	   &  ($\degr$)	& 	\\
\hline
B335    		& 90		&	55$\pm$3			& 35\\
SVS13-A			& 148		& 	140$\pm$3			& 8 \\
SVS13-B	 		& 160		& 	19$\pm$6			& 39\\
HH797			& 150		& 	113$\pm$2			& 37 \\
L1448C			& 161		& 	-					& - \\
L1448N-B		& 105		&	23$\pm$5			& 82\\
L1448-2A		& 138		& 	-					& - \\
IRAS03282		& 120		& 	-					& - \\
NGC~1333 IRAS4A	& 170		& 	57$\pm$13			& 67\\
NGC~1333 IRAS4B	& 0			& 	58$\pm$49			& 58\\
IRAS16293-A		& 145		& 	163$\pm$15			& 18 \\
IRAS16293-B		& 130		&	114$\pm$17			& 16\\
L1157			& 146		& 	146$\pm$4			& 0\\
CB230			& 172		& 	86$\pm$3			& 86\\
\hline
\end{tabular}
\begin{list}{}{}
\item[$^a$] Position angles are provided east of north. 
References: \citet{Bachiller1998}, \citet{Bachiller2001}, \citet{Choi2001},
\citet{Hull2013}, \citet{Hull2014}, \citet{Rodriguez1997}, \citet{Tafalla2006}, 
\citet{Rao2009}, \citet{Santangelo2015}, \citet{Yen2015}. 
\end{list}
\label{Angles}
\end{table}

\subsubsection{A relation between misaligned magnetic field, strong rotation and fragmentation?}

Details on the various velocity gradients (orientation, strength) measured in 
our sources are provided in appendix. Velocity fields in protostars can trace 
many processes (turbulence, outflow motions, etc.). In particular, gradients detected in the 
equatorial plane of the envelope can trace both rotation and infall motions. 
Organized velocity gradients with radius are often interpreted as a sign that 
the rotational motions dominate the velocity field. In this section, we compare 
the B orientation and the misalignment $\cal{A}$$_{B-O}$ with velocity gradients
associated with rotation in our envelopes. \\

\noindent{\it $\cal{A}$$_{B-O}$ $>$ 45$\degr$ case - }
\noindent L1448N-B and CB230 have a $\cal{A}$$_{B-O}$ close to 90$\degr$.  
They both have large envelope masses and velocity gradients perpendicular to 
their outflow direction detected at hundreds of au scales.
A high rotational to magnetic energy could lead to a twist of the field lines 
in the main rotation plane. Their high mass-to-flux ratio 
$\mu=M/\Phi \sim E_{grav}/E_{mag}$ could also favor a gravitational 
pull of the equatorial field lines. Both scenarios can efficiently produce a 
toroidal/radial field from an initially poloidal field that would explain 
the main field component we observe perpendicular to the outflow/rotation axis. 
The opposite causal relationship is also possible: an initially misaligned 
magnetic field configuration could be less efficient at braking the rotation 
and lead to the large rotational motions detected at envelope scales.
Additional observations of Class 0 from the literature show 
similar misalignments. In NGC~1333 IRAS2A, the B direction 
is misaligned with the outflow axis \citep{Hull2014} but aligned
with the velocity gradient observed in the combined PdBI+IRAM30 
C$^{18}$O map by Gaudel et al. (in prep).
In L1527, $\cal{A}$$_{B-O}$ is nearly 90$\degr$ \citep{SeguraCox2015}
but the B direction follows the clear north-south velocity gradient 
tracing the Keplerian rotation of a rather large $\sim$60 au disk 
\citep{Ohashi2014}. We stress out, however, that polarization in disks 
could also be produced by the self-scattering of dust grains 
\citep{Kataoka2015} and not tracing B. 
Finally, our source NGC~1333 IRAS4A has a B orientation 
misaligned compared to the small scale north-south outflow direction 
\citep[see also][]{Girart2006} but aligned with the 
large scale 45$\deg$ outflow direction (bended outflow; see appendix). 
This is consistent with the position angle of the velocity 
gradient found by \citet{Belloche2006}, even if the interpretation 
of gradients is difficult in IRAS4A because strong infall 
motions probably dominate the velocity field.\\

\noindent{\it $\cal{A}$$_{B-O}$ $<$ 45$\degr$ case - }
\noindent We find only small misalignments in L1157 and B335, 
sources that show low to no velocity gradient at 1000 au scales \citep[][Gaudel et al. in prep]{Tobin2011,Yen2010}. 
The B orientation of the off-center detection in L1448C (see Fig.~\ref{StokesI_Borientation}) 
is also aligned with the outflow axis. The source has one of the smallest velocity 
gradient perpendicular to the outflow axis determined by \citet{Yen2015}. 
Our wide-multiple systems IRAS16293 (A and B) and SVS13 (A and B) 
present good alignments.
IRAS16293-B does not show rotation but the source is face-on, 
which might explain the absence of rotation signatures.
In IRAS16293-A, the magnetic field has an ``hourglass'' 
shape at smaller scales \citep{Rao2009}, a signature of 
strong magnetic fields in the source, but the source 
also has a strong velocity gradient perpendicular 
to the outflow direction that could be responsible
for some of the misalignment. 
Finally, the B direction is slightly tilted in the center of SVS13B. 
We note that \citet{Chen2007} detected an extended velocity 
gradient detected along the SVS13-A/SVS13-B axis across the 
whole SVS13 system that has been interpreted as core rotation.  \\

Even if our sample is limited, the coincidence of a misalignment 
of B with the outflow direction when large perpendicular velocity 
gradients are present or the alignment of B with the velocity gradient
itself strongly suggests that the orientation of B at the envelope 
scales traced by the SMA can be affected by the strong rotational 
energy in the envelope. These results are consistent with predictions 
from numerical simulations that show a strong relation between the 
angular velocity of the envelope and the magnetic field direction 
\citep{Machida2005,Machida2007}. 
Rotationally twisted magnetic fields have already been observed at 
larger scales \citep{Girart2013,Qiu2013}. After a given number of periods, 
rotation could be distorting or even twisting the field lines, producing 
a toroidal component in the equatorial plane. 
Using MHD simulations, \citet{CiardiHennebelle2010} and \citet{Joos2012} showed that 
the mass ejection via outflows could be less efficient when the rotation axis 
and B direction are misaligned. The presence of outflows in sources with 
$\cal{A}$$_{B-O}$ $\sim$ 90$\degr$ might favour a twisted/pinched 
magnetic field scenario compared to an initially misaligned magnetic field configuration. \\

Reinforcing the possible correlation between the field misalignment and 
the presence of strong rotational motions, our SMA observations also 
suggest that a possible correlation could be present between the 
envelope-scale main magnetic field direction and the multiplicity 
observed in the inner envelopes. Indeed, 
sources that present a strong misalignment (IRAS03282, NGC~1333 IRAS4A, 
L1448N, L1448-2A or CB230) are all close multiples 
\citep[][Maury et al. in prep]{Launhardt2004,Chandler2005,Choi2005,Tobin2013,Tobin2016}.
SVS13-A does host a close multiple but it is a Class I: we are thus not witnessing 
the initial conditions leading to fragmentation, and the magnetic field we
observe could have been severely affected by the evolution of both the 
environment and the protostar itself.
On the contrary, sources that show an alignment between the B 
direction and the outflow direction (e.g. L1157, B335, SVS13-B and L1448C) 
have been studied with ALMA or NOEMA high-resolution observations 
and stand as robust single protostars at envelope scales of $\sim$5000-10000 au. 
No large disk has been detected so far 
in these sources \citep[][]{Saito1999,Yen2010,Chiang2012,Tobin2015}
while L1448N-B for instance was suggested recently to harbor a quite 
large disk encompassing several multiple sources at scales $\sim 100$ au \citep{Tobin2016}. 
These results support those of \citet{Zhang2014}: the possible link 
between fragmentation and magnetic field topology points towards a 
less efficient magnetic braking and redistribution of angular momentum
when the B-field is misaligned.  \\


\section{Summary}

We perform a survey of 0.87mm continuum and polarized emission from dust 
toward 12 low-mass Class 0 protostars. 

\vspace{5pt}
\noindent {\it (i)} 
By comparing the 0.87mm SMA continuum fluxes with single-dish 
observations, we find that our interferometric observations 
recover $<$20\% of the total fluxes. The fluxes derived directly 
from the visibility data are also about twice larger than those 
derived on the reconstructed maps. For 5 objects, we observed 
baselines above 40 k$\lambda$ that allow us to separate the most compact components. 
The low flux fractions contained in the compact component are consistent 
with the classification as Class 0 of our sources.

\vspace{5pt}
\noindent {\it (ii)} 
We report the detection of linearly polarized 
dust emission in all the objects of the sample, 
with mean polarization fractions ranging from 2 to 10\%. 

\vspace{5pt}
\noindent {\it (iii)} 
We find a decrease of the polarization fraction with the estimated H$_2$ column 
density. Our polarization intensity profiles are relatively flat in most of our 
sources and increases toward the center in B335, SVS13 or IRAS16293, suggesting
a decrease in the dust alignment efficiency rather than a polarization cancellation in 
these sources. Two-lobes structures in the polarized intensity distributions 
are observed in NGC~1333 IRAS4A and IRAS4B.

\vspace{5pt}
\noindent {\it (iv)} 
The orientations observed at 0.87mm with SMA and 1.3mm with CARMA are 
consistent with each other on the 1000 au scales probed in this analysis. 

\vspace{5pt}
\noindent {\it (v)} 
Like in \citet{Hull2013}, we find that sources with a high polarization 
fraction have consistent large-to-small-scale fields. We do not observe, 
however, a relation between the misalignment of the magnetic field orientation with respect to the outflow axis 
and the polarization intensity or fraction. 

\vspace{5pt}
\noindent {\it (vi)}
We find clues that large misalignment of the magnetic field 
orientation with the outflow orientation could be found 
preferentially in protostars with a higher rotational energy. 
The misalignment of 90$\degr$ observed in some objects 
could thus be a signature of winding of the magnetic field 
lines in the equatorial plane when the rotational energy is significant. Strengthening this possible 
link, there are also hints that a B-outflow misalignment is found 
preferentially in protostars that are close multiple and/or harbor 
a larger disk, while single objects seem to show a good agreement between 
the magnetic field direction at envelope scales and the direction 
of their protostellar outflow. This suggests that the topology and strength of magnetic fields at envelope 
scale may significantly impact the outcome of protostellar collapse, eventually playing a major role in the formation of disks and multiple systems.\\

Our results are aligned with previous studies of B in massive cores \citep{Zhang2014}
as well as with theoretical predictions from magnetic collapse models \citep{Machida2005,Joos2012}.
More observations of the magnetic field of low-mass protostars at envelope 
scale would be necessary to observationally confirm these `tentative' but promising 
results and re-enforce our interpretations.


\section*{Acknowledgments}
We thank Baobab Liu for fruitful discussions on this project and Bilal 
Ladjelate for providing the internal luminosities from the {\it{Herschel}} 
Gould Belt Survey dataset, for the sources belonging to the CALYPSO sample.
This project has received funding from the European Research Council (ERC) 
under the European Union Horizon 2020 research and innovation programme 
(MagneticYSOs project, grant agreement N$\degr$ 679937). 
JMG is supported by the MINECO (Spain) AYA2014-57369-C3 and AYA2017-84390-C2 
grants. 
SPL acknowledges support from the Ministry of Science and Technology of Taiwan
with Grant MOST 106-2119-M-007-021-MY3.
This publication is based on data of the Submillimeter Array. The SMA is a joint 
project between the Smithsonian Astrophysical Observatory and the Academia 
Sinica Institute of Astronomy and Astrophysics, and is funded by the Smithsonian 
Institution and the Academia Sinica.

\vspace{-10pt}
\bibliographystyle{aa}
\bibliography{mybiblio_MY.bib}

\begin{thebibliography}{172}
\expandafter\ifx\csname natexlab\endcsname\relax\def\natexlab#1{#1}\fi

\bibitem[{{Alves} {et~al.}(2014){Alves}, {Frau}, {Girart}, {Franco}, {Santos},
  \& {Wiesemeyer}}]{Alves2014}
{Alves}, F.~O., {Frau}, P., {Girart}, J.~M., {et~al.} 2014, \aap, 569, L1

\bibitem[{{Andersson} {et~al.}(2015){Andersson}, {Lazarian}, \&
  {Vaillancourt}}]{Andersson2015}
{Andersson}, B.-G., {Lazarian}, A., \& {Vaillancourt}, J.~E. 2015, \araa, 53,
  501

\bibitem[{{Andre} {et~al.}(1993){Andre}, {Ward-Thompson}, \&
  {Barsony}}]{Andre1993}
{Andre}, P., {Ward-Thompson}, D., \& {Barsony}, M. 1993, \apj, 406, 122

\bibitem[{{Andre} {et~al.}(2000){Andre}, {Ward-Thompson}, \&
  {Barsony}}]{Andre2000}
{Andre}, P., {Ward-Thompson}, D., \& {Barsony}, M. 2000, Protostars and Planets
  IV, 59

\bibitem[{{Attard} {et~al.}(2009){Attard}, {Houde}, {Novak}, {Li},
  {Vaillancourt}, {Dowell}, {Davidson}, \& {Shinnaga}}]{Attard2009}
{Attard}, M., {Houde}, M., {Novak}, G., {et~al.} 2009, \apj, 702, 1584

\bibitem[{{Bachiller} {et~al.}(2000){Bachiller}, {Gueth}, {Guilloteau},
  {Tafalla}, \& {Dutrey}}]{Bachiller2000}
{Bachiller}, R., {Gueth}, F., {Guilloteau}, S., {Tafalla}, M., \& {Dutrey}, A.
  2000, \aap, 362, L33

\bibitem[{{Bachiller} {et~al.}(1995){Bachiller}, {Guilloteau}, {Dutrey},
  {Planesas}, \& {Martin-Pintado}}]{Bachiller1995}
{Bachiller}, R., {Guilloteau}, S., {Dutrey}, A., {Planesas}, P., \&
  {Martin-Pintado}, J. 1995, \aap, 299, 857

\bibitem[{{Bachiller} {et~al.}(1998){Bachiller}, {Guilloteau}, {Gueth},
  {Tafalla}, {Dutrey}, {Codella}, \& {Castets}}]{Bachiller1998}
{Bachiller}, R., {Guilloteau}, S., {Gueth}, F., {et~al.} 1998, \aap, 339, L49

\bibitem[{{Bachiller} {et~al.}(1991){Bachiller}, {Martin-Pintado}, \&
  {Planesas}}]{Bachiller1991}
{Bachiller}, R., {Martin-Pintado}, J., \& {Planesas}, P. 1991, \aap, 251, 639

\bibitem[{{Bachiller} {et~al.}(2001){Bachiller}, {P{\'e}rez Guti{\'e}rrez},
  {Kumar}, \& {Tafalla}}]{Bachiller2001}
{Bachiller}, R., {P{\'e}rez Guti{\'e}rrez}, M., {Kumar}, M.~S.~N., \&
  {Tafalla}, M. 2001, \aap, 372, 899

\bibitem[{{Bally}(2016)}]{Bally2016}
{Bally}, J. 2016, \araa, 54, 491

\bibitem[{{Barsony} {et~al.}(1998){Barsony}, {Ward-Thompson}, {Andr{\'e}}, \&
  {O'Linger}}]{Barsony1998}
{Barsony}, M., {Ward-Thompson}, D., {Andr{\'e}}, P., \& {O'Linger}, J. 1998,
  \apj, 509, 733

\bibitem[{{Belloche}(2013)}]{Belloche2013}
{Belloche}, A. 2013, in EAS Publications Series, Vol.~62, EAS Publications
  Series, ed. P.~{Hennebelle} \& C.~{Charbonnel}, 25--66

\bibitem[{{Belloche} {et~al.}(2006){Belloche}, {Hennebelle}, \&
  {Andr{\'e}}}]{Belloche2006}
{Belloche}, A., {Hennebelle}, P., \& {Andr{\'e}}, P. 2006, \aap, 453, 145

\bibitem[{{Bertrang} {et~al.}(2014){Bertrang}, {Wolf}, \& {Das}}]{Bertrang2014}
{Bertrang}, G., {Wolf}, S., \& {Das}, H.~S. 2014, \aap, 565, A94

\bibitem[{{Bethell} {et~al.}(2007){Bethell}, {Chepurnov}, {Lazarian}, \&
  {Kim}}]{Bethell2007}
{Bethell}, T.~J., {Chepurnov}, A., {Lazarian}, A., \& {Kim}, J. 2007, \apj,
  663, 1055

\bibitem[{{Bock} {et~al.}(2006){Bock}, {Bolatto}, {Hawkins}, {Kemball}, {Lamb},
  {Plambeck}, {Pound}, {Scott}, {Woody}, \& {Wright}}]{Bock2006}
{Bock}, D.~C.-J., {Bolatto}, A.~D., {Hawkins}, D.~W., {et~al.} 2006, in
  \procspie, Vol. 6267, Society of Photo-Optical Instrumentation Engineers
  (SPIE) Conference Series, 626713

\bibitem[{{Bodenheimer}(1995)}]{Bodenheimer1995}
{Bodenheimer}, P. 1995, \araa, 33, 199

\bibitem[{{Bontemps} {et~al.}(1996){Bontemps}, {Andre}, {Terebey}, \&
  {Cabrit}}]{Bontemps1996}
{Bontemps}, S., {Andre}, P., {Terebey}, S., \& {Cabrit}, S. 1996, \aap, 311,
  858

\bibitem[{{Brauer} {et~al.}(2016){Brauer}, {Wolf}, \& {Reissl}}]{Brauer2016}
{Brauer}, R., {Wolf}, S., \& {Reissl}, S. 2016, \aap, 588, A129

\bibitem[{{Caselli} {et~al.}(2002){Caselli}, {Benson}, {Myers}, \&
  {Tafalla}}]{Caselli02}
{Caselli}, P., {Benson}, P.~J., {Myers}, P.~C., \& {Tafalla}, M. 2002, \apj,
  572, 238

\bibitem[{{Chac{\'o}n-Tanarro} {et~al.}(2017){Chac{\'o}n-Tanarro}, {Caselli},
  {Bizzocchi}, {Pineda}, {Harju}, {Spaans}, \&
  {D{\'e}sert}}]{Chacon-Tanarro2017}
{Chac{\'o}n-Tanarro}, A., {Caselli}, P., {Bizzocchi}, L., {et~al.} 2017, \aap,
  606, A142

\bibitem[{{Chandler} {et~al.}(2005){Chandler}, {Brogan}, {Shirley}, \&
  {Loinard}}]{Chandler2005}
{Chandler}, C.~J., {Brogan}, C.~L., {Shirley}, Y.~L., \& {Loinard}, L. 2005,
  \apj, 632, 371

\bibitem[{{Chandler} \& {Richer}(2000)}]{Chandler2000}
{Chandler}, C.~J. \& {Richer}, J.~S. 2000, \apj, 530, 851

\bibitem[{{Chapman} {et~al.}(2013){Chapman}, {Davidson}, {Goldsmith}, {Houde},
  {Kwon}, {Li}, {Looney}, {Matthews}, {Matthews}, {Novak}, {Peng},
  {Vaillancourt}, \& {Volgenau}}]{Chapman2013}
{Chapman}, N.~L., {Davidson}, J.~A., {Goldsmith}, P.~F., {et~al.} 2013, \apj,
  770, 151

\bibitem[{{Chen} {et~al.}(2013){Chen}, {Arce}, {Zhang}, {Bourke}, {Launhardt},
  {J{\o}rgensen}, {Lee}, {Foster}, {Dunham}, {Pineda}, \& {Henning}}]{Chen2013}
{Chen}, X., {Arce}, H.~G., {Zhang}, Q., {et~al.} 2013, \apj, 768, 110

\bibitem[{{Chen} {et~al.}(2007){Chen}, {Launhardt}, \& {Henning}}]{Chen2007}
{Chen}, X., {Launhardt}, R., \& {Henning}, T. 2007, \apj, 669, 1058

\bibitem[{{Chen} {et~al.}(2009){Chen}, {Launhardt}, \& {Henning}}]{Chen2009}
{Chen}, X., {Launhardt}, R., \& {Henning}, T. 2009, \apj, 691, 1729

\bibitem[{{Chiang} {et~al.}(2012){Chiang}, {Looney}, \& {Tobin}}]{Chiang2012}
{Chiang}, H.-F., {Looney}, L.~W., \& {Tobin}, J.~J. 2012, \apj, 756, 168

\bibitem[{{Ching} {et~al.}(2016){Ching}, {Lai}, {Zhang}, {Yang}, {Girart}, \&
  {Rao}}]{Ching2016}
{Ching}, T.-C., {Lai}, S.-P., {Zhang}, Q., {et~al.} 2016, \apj, 819, 159

\bibitem[{{Choi}(2001)}]{Choi2001}
{Choi}, M. 2001, \apj, 553, 219

\bibitem[{{Choi}(2005)}]{Choi2005}
{Choi}, M. 2005, \apj, 630, 976

\bibitem[{{Ciardi} \& {Hennebelle}(2010)}]{CiardiHennebelle2010}
{Ciardi}, A. \& {Hennebelle}, P. 2010, \mnras, 409, L39

\bibitem[{{Ciardi} {et~al.}(2003){Ciardi}, {Telesco}, {Williams}, {Fisher},
  {Packham}, {Pi{\~n}a}, \& {Radomski}}]{Ciardi2003}
{Ciardi}, D.~R., {Telesco}, C.~M., {Williams}, J.~P., {et~al.} 2003, \apj, 585,
  392

\bibitem[{{Correia} {et~al.}(2004){Correia}, {Griffin}, \&
  {Saraceno}}]{Correia2004}
{Correia}, J.~C., {Griffin}, M., \& {Saraceno}, P. 2004, \aap, 418, 607

\bibitem[{{Crimier} {et~al.}(2010){Crimier}, {Ceccarelli}, {Maret},
  {Bottinelli}, {Caux}, {Kahane}, {Lis}, \& {Olofsson}}]{Crimier2010}
{Crimier}, N., {Ceccarelli}, C., {Maret}, S., {et~al.} 2010, \aap, 519, A65

\bibitem[{{Crutcher}(2012)}]{Crutcher2012}
{Crutcher}, R.~M. 2012, \araa, 50, 29

\bibitem[{{Crutcher} {et~al.}(2004){Crutcher}, {Nutter}, {Ward-Thompson}, \&
  {Kirk}}]{Crutcher2004}
{Crutcher}, R.~M., {Nutter}, D.~J., {Ward-Thompson}, D., \& {Kirk}, J.~M. 2004,
  \apj, 600, 279

\bibitem[{{Curiel} {et~al.}(1990){Curiel}, {Raymond}, {Moran}, {Rodriguez}, \&
  {Canto}}]{Curiel1990}
{Curiel}, S., {Raymond}, J.~C., {Moran}, J.~M., {Rodriguez}, L.~F., \& {Canto},
  J. 1990, \apjl, 365, L85

\bibitem[{{Curiel} {et~al.}(1999){Curiel}, {Torrelles}, {Rodr{\'{\i}}guez},
  {G{\'o}mez}, \& {Anglada}}]{Curiel1999}
{Curiel}, S., {Torrelles}, J.~M., {Rodr{\'{\i}}guez}, L.~F., {G{\'o}mez},
  J.~F., \& {Anglada}, G. 1999, \apj, 527, 310

\bibitem[{{Curran} \& {Chrysostomou}(2007)}]{Curran2007}
{Curran}, R.~L. \& {Chrysostomou}, A. 2007, \mnras, 382, 699

\bibitem[{{Davidson} {et~al.}(2011){Davidson}, {Novak}, {Matthews}, {Matthews},
  {Goldsmith}, {Chapman}, {Volgenau}, {Vaillancourt}, \&
  {Attard}}]{Davidson2011}
{Davidson}, J.~A., {Novak}, G., {Matthews}, T.~G., {et~al.} 2011, \apj, 732, 97

\bibitem[{{Di Francesco} {et~al.}(2008){Di Francesco}, {Johnstone}, {Kirk},
  {MacKenzie}, \& {Ledwosinska}}]{DiFrancesco2008}
{Di Francesco}, J., {Johnstone}, D., {Kirk}, H., {MacKenzie}, T., \&
  {Ledwosinska}, E. 2008, \apjs, 175, 277

\bibitem[{{Di Francesco} {et~al.}(2001){Di Francesco}, {Myers}, {Wilner},
  {Ohashi}, \& {Mardones}}]{DiFrancesco2001}
{Di Francesco}, J., {Myers}, P.~C., {Wilner}, D.~J., {Ohashi}, N., \&
  {Mardones}, D. 2001, \apj, 562, 770

\bibitem[{{Dotson}(1996)}]{Dotson1996}
{Dotson}, J.~L. 1996, \apj, 470, 566

\bibitem[{{Enoch} {et~al.}(2011){Enoch}, {Corder}, {Duch{\^e}ne}, {Bock},
  {Bolatto}, {Culverhouse}, {Kwon}, {Lamb}, {Leitch}, {Marrone}, {Muchovej},
  {P{\'e}rez}, {Scott}, {Teuben}, {Wright}, \& {Zauderer}}]{Enoch2011}
{Enoch}, M.~L., {Corder}, S., {Duch{\^e}ne}, G., {et~al.} 2011, \apjs, 195, 21

\bibitem[{{Evans} {et~al.}(2015){Evans}, {Di Francesco}, {Lee}, {J{\o}rgensen},
  {Choi}, {Myers}, \& {Mardones}}]{Evans2015}
{Evans}, II, N.~J., {Di Francesco}, J., {Lee}, J.-E., {et~al.} 2015, \apj, 814,
  22

\bibitem[{{Evans} {et~al.}(2009){Evans}, {Dunham}, {J{\o}rgensen}, {Enoch},
  {Mer{\'{\i}}n}, {van Dishoeck}, {Alcal{\'a}}, {Myers}, {Stapelfeldt},
  {Huard}, {Allen}, {Harvey}, {van Kempen}, {Blake}, {Koerner}, {Mundy},
  {Padgett}, \& {Sargent}}]{Evans2009}
{Evans}, II, N.~J., {Dunham}, M.~M., {J{\o}rgensen}, J.~K., {et~al.} 2009,
  \apjs, 181, 321

\bibitem[{{Falceta-Gon{\c c}alves} {et~al.}(2008){Falceta-Gon{\c c}alves},
  {Lazarian}, \& {Kowal}}]{FalcetaGoncalves2008}
{Falceta-Gon{\c c}alves}, D., {Lazarian}, A., \& {Kowal}, G. 2008, \apj, 679,
  537

\bibitem[{{Frau} {et~al.}(2011){Frau}, {Galli}, \& {Girart}}]{Frau2011}
{Frau}, P., {Galli}, D., \& {Girart}, J.~M. 2011, \aap, 535, A44

\bibitem[{{G{\^a}lfalk} \& {Olofsson}(2007)}]{Galfalk2007}
{G{\^a}lfalk}, M. \& {Olofsson}, G. 2007, \aap, 475, 281

\bibitem[{{Galli} {et~al.}(2006){Galli}, {Lizano}, {Shu}, \&
  {Allen}}]{Galli2006}
{Galli}, D., {Lizano}, S., {Shu}, F.~H., \& {Allen}, A. 2006, \apj, 647, 374

\bibitem[{{Girart} {et~al.}(1999){Girart}, {Crutcher}, \& {Rao}}]{Girart1999}
{Girart}, J.~M., {Crutcher}, R.~M., \& {Rao}, R. 1999, \apjl, 525, L109

\bibitem[{{Girart} {et~al.}(2014){Girart}, {Estalella}, {Palau}, {Torrelles},
  \& {Rao}}]{Girart2014}
{Girart}, J.~M., {Estalella}, R., {Palau}, A., {Torrelles}, J.~M., \& {Rao}, R.
  2014, \apjl, 780, L11

\bibitem[{{Girart} {et~al.}(2013){Girart}, {Frau}, {Zhang}, {Koch}, {Qiu},
  {Tang}, {Lai}, \& {Ho}}]{Girart2013}
{Girart}, J.~M., {Frau}, P., {Zhang}, Q., {et~al.} 2013, \apj, 772, 69

\bibitem[{{Girart} {et~al.}(2006){Girart}, {Rao}, \& {Marrone}}]{Girart2006}
{Girart}, J.~M., {Rao}, R., \& {Marrone}, D.~P. 2006, Science, 313, 812

\bibitem[{{Gon{\c c}alves} {et~al.}(2008){Gon{\c c}alves}, {Galli}, \&
  {Girart}}]{Goncalves2008}
{Gon{\c c}alves}, J., {Galli}, D., \& {Girart}, J.~M. 2008, \aap, 490, L39

\bibitem[{{Goodman} {et~al.}(1993){Goodman}, {Benson}, {Fuller}, \&
  {Myers}}]{Goodman93}
{Goodman}, A.~A., {Benson}, P.~J., {Fuller}, G.~A., \& {Myers}, P.~C. 1993,
  \apj, 406, 528

\bibitem[{{Gueth} {et~al.}(1996){Gueth}, {Guilloteau}, \&
  {Bachiller}}]{Gueth1996}
{Gueth}, F., {Guilloteau}, S., \& {Bachiller}, R. 1996, \aap, 307, 891

\bibitem[{{Harvey} {et~al.}(2003){Harvey}, {Wilner}, {Myers}, \&
  {Tafalla}}]{Harvey2003}
{Harvey}, D.~W.~A., {Wilner}, D.~J., {Myers}, P.~C., \& {Tafalla}, M. 2003,
  \apj, 596, 383

\bibitem[{{Hennebelle} \& {Ciardi}(2009)}]{HennebelleCiardi2009}
{Hennebelle}, P. \& {Ciardi}, A. 2009, \aap, 506, L29

\bibitem[{{Henning} {et~al.}(2001){Henning}, {Wolf}, {Launhardt}, \&
  {Waters}}]{Henning2001}
{Henning}, T., {Wolf}, S., {Launhardt}, R., \& {Waters}, R. 2001, \apj, 561,
  871

\bibitem[{{Hildebrand}(1983)}]{Hildebrand1983}
{Hildebrand}, R.~H. 1983, \qjras, 24, 267

\bibitem[{{Hirano} {et~al.}(2010){Hirano}, {Ho}, {Liu}, {Shang}, {Lee}, \&
  {Bourke}}]{Hirano2010}
{Hirano}, N., {Ho}, P.~P.~T., {Liu}, S.-Y., {et~al.} 2010, \apj, 717, 58

\bibitem[{{Hirano} {et~al.}(1988){Hirano}, {Kameya}, {Nakayama}, \&
  {Takakubo}}]{Hirano1988}
{Hirano}, N., {Kameya}, O., {Nakayama}, M., \& {Takakubo}, K. 1988, \apjl, 327,
  L69

\bibitem[{{Hirota} {et~al.}(2008){Hirota}, {Bushimata}, {Choi}, {Honma},
  {Imai}, {Iwadate}, {Jike}, {Kameya}, {Kamohara}, {Kan-Ya}, {Kawaguchi},
  {et~al.}}]{Hirota2008}
{Hirota}, T., {Bushimata}, T., {Choi}, Y.~K., {et~al.} 2008, \pasj, 60, 37

\bibitem[{{Hirota} {et~al.}(2011){Hirota}, {Honma}, {Imai}, {Sunada}, {Ueno},
  {Kobayashi}, \& {Kawaguchi}}]{Hirota2011}
{Hirota}, T., {Honma}, M., {Imai}, H., {et~al.} 2011, \pasj, 63, 1

\bibitem[{{Ho} {et~al.}(2004){Ho}, {Moran}, \& {Lo}}]{Ho2004}
{Ho}, P.~T.~P., {Moran}, J.~M., \& {Lo}, K.~Y. 2004, \apjl, 616, L1

\bibitem[{{Hoang} \& {Lazarian}(2009)}]{Hoang2009}
{Hoang}, T. \& {Lazarian}, A. 2009, \apj, 695, 1457

\bibitem[{{Holland} {et~al.}(1999){Holland}, {Robson}, {Gear}, {Cunningham},
  {Lightfoot}, {Jenness}, {Ivison}, {Stevens}, {Ade}, {Griffin}, {Duncan},
  {Murphy}, \& {Naylor}}]{Holland1999}
{Holland}, W.~S., {Robson}, E.~I., {Gear}, W.~K., {et~al.} 1999, \mnras, 303,
  659

\bibitem[{{Hull} {et~al.}(2017){Hull}, {Girart}, {Tychoniec}, {Rao},
  {Cort{\'e}s}, {Pokhrel}, {Zhang}, {Houde}, {Dunham}, {Kristensen}, {Lai},
  {Li}, \& {Plambeck}}]{Hull2017}
{Hull}, C.~L.~H., {Girart}, J.~M., {Tychoniec}, {\L}., {et~al.} 2017, \apj,
  847, 92

\bibitem[{{Hull} {et~al.}(2013){Hull}, {Plambeck}, {Bolatto}, {Bower},
  {Carpenter}, {Crutcher}, {Fiege}, {Franzmann}, {Hakobian}, {Heiles}, {Houde},
  {Hughes}, {Jameson}, {Kwon}, {Lamb}, {Looney}, {Matthews}, {Mundy}, {Pillai},
  {Pound}, {Stephens}, {Tobin}, {Vaillancourt}, {Volgenau}, \&
  {Wright}}]{Hull2013}
{Hull}, C.~L.~H., {Plambeck}, R.~L., {Bolatto}, A.~D., {et~al.} 2013, \apj,
  768, 159

\bibitem[{{Hull} {et~al.}(2014){Hull}, {Plambeck}, {Kwon}, {Bower},
  {Carpenter}, {Crutcher}, {Fiege}, {Franzmann}, {Hakobian}, {Heiles}, {Houde},
  {Hughes}, {et~al.}}]{Hull2014}
{Hull}, C.~L.~H., {Plambeck}, R.~L., {Kwon}, W., {et~al.} 2014, \apjs, 213, 13

\bibitem[{{Jennings} {et~al.}(1987){Jennings}, {Cameron}, {Cudlip}, \&
  {Hirst}}]{Jennings1987}
{Jennings}, R.~E., {Cameron}, D.~H.~M., {Cudlip}, W., \& {Hirst}, C.~J. 1987,
  \mnras, 226, 461

\bibitem[{{Jones} {et~al.}(2015){Jones}, {Bagley}, {Krejny}, {Andersson}, \&
  {Bastien}}]{Jones2015}
{Jones}, T.~J., {Bagley}, M., {Krejny}, M., {Andersson}, B.-G., \& {Bastien},
  P. 2015, \aj, 149, 31

\bibitem[{{Jones} {et~al.}(2016){Jones}, {Gordon}, {Shenoy}, {Gehrz},
  {Vaillancourt}, \& {Krejny}}]{Jones2016}
{Jones}, T.~J., {Gordon}, M., {Shenoy}, D., {et~al.} 2016, \aj, 151, 156

\bibitem[{{Joos} {et~al.}(2012){Joos}, {Hennebelle}, \& {Ciardi}}]{Joos2012}
{Joos}, M., {Hennebelle}, P., \& {Ciardi}, A. 2012, \aap, 543, A128

\bibitem[{{J{\o}rgensen} {et~al.}(2007){J{\o}rgensen}, {Bourke}, {Myers}, {Di
  Francesco}, {van Dishoeck}, {Lee}, {Ohashi}, {Sch{\"o}ier}, {Takakuwa},
  {Wilner}, \& {Zhang}}]{Jorgensen2007}
{J{\o}rgensen}, J.~K., {Bourke}, T.~L., {Myers}, P.~C., {et~al.} 2007, \apj,
  659, 479

\bibitem[{{J{\o}rgensen} {et~al.}(2006){J{\o}rgensen}, {Harvey}, {Evans},
  {Huard}, {Allen}, {Porras}, {Blake}, {Bourke}, {Chapman}, {Cieza}, {Koerner},
  {Lai}, {Mundy}, {Myers}, {Padgett}, {Rebull}, {Sargent}, {Spiesman},
  {Stapelfeldt}, {van Dishoeck}, {Wahhaj}, \& {Young}}]{Jorgensen2006}
{J{\o}rgensen}, J.~K., {Harvey}, P.~M., {Evans}, II, N.~J., {et~al.} 2006,
  \apj, 645, 1246

\bibitem[{{J{\o}rgensen} {et~al.}(2016){J{\o}rgensen}, {van der Wiel},
  {Coutens}, {Lykke}, {M{\"u}ller}, {van Dishoeck}, {Calcutt}, {Bjerkeli},
  {Bourke}, {Drozdovskaya}, {Favre}, {Fayolle}, {Garrod}, {Jacobsen},
  {{\"O}berg}, {Persson}, \& {Wampfler}}]{Jorgensen2016}
{J{\o}rgensen}, J.~K., {van der Wiel}, M.~H.~D., {Coutens}, A., {et~al.} 2016,
  \aap, 595, A117

\bibitem[{{Kataoka} {et~al.}(2012){Kataoka}, {Machida}, \&
  {Tomisaka}}]{Kataoka2012}
{Kataoka}, A., {Machida}, M.~N., \& {Tomisaka}, K. 2012, \apj, 761, 40

\bibitem[{{Kataoka} {et~al.}(2015){Kataoka}, {Muto}, {Momose}, {Tsukagoshi},
  {Fukagawa}, {Shibai}, {Hanawa}, {Murakawa}, \& {Dullemond}}]{Kataoka2015}
{Kataoka}, A., {Muto}, T., {Momose}, M., {et~al.} 2015, \apj, 809, 78

\bibitem[{{Kauffmann} {et~al.}(2008){Kauffmann}, {Bertoldi}, {Bourke}, {Evans},
  \& {Lee}}]{Kauffmann2008}
{Kauffmann}, J., {Bertoldi}, F., {Bourke}, T.~L., {Evans}, II, N.~J., \& {Lee},
  C.~W. 2008, \aap, 487, 993

\bibitem[{{Keene} {et~al.}(1983){Keene}, {Davidson}, {Harper}, {Hildebrand},
  {Jaffe}, {Loewenstein}, {Low}, \& {Pernic}}]{Keene1983}
{Keene}, J., {Davidson}, J.~A., {Harper}, D.~A., {et~al.} 1983, \apjl, 274, L43

\bibitem[{{Knude} \& {Hog}(1998)}]{Knude_Hog_1998}
{Knude}, J. \& {Hog}, E. 1998, \aap, 338, 897

\bibitem[{{Koumpia} {et~al.}(2016){Koumpia}, {van der Tak}, {Kwon}, {Tobin},
  {Fuller}, \& {Plume}}]{Koumpia2016}
{Koumpia}, E., {van der Tak}, F.~F.~S., {Kwon}, W., {et~al.} 2016, \aap, 595,
  A51

\bibitem[{{Krumholz} {et~al.}(2013){Krumholz}, {Crutcher}, \&
  {Hull}}]{Krumholz2013}
{Krumholz}, M.~R., {Crutcher}, R.~M., \& {Hull}, C.~L.~H. 2013, \apjl, 767, L11

\bibitem[{{Kwon} {et~al.}(2015){Kwon}, {Fern{\'a}ndez-L{\'o}pez}, {Stephens},
  \& {Looney}}]{Kwon2015}
{Kwon}, W., {Fern{\'a}ndez-L{\'o}pez}, M., {Stephens}, I.~W., \& {Looney},
  L.~W. 2015, \apj, 814, 43

\bibitem[{{Kwon} {et~al.}(2006){Kwon}, {Looney}, {Crutcher}, \&
  {Kirk}}]{Kwon2006}
{Kwon}, W., {Looney}, L.~W., {Crutcher}, R.~M., \& {Kirk}, J.~M. 2006, \apj,
  653, 1358

\bibitem[{{Kwon} {et~al.}(2009){Kwon}, {Looney}, {Mundy}, {Chiang}, \&
  {Kemball}}]{Kwon2009}
{Kwon}, W., {Looney}, L.~W., {Mundy}, L.~G., {Chiang}, H.-F., \& {Kemball},
  A.~J. 2009, \apj, 696, 841

\bibitem[{{Lai} {et~al.}(2002){Lai}, {Crutcher}, {Girart}, \& {Rao}}]{Lai2002}
{Lai}, S.-P., {Crutcher}, R.~M., {Girart}, J.~M., \& {Rao}, R. 2002, \apj, 566,
  925

\bibitem[{{Launhardt}(2001)}]{Launhardt2001}
{Launhardt}, R. 2001, in IAU Symposium, Vol. 200, The Formation of Binary
  Stars, ed. H.~{Zinnecker} \& R.~{Mathieu}, 117

\bibitem[{{Launhardt}(2004)}]{Launhardt2004}
{Launhardt}, R. 2004, in IAU Symposium, Vol. 221, Star Formation at High
  Angular Resolution, ed. M.~G. {Burton}, R.~{Jayawardhana}, \& T.~L. {Bourke},
  213

\bibitem[{{Launhardt} {et~al.}(2013){Launhardt}, {Stutz}, {Schmiedeke},
  {Henning}, {Krause}, {Balog}, {Beuther}, {Birkmann}, {Hennemann},
  {Kainulainen}, {Khanzadyan}, {Linz}, {Lippok}, {Nielbock}, {Pitann}, {Ragan},
  {Risacher}, {Schmalzl}, {Shirley}, {Stecklum}, {Steinacker}, \&
  {Tackenberg}}]{Launhardt2013}
{Launhardt}, R., {Stutz}, A.~M., {Schmiedeke}, A., {et~al.} 2013, \aap, 551,
  A98

\bibitem[{{Lazarian}(2007)}]{Lazarian2007}
{Lazarian}, A. 2007, \jqsrt, 106, 225

\bibitem[{{Lee} {et~al.}(2017){Lee}, {Hull}, \& {Offner}}]{Lee2017}
{Lee}, J.~W.~Y., {Hull}, C.~L.~H., \& {Offner}, S.~S.~R. 2017, \apj, 834, 201

\bibitem[{{Lee} {et~al.}(2015){Lee}, {Dunham}, {Myers}, {Tobin}, {Kristensen},
  {Pineda}, {Vorobyov}, {Offner}, {Arce}, {Li}, {Bourke}, {J{\o}rgensen},
  {Goodman}, {Sadavoy}, {Chandler}, {Harris}, {Kratter}, {Looney}, {Melis},
  {Perez}, \& {Segura-Cox}}]{Lee2015}
{Lee}, K.~I., {Dunham}, M.~M., {Myers}, P.~C., {et~al.} 2015, \apj, 814, 114

\bibitem[{{Lefloch} {et~al.}(1998){Lefloch}, {Castets}, {Cernicharo}, {Langer},
  \& {Zylka}}]{Lefloch1998}
{Lefloch}, B., {Castets}, A., {Cernicharo}, J., {Langer}, W.~D., \& {Zylka}, R.
  1998, \aap, 334, 269

\bibitem[{{Li} {et~al.}(2014){Li}, {Goodman}, {Sridharan}, {Houde}, {Li},
  {Novak}, \& {Tang}}]{Li2014}
{Li}, H.-B., {Goodman}, A., {Sridharan}, T.~K., {et~al.} 2014, Protostars and
  Planets VI, 101

\bibitem[{{Li} {et~al.}(2013){Li}, {Krasnopolsky}, \& {Shang}}]{Li2013}
{Li}, Z.-Y., {Krasnopolsky}, R., \& {Shang}, H. 2013, \apj, 774, 82

\bibitem[{{Liu} {et~al.}(2016){Liu}, {Lai}, {Hasegawa}, {Hirano}, {Rao}, {Li},
  {Fukagawa}, {Girart}, {Carrasco-Gonz{\'a}lez}, \&
  {Rodr{\'{\i}}guez}}]{Liu2016}
{Liu}, H.~B., {Lai}, S.-P., {Hasegawa}, Y., {et~al.} 2016, \apj, 821, 41

\bibitem[{{Loinard} {et~al.}(2013){Loinard}, {Zapata}, {Rodr{\'{\i}}guez},
  {Pech}, {Chandler}, {Brogan}, {Wilner}, {Ho}, {Parise}, {Hartmann}, {Zhu},
  {Takahashi}, \& {Trejo}}]{Loinard2013}
{Loinard}, L., {Zapata}, L.~A., {Rodr{\'{\i}}guez}, L.~F., {et~al.} 2013,
  \mnras, 430, L10

\bibitem[{{Looney} {et~al.}(2000){Looney}, {Mundy}, \& {Welch}}]{Looney2000}
{Looney}, L.~W., {Mundy}, L.~G., \& {Welch}, W.~J. 2000, \apj, 529, 477

\bibitem[{{Looney} {et~al.}(2003){Looney}, {Mundy}, \& {Welch}}]{Looney2003}
{Looney}, L.~W., {Mundy}, L.~G., \& {Welch}, W.~J. 2003, \apj, 592, 255

\bibitem[{{Looney} {et~al.}(2007){Looney}, {Tobin}, \& {Kwon}}]{Looney2007}
{Looney}, L.~W., {Tobin}, J.~J., \& {Kwon}, W. 2007, \apjl, 670, L131

\bibitem[{{L{\'o}pez-Sepulcre} {et~al.}(2017){L{\'o}pez-Sepulcre}, {Sakai},
  {Neri}, {Imai}, {Oya}, {Ceccarelli}, {Higuchi}, {Aikawa}, {Bottinelli},
  {Caux}, {Hirota}, {Kahane}, {Lefloch}, {Vastel}, {Watanabe}, \&
  {Yamamoto}}]{Lopez-Sepulcre2017}
{L{\'o}pez-Sepulcre}, A., {Sakai}, N., {Neri}, R., {et~al.} 2017, \aap, 606,
  A121

\bibitem[{{Machida} {et~al.}(2007){Machida}, {Inutsuka}, \&
  {Matsumoto}}]{Machida2007}
{Machida}, M.~N., {Inutsuka}, S.-i., \& {Matsumoto}, T. 2007, \apj, 670, 1198

\bibitem[{{Machida} {et~al.}(2005){Machida}, {Matsumoto}, {Tomisaka}, \&
  {Hanawa}}]{Machida2005}
{Machida}, M.~N., {Matsumoto}, T., {Tomisaka}, K., \& {Hanawa}, T. 2005,
  \mnras, 362, 369

\bibitem[{{Marrone}(2006)}]{Marrone2006}
{Marrone}, D.~P. 2006, PhD thesis, Harvard University

\bibitem[{{Marrone} \& {Rao}(2008)}]{Marrone2008}
{Marrone}, D.~P. \& {Rao}, R. 2008, in \procspie, Vol. 7020, Millimeter and
  Submillimeter Detectors and Instrumentation for Astronomy IV, 70202B

\bibitem[{{Marvel} {et~al.}(2008){Marvel}, {Wilking}, {Claussen}, \&
  {Wootten}}]{Marvel2008}
{Marvel}, K.~B., {Wilking}, B.~A., {Claussen}, M.~J., \& {Wootten}, A. 2008,
  \apj, 685, 285

\bibitem[{{Massi} {et~al.}(2008){Massi}, {Codella}, {Brand}, {di Fabrizio}, \&
  {Wouterloot}}]{Massi2008}
{Massi}, F., {Codella}, C., {Brand}, J., {di Fabrizio}, L., \& {Wouterloot},
  J.~G.~A. 2008, \aap, 490, 1079

\bibitem[{{Matthews} {et~al.}(2009){Matthews}, {McPhee}, {Fissel}, \&
  {Curran}}]{Matthews2009}
{Matthews}, B.~C., {McPhee}, C.~A., {Fissel}, L.~M., \& {Curran}, R.~L. 2009,
  \apjs, 182, 143

\bibitem[{{Matthews} \& {Wilson}(2000)}]{Matthews_Wilson_2000}
{Matthews}, B.~C. \& {Wilson}, C.~D. 2000, \apj, 531, 868

\bibitem[{{Maury} {et~al.}(2010){Maury}, {Andr{\'e}}, {Hennebelle}, {Motte},
  {Stamatellos}, {Bate}, {Belloche}, {Duch{\^e}ne}, \& {Whitworth}}]{Maury2010}
{Maury}, A.~J., {Andr{\'e}}, P., {Hennebelle}, P., {et~al.} 2010, \aap, 512,
  A40+

\bibitem[{{Maury} {et~al.}(2011){Maury}, {Andr{\'e}}, {Men'shchikov},
  {K{\"o}nyves}, \& {Bontemps}}]{Maury2011}
{Maury}, A.~J., {Andr{\'e}}, P., {Men'shchikov}, A., {K{\"o}nyves}, V., \&
  {Bontemps}, S. 2011, \aap, 535, A77

\bibitem[{{Maury} {et~al.}(2014){Maury}, {Belloche}, {Andr{\'e}}, {Maret},
  {Gueth}, {Codella}, {Cabrit}, {Testi}, \& {Bontemps}}]{Maury2014}
{Maury}, A.~J., {Belloche}, A., {Andr{\'e}}, P., {et~al.} 2014, \aap, 563, L2

\bibitem[{{Maury} {et~al.}(2018){Maury}, {Girart}, {Zhang}, {Hennebelle},
  {Keto}, {Rao}, {Lai}, {Ohashi}, \& {Galametz}}]{Maury2018}
{Maury}, A.~J., {Girart}, J.~M., {Zhang}, Q., {et~al.} 2018, accepted for
  publication in \mnras

\bibitem[{{Motte} \& {Andr{\'e}}(2001)}]{Motte2001}
{Motte}, F. \& {Andr{\'e}}, P. 2001, \aap, 365, 440

\bibitem[{{Ohashi} {et~al.}(2014){Ohashi}, {Saigo}, {Aso}, {Aikawa},
  {Koyamatsu}, {Machida}, {Saito}, {Takahashi}, {Takakuwa}, {Tomida},
  {Tomisaka}, \& {Yen}}]{Ohashi2014}
{Ohashi}, N., {Saigo}, K., {Aso}, Y., {et~al.} 2014, \apj, 796, 131

\bibitem[{{O'Linger} {et~al.}(1999){O'Linger}, {Wolf-Chase}, {Barsony}, \&
  {Ward-Thompson}}]{OLinger1999}
{O'Linger}, J., {Wolf-Chase}, G., {Barsony}, M., \& {Ward-Thompson}, D. 1999,
  \apj, 515, 696

\bibitem[{{Ossenkopf} \& {Henning}(1994)}]{Ossenkopf1994}
{Ossenkopf}, V. \& {Henning}, T. 1994, \aap, 291, 943

\bibitem[{{Oya} {et~al.}(2016){Oya}, {Sakai}, {L{\'o}pez-Sepulcre}, {Watanabe},
  {Ceccarelli}, {Lefloch}, {Favre}, \& {Yamamoto}}]{Oya2016}
{Oya}, Y., {Sakai}, N., {L{\'o}pez-Sepulcre}, A., {et~al.} 2016, \apj, 824, 88

\bibitem[{{Padoan} {et~al.}(2001){Padoan}, {Goodman}, {Draine}, {Juvela},
  {Nordlund}, \& {R{\"o}gnvaldsson}}]{Padoan2001}
{Padoan}, P., {Goodman}, A., {Draine}, B.~T., {et~al.} 2001, \apj, 559, 1005

\bibitem[{{Palau} {et~al.}(2014){Palau}, {Zapata}, {Rodr{\'{\i}}guez}, {Bouy},
  {Barrado}, {Morales-Calder{\'o}n}, {Myers}, {Chapman}, {Ju{\'a}rez}, \&
  {Li}}]{Palau2014}
{Palau}, A., {Zapata}, L.~A., {Rodr{\'{\i}}guez}, L.~F., {et~al.} 2014, \mnras,
  444, 833

\bibitem[{{Pech} {et~al.}(2012){Pech}, {Zapata}, {Loinard}, \&
  {Rodr{\'{\i}}guez}}]{Pech2012}
{Pech}, G., {Zapata}, L.~A., {Loinard}, L., \& {Rodr{\'{\i}}guez}, L.~F. 2012,
  \apj, 751, 78

\bibitem[{{Pelkonen} {et~al.}(2009){Pelkonen}, {Juvela}, \&
  {Padoan}}]{Pelkonen2009}
{Pelkonen}, V.-M., {Juvela}, M., \& {Padoan}, P. 2009, \aap, 502, 833

\bibitem[{{Pineda} {et~al.}(2012){Pineda}, {Maury}, {Fuller}, {Testi},
  {Garc{\'{\i}}a-Appadoo}, {Peck}, {Villard}, {Corder}, {van Kempen}, {Turner},
  {Tachihara}, \& {Dent}}]{Pineda2012}
{Pineda}, J.~E., {Maury}, A.~J., {Fuller}, G.~A., {et~al.} 2012, \aap, 544, L7

\bibitem[{{Planck Collaboration}(2016)}]{PlanckCollaboration2016_MF}
{Planck Collaboration}. 2016, \aap, 586, A138

\bibitem[{{Plunkett} {et~al.}(2013){Plunkett}, {Arce}, {Corder}, {Mardones},
  {Sargent}, \& {Schnee}}]{Plunkett2013}
{Plunkett}, A.~L., {Arce}, H.~G., {Corder}, S.~A., {et~al.} 2013, \apj, 774, 22

\bibitem[{{Podio} {et~al.}(2016){Podio}, {Codella}, {Gueth}, {Cabrit}, {Maury},
  {Tabone}, {Lef{\`e}vre}, {Anderl}, {Andr{\'e}}, {Belloche}, {Bontemps},
  {Hennebelle}, {Lefloch}, {Maret}, \& {Testi}}]{Podio2016}
{Podio}, L., {Codella}, C., {Gueth}, F., {et~al.} 2016, \aap, 593, L4

\bibitem[{{Poidevin} {et~al.}(2010){Poidevin}, {Bastien}, \&
  {Matthews}}]{Poidevin2010}
{Poidevin}, F., {Bastien}, P., \& {Matthews}, B.~C. 2010, \apj, 716, 893

\bibitem[{{Pudritz} \& {Norman}(1983)}]{Pudritz1983}
{Pudritz}, R.~E. \& {Norman}, C.~A. 1983, \apj, 274, 677

\bibitem[{{Qiu} {et~al.}(2013){Qiu}, {Zhang}, {Menten}, {Liu}, \&
  {Tang}}]{Qiu2013}
{Qiu}, K., {Zhang}, Q., {Menten}, K.~M., {Liu}, H.~B., \& {Tang}, Y.-W. 2013,
  \apj, 779, 182

\bibitem[{{Rao} {et~al.}(1998){Rao}, {Crutcher}, {Plambeck}, \&
  {Wright}}]{Rao1998}
{Rao}, R., {Crutcher}, R.~M., {Plambeck}, R.~L., \& {Wright}, M.~C.~H. 1998,
  \apjl, 502, L75

\bibitem[{{Rao} {et~al.}(2009){Rao}, {Girart}, {Marrone}, {Lai}, \&
  {Schnee}}]{Rao2009}
{Rao}, R., {Girart}, J.~M., {Marrone}, D.~P., {Lai}, S.-P., \& {Schnee}, S.
  2009, \apj, 707, 921

\bibitem[{{Reipurth} {et~al.}(1992){Reipurth}, {Heathcote}, \&
  {Vrba}}]{Reipurth1992}
{Reipurth}, B., {Heathcote}, S., \& {Vrba}, F. 1992, \aap, 256, 225

\bibitem[{{Reipurth} {et~al.}(2002){Reipurth}, {Rodr{\'{\i}}guez}, {Anglada},
  \& {Bally}}]{Reipurth2002}
{Reipurth}, B., {Rodr{\'{\i}}guez}, L.~F., {Anglada}, G., \& {Bally}, J. 2002,
  \aj, 124, 1045

\bibitem[{{Rodr{\'{\i}}guez} {et~al.}(1997){Rodr{\'{\i}}guez}, {Anglada}, \&
  {Curiel}}]{Rodriguez1997}
{Rodr{\'{\i}}guez}, L.~F., {Anglada}, G., \& {Curiel}, S. 1997, \apjl, 480,
  L125

\bibitem[{{Sadavoy} {et~al.}(2014){Sadavoy}, {Di Francesco}, {Andre},
  {Pezzuto}, {Bernard}, {Maury}, {Men'shchikov}, {Motte}, {Nguyen-Lu'o'ng},
  {Schneider}, {Arzoumanian}, {Benedettini}, {Bontemps}, {Elia}, {Hennemann},
  {Hill}, {Konyves}, {Louvet}, {Peretto}, {Roy}, \& {White}}]{Sadavoy2014}
{Sadavoy}, S.~I., {Di Francesco}, J., {Andre}, P., {et~al.} 2014, \apjl, 787,
  L18

\bibitem[{{Saito} {et~al.}(1999){Saito}, {Sunada}, {Kawabe}, {Kitamura}, \&
  {Hirano}}]{Saito1999}
{Saito}, M., {Sunada}, K., {Kawabe}, R., {Kitamura}, Y., \& {Hirano}, N. 1999,
  \apj, 518, 334

\bibitem[{{Santangelo} {et~al.}(2015){Santangelo}, {Codella}, {Cabrit},
  {Maury}, {Gueth}, {Maret}, {Lefloch}, {Belloche}, {Andr{\'e}}, {Hennebelle},
  {Anderl}, {Podio}, \& {Testi}}]{Santangelo2015}
{Santangelo}, G., {Codella}, C., {Cabrit}, S., {et~al.} 2015, \aap, 584, A126

\bibitem[{{Schuller} {et~al.}(2009){Schuller}, {Menten}, {Contreras},
  {Wyrowski}, {Schilke}, {Bronfman}, {Henning}, {Walmsley}, {Beuther},
  {Bontemps}, {Cesaroni}, {Deharveng}, {Garay}, {Herpin}, {Lefloch}, {Linz},
  {Mardones}, {Minier}, {Molinari}, {Motte}, {Nyman}, {Reveret}, {Risacher},
  {Russeil}, {Schneider}, {Testi}, {Troost}, {Vasyunina}, {Wienen}, {Zavagno},
  {Kovacs}, {Kreysa}, {Siringo}, \& {Wei{\ss}}}]{Schuller2009}
{Schuller}, F., {Menten}, K.~M., {Contreras}, Y., {et~al.} 2009, \aap, 504, 415

\bibitem[{{Segura-Cox} {et~al.}(2016){Segura-Cox}, {Harris}, {Tobin}, {Looney},
  {Li}, {Chandler}, {Kratter}, {Dunham}, {Sadavoy}, {Perez}, \&
  {Melis}}]{SeguraCox2016}
{Segura-Cox}, D.~M., {Harris}, R.~J., {Tobin}, J.~J., {et~al.} 2016, \apjl,
  817, L14

\bibitem[{{Segura-Cox} {et~al.}(2015){Segura-Cox}, {Looney}, {Stephens},
  {Fern{\'a}ndez-L{\'o}pez}, {Kwon}, {Tobin}, {Li}, \&
  {Crutcher}}]{SeguraCox2015}
{Segura-Cox}, D.~M., {Looney}, L.~W., {Stephens}, I.~W., {et~al.} 2015, \apjl,
  798, L2

\bibitem[{{Shu} {et~al.}(2000){Shu}, {Najita}, {Shang}, \& {Li}}]{Shu2000}
{Shu}, F.~H., {Najita}, J.~R., {Shang}, H., \& {Li}, Z.-Y. 2000, Protostars and
  Planets IV, 789

\bibitem[{{Stephens} {et~al.}(2013){Stephens}, {Looney}, {Kwon}, {Hull},
  {Plambeck}, {Crutcher}, {Chapman}, {Novak}, {Davidson}, {Vaillancourt},
  {Shinnaga}, \& {Matthews}}]{Stephens2013}
{Stephens}, I.~W., {Looney}, L.~W., {Kwon}, W., {et~al.} 2013, \apjl, 769, L15

\bibitem[{{Straizys} {et~al.}(1992){Straizys}, {Cernis}, {Kazlauskas}, \&
  {Meistas}}]{Straizys1992}
{Straizys}, V., {Cernis}, K., {Kazlauskas}, A., \& {Meistas}, E. 1992, Baltic
  Astronomy, 1, 149

\bibitem[{{Stutz} {et~al.}(2008){Stutz}, {Rubin}, {Werner}, {Rieke}, {Bieging},
  {Keene}, {Kang}, {Shirley}, {Su}, {Velusamy}, \& {Wilner}}]{Stutz2008}
{Stutz}, A.~M., {Rubin}, M., {Werner}, M.~W., {et~al.} 2008, \apj, 687, 389

\bibitem[{{Tafalla} \& {Bachiller}(1995)}]{Tafalla1995}
{Tafalla}, M. \& {Bachiller}, R. 1995, \apjl, 443, L37

\bibitem[{{Tafalla} {et~al.}(2006){Tafalla}, {Kumar}, \&
  {Bachiller}}]{Tafalla2006}
{Tafalla}, M., {Kumar}, M.~S.~N., \& {Bachiller}, R. 2006, \aap, 456, 179

\bibitem[{{Tang} {et~al.}(2013){Tang}, {Ho}, {Koch}, {Guilloteau}, \&
  {Dutrey}}]{Tang2013}
{Tang}, Y.-W., {Ho}, P.~T.~P., {Koch}, P.~M., {Guilloteau}, S., \& {Dutrey}, A.
  2013, \apj, 763, 135

\bibitem[{{Terebey} {et~al.}(1993){Terebey}, {Chandler}, \&
  {Andre}}]{Terebey1993}
{Terebey}, S., {Chandler}, C.~J., \& {Andre}, P. 1993, \apj, 414, 759

\bibitem[{{Terebey} \& {Padgett}(1997)}]{Terebey1997}
{Terebey}, S. \& {Padgett}, D.~L. 1997, in IAU Symposium, Vol. 182, Herbig-Haro
  Flows and the Birth of Stars, ed. B.~{Reipurth} \& C.~{Bertout}, 507--514

\bibitem[{{Tobin} {et~al.}(2013){Tobin}, {Chandler}, {Wilner}, {Looney},
  {Loinard}, {Chiang}, {Hartmann}, {Calvet}, {D'Alessio}, {Bourke}, \&
  {Kwon}}]{Tobin2013}
{Tobin}, J.~J., {Chandler}, C.~J., {Wilner}, D.~J., {et~al.} 2013, \apj, 779,
  93

\bibitem[{{Tobin} {et~al.}(2011){Tobin}, {Hartmann}, {Chiang}, {Looney},
  {Bergin}, {Chandler}, {Masqu{\'e}}, {Maret}, \& {Heitsch}}]{Tobin2011}
{Tobin}, J.~J., {Hartmann}, L., {Chiang}, H.-F., {et~al.} 2011, \apj, 740, 45

\bibitem[{{Tobin} {et~al.}(2016){Tobin}, {Looney}, {Li}, {Chandler}, {Dunham},
  {Segura-Cox}, {Sadavoy}, {Melis}, {Harris}, {Kratter}, \&
  {Perez}}]{Tobin2016}
{Tobin}, J.~J., {Looney}, L.~W., {Li}, Z.-Y., {et~al.} 2016, \apj, 818, 73

\bibitem[{{Tobin} {et~al.}(2007){Tobin}, {Looney}, {Mundy}, {Kwon}, \&
  {Hamidouche}}]{Tobin2007}
{Tobin}, J.~J., {Looney}, L.~W., {Mundy}, L.~G., {Kwon}, W., \& {Hamidouche},
  M. 2007, \apj, 659, 1404

\bibitem[{{Tobin} {et~al.}(2015){Tobin}, {Looney}, {Wilner}, {Kwon},
  {Chandler}, {Bourke}, {Loinard}, {Chiang}, {Schnee}, \& {Chen}}]{Tobin2015}
{Tobin}, J.~J., {Looney}, L.~W., {Wilner}, D.~J., {et~al.} 2015, \apj, 805, 125

\bibitem[{{Watson} {et~al.}(2007){Watson}, {Bohac}, {Hull}, {Forrest},
  {Furlan}, {Najita}, {Calvet}, {D'Alessio}, {Hartmann}, {Sargent}, {Green},
  {Kim}, \& {Houck}}]{Watson2007}
{Watson}, D.~M., {Bohac}, C.~J., {Hull}, C., {et~al.} 2007, \nat, 448, 1026

\bibitem[{{Welch} {et~al.}(1996){Welch}, {Thornton}, {Plambeck}, {Wright},
  {Lugten}, {Urry}, {Fleming}, {Hoffman}, {Hudson}, {Lum}, {Forster}, {Thatte},
  {Zhang}, {Zivanovic}, {Snyder}, {Crutcher}, {Lo}, {Wakker}, {Stupar},
  {Sault}, {Miao}, {Rao}, {Wan}, {Dickel}, {Blitz}, {Vogel}, {Mundy},
  {Erickson}, {Teuben}, {Morgan}, {Helfer}, {Looney}, {de Gues}, {Grossman},
  {Howe}, {Pound}, \& {Regan}}]{Welch1996}
{Welch}, W.~J., {Thornton}, D.~D., {Plambeck}, R.~L., {et~al.} 1996, \pasp,
  108, 93

\bibitem[{{Whittet} {et~al.}(2008){Whittet}, {Hough}, {Lazarian}, \&
  {Hoang}}]{Whittet2008}
{Whittet}, D.~C.~B., {Hough}, J.~H., {Lazarian}, A., \& {Hoang}, T. 2008, \apj,
  674, 304

\bibitem[{Wolf {et~al.}(2003)Wolf, Launhardt, \& Henning}]{Wolf2003}
Wolf, S., Launhardt, R., \& Henning, T. 2003, The Astrophysical Journal, 592,
  233

\bibitem[{{Wolf-Chase} {et~al.}(2000){Wolf-Chase}, {Barsony}, \&
  {O'Linger}}]{WolfChase2000}
{Wolf-Chase}, G.~A., {Barsony}, M., \& {O'Linger}, J. 2000, \aj, 120, 1467

\bibitem[{{Wu} {et~al.}(2007){Wu}, {Dunham}, {Evans}, {Bourke}, \&
  {Young}}]{Wu2007}
{Wu}, J., {Dunham}, M.~M., {Evans}, II, N.~J., {Bourke}, T.~L., \& {Young},
  C.~H. 2007, \aj, 133, 1560

\bibitem[{{Yeh} {et~al.}(2008){Yeh}, {Hirano}, {Bourke}, {Ho}, {Lee}, {Ohashi},
  \& {Takakuwa}}]{Yeh2008}
{Yeh}, S.~C.~C., {Hirano}, N., {Bourke}, T.~L., {et~al.} 2008, \apj, 675, 454

\bibitem[{{Yen} {et~al.}(2015){Yen}, {Koch}, {Takakuwa}, {Ho}, {Ohashi}, \&
  {Tang}}]{Yen2015}
{Yen}, H.-W., {Koch}, P.~M., {Takakuwa}, S., {et~al.} 2015, \apj, 799, 193

\bibitem[{{Yen} {et~al.}(2010){Yen}, {Takakuwa}, \& {Ohashi}}]{Yen2010}
{Yen}, H.-W., {Takakuwa}, S., \& {Ohashi}, N. 2010, \apj, 710, 1786

\bibitem[{{Yen} {et~al.}(2011){Yen}, {Takakuwa}, \& {Ohashi}}]{Yen2011}
{Yen}, H.-W., {Takakuwa}, S., \& {Ohashi}, N. 2011, \apj, 742, 57

\bibitem[{{Yen} {et~al.}(2013){Yen}, {Takakuwa}, {Ohashi}, \& {Ho}}]{Yen2013}
{Yen}, H.-W., {Takakuwa}, S., {Ohashi}, N., \& {Ho}, P.~T.~P. 2013, \apj, 772,
  22

\bibitem[{{Zhang} {et~al.}(1995){Zhang}, {Ho}, {Wright}, \&
  {Wilner}}]{Zhang1995}
{Zhang}, Q., {Ho}, P.~T.~P., {Wright}, M.~C.~H., \& {Wilner}, D.~J. 1995,
  \apjl, 451, L71

\bibitem[{{Zhang} {et~al.}(2014){Zhang}, {Qiu}, {Girart}, {(Baobab Liu},
  {Tang}, {Koch}, {Li}, {Keto}, {Ho}, {Rao}, {Lai}, {Ching}, {Frau}, {Chen},
  {Li}, {Padovani}, {Bontemps}, {Csengeri}, \& {Ju{\'a}rez}}]{Zhang2014}
{Zhang}, Q., {Qiu}, K., {Girart}, J.~M., {et~al.} 2014, \apj, 792, 116

\end{thebibliography}

\appendix

\section{Source description}

 In this section, we provide a quick description of the 
 objects and of the 0.87mm continuum observed with the SMA. \\

\noindent{\bf B335}
\vspace{2pt}

\noindent B335 is an isolated Bok globule hosting a Class 0 protostar. 
Its luminosity is $\sim$1$L_{\sun}$ \citep{Keene1983} and its mass $<$2$M_{\sun}$.

\noindent{\it Ouflows -} It possesses an east-west elongated, 
conical-shaped molecular outflow \citep[PA: 80$\degr$;][]{Hirano1988}.
Collimated $^{12}$CO (2$-$1) jets \citep{Yen2010} and 
Herbig-Haro (HH) objects \citep[HH 119 A$-$F;][]{Reipurth1992,Galfalk2007}  
are detected along this outflow axis. 

\noindent{\it Velocity field -} C$^{18}$O (2$-$1) and H$^{13}$CO$^{+}$ (1$-$0) interferometric 
observations have traced the rotational infalling motion of the 
envelope from radii of $\sim20 000$ down to $\sim1000$ au. 
No clear rotational motion is detected on 100-500 au scales \citep{Saito1999,Harvey2003,Yen2011} and the central circumstellar 
disk radius is estimated to be smaller than 100 au. 

\noindent{\it Our SMA observations -} The SMA 0.87mm continuum emission 
has a north-south elongated structure coherent with previous 
1.3mm observations of the object by \citet{Motte2001}. The two 
protuberances observed in the north and southwest regions trace 
the edge of the cavity produced by the horizontal outflow \citep{Yen2010}.\\

\noindent {\bf SVS13} 
\vspace{2pt}

\noindent SVS13 is a multiple system in the NGC~1333 
star-forming region. It possesses 3 main submm continuum sources 
aligned in the northeast-southwest direction named A, B and C \citep{Looney2003}. 
SVS13-A is a Class I protostar while SVS13-B and C are Class 0. 
A large filamentary structure encompasses the three sources at 
450 \mic\ \citep{Chandler2000} and in mm \citep{Hull2014}.
The combined luminosity for SVS13-A, B, and C is 45$L_{\sun}$ 
\citep{Jennings1987}. 70 \mic\ emission is detected toward SVS13-A 
and C but not toward SVS13-B, suggesting that the source is deeply 
embedded \citep{Chen2009}. A radio source called VLA3 has also 
been detected southwest of SVS13-A \citep{Rodriguez1997}. 

\noindent{\it Ouflows -} SVS13-A is likely powering the strong 
northwest-southeast HH7-11 outflow studied by \citet{Bachiller2000} and
\citet{Plunkett2013}. SVS13-B is known to drive a highly collimated 
SiO jet \citep{Bachiller1998}.

\noindent{\it Velocity field -} A symmetric velocity 
gradient of 28km~s$^{-1}~$pc$^{-1}$ across SVS13-B and 
SVS13A/VLA3 has been detected using N$_2$H$^+$ observations. 
This large gradient could suggest that the wide 
binary system is physically bound \citep{Chen2009}. 

\noindent{\it Our SMA observations -} The SMA 0.87mm observations 
trace the extended envelope around SVS13-A and SVS13-B. We resolve 
the two sources while SVS13-C is only marginally detected in the southwest. 
The continuum emission of SVS13-A is slightly 
elongated in the southwest direction, toward the 
position of VLA3.
The 0.87mm emission of SVS13-B extends in the north-northwest direction. 
That extension aligns with that of the SiO jet driven by the protostar. 
The continuum emission also extends in the southwest direction 
of SVS13-B: the extension is also observed in the elongated 
N$_2$H$^+$ map but not in the PdBI 1.4 and 3 mm observations, which could 
be linked with filtering effects affecting the PdBI maps \citep{Chen2009}. \\

\noindent {\bf IC348-SMM2 / HH797} 
\vspace{2pt}

\noindent SMM2 is a low-mass class 0 belonging to the IC 348 
star-forming region. SMA observations also reveal 
a companion source, SMM2E, at $\sim$2400 au in the northeast 
direction \citep{Palau2014} whose Spectral Energy 
Distribution is typical of Class 0 objects.

\noindent{\it Ouflows -} SMM2 probably drives the HH797 jet, 
a very extended and collimated northwest-southeast outflow \citep{Chen2013}. 
Velocity asymmetries along the length of the flow were 
interpreted as jet rotation \citep{Pech2012}. A compact 
low-velocity outflow seems to also be associated with SMM2E. 

\noindent{\it Velocity field -} 
C$^{18}$O observations of the companion source SMM2E 
suggest that the envelope is probably rotating with a 
north-south axis close to the outflow direction \citep{Palau2014}. 

\noindent{\it Our SMA observations -} 
Our 0.87mm continuum map is consistent with that previously 
obtained by \citet{Palau2014} with SMA as well as far-IR 
continuum emission observed with {\it Herschel}. Both SMM2 and 
SMM2E are detected at 0.87mm, but SMME2 is not 
resolved. \\

\noindent {\bf L1448C} 
\vspace{2pt}

\noindent The L1448C complex (also called L1448-mm) hosts 
several embedded Class 0 protostars \citep{Tobin2007}. 

\noindent{\it Ouflows -} 
It possesses well collimated outflows \citep{Hirano2010}. 
The northern CO lobe is probably interacting 
with the neighboring source L1448N \citep{Bachiller1995}. 

\noindent{\it Velocity field -} 
VLA observations of NH$_3$ reveal velocity gradients perpendicular 
to the outflow as well as along the outflow axis. 
This suggests the presence of a self-gravitating structure 
around L1448C, with an envelope that is both rotating and 
contracting at comparable velocities \citep{Curiel1999}. 

\noindent{\it Our SMA observations -} 
Our SMA observations show that L1448C is elongated in the northwest-southeast 
direction, which is consistent with the elongation observed at 
1.3mm \citep{Barsony1998}. The southeast elongation encompasses the 
companion source L1448-mm B (2000 au separation) observed with 
{\it Spitzer} \citep{Tobin2007} 
\citep[also called L1448C(S);][]{Jorgensen2006}.
A distortion is also observed in the southwest direction where a 0.80mm
continuum emission has been observed by \citet{Barsony1998}. 
This distortion as well as the northern elongation could also be 
tracing the edge of the outflow cavities.\\

\noindent {\bf L1448N} 
\vspace{2pt}

\noindent L1448N (or L1448 IRS3) is a multiple system. 
L1448N-A is located 2100 au north to L1448N-B. B is the 
strongest source at mm wavelengths \citep{Looney2000}
while A dominates at mid-IR \citep{Ciardi2003} and cm wavelengths 
\citep{Curiel1990}. The system is also connected 
to L1448NW located 20\arcsec\ north. \citet{Barsony1998} suggest 
that star formation in L1448N could have been induced by the strong 
outflow driven by L1448C \citep[see also][]{Kwon2006}. 

\noindent{\it Ouflows -} Two bipolar outflows originate from 
the system. The outflow driven by L1448N-A is nearly 
perpendicular to the line-of-sight. \citet{Kwon2006} show that 
the two outflows are probably interacting with each other.

\noindent{\it Velocity field -}
The inner 5-10\arcsec\ of L1448N presents a solid-body 
rotation while the outer envelope shows a flatter profile \citep{Terebey1997}. 
C$^{18}$O observations also reveal a strong velocity gradient 
\citep[$>$100 km~s~$^{-1}$~pc$^{-1}$; Gaudel et al. in prep;][]{Yen2015} 
perpendicular to the outflow direction within 20$\degr$ \citep{Lee2015}.

\noindent{\it Our SMA observations -}
L1448N-A and B are resolved by the SMA while L1448NW is 
marginally detected. Both sources are elongated in 
the northeast-southwest direction, which is consistent with previous 230 GHz 
and 0.87mm SMA observations by \citet{Lee2015}.\\

\noindent {\bf L1448-2A} 
\vspace{2pt}

\noindent L1448-2A (or L1448 IRS 2) is a low luminosity (5.2 \lsun) Class 0 protostar located toward the western edge of the L1448 complex that seems to be evolving in isolation from L1448N and C. 

\noindent{\it Ouflows -} 
High-velocity CO maps show that L1448-2A drives an outflow in the 
northwest-southeast direction \citep{OLinger1999}. Additional studies from 
\citet{WolfChase2000} suggest that the source is actually driving two distinct outflows, 
which could be a signature of an (unresolved) binary system. 

\noindent{\it Velocity field -} Using C$^{18}$O observations, 
\citet{Yen2015} found a velocity gradient nearly perpendicular 
to the outflow axis.

\noindent{\it Our SMA observations -}
The northeast-southwest extension of the envelope is confirmed 
by our SMA observation. Contrary to the CARMA 1.3 mm dust continuum 
presented in \citet{Hull2014}, we do not observe elongation in the 
northwest-southeast direction. As this elongation is not observed at 
other wavelengths, we conclude that the elongation observed at 1.3 mm 
might be due to contamination by the outflow. The `horns' detected in the 
east and west direction trace the edges of the large outflow cavity.\\

\noindent {\bf IRAS03282} 
\vspace{2pt}

\noindent IRAS03282 is a highly-embedded Class 0 object lying at 1$\degr$ 
southeast of NGC~1333 \citep{Jorgensen2006}. IRAS~03282 possesses 
two sources separated by 1.5\arcsec\ \citep{Launhardt2004}.

\noindent{\it Outflows -}
The source drives an outflow in which various velocity components 
are observed, in particular a high-velocity ($\sim$60km~s$^{-1}$) 
jet with fast molecular clumps detected along the outflow axis 
surrounded by a less collimated and more standard velocity 
($\sim$20km~s$^{-1}$) outflow \citep{Bachiller1991}.

\noindent{\it Velocity field -}
N$_2$H$^+$ and NH$_3$ observations were used to trace the 
large-scale velocity gradient of the envelope that is mostly 
distributed along the outflow direction \citep{Tobin2011}. 

\noindent{\it Our SMA observations -}
The N$_2$H$^+$ and NH$_3$ maps present an elongation along the 
north-south direction. Our SMA 0.87mm map presents a similar 
elongation. The southern elongation is also very similar to that 
observed at 350\mic\ by \citet{Wu2007} with SHARC-II. \\

\noindent {\bf NGC~1333 IRAS4A} 
\vspace{2pt}

\noindent NGC~1333 IRAS4A is a Class 0 object located in the south of NGC~1333. 
It hosts two sources (IRAS4A1 and IRAS4A2) in a close (1.8\arcsec) 
binary pair that was resolved with interferometers at submm 
and mm wavelengths using the SMA or BIMA
\citep{Looney2000,Girart2006,Jorgensen2007} and more recently 
ALMA \citep{Lopez-Sepulcre2017}: IRAS4A1 dominates the 
submm and mm emission. 

\noindent{\it Outflows -}
An arcmin-long outflow, seen in high-velocity CO(3-2) emission, is associated with IRAS4A2 and bends 
from a north-south orientation on small scales \citep{Santangelo2015}
to a position angle of 45$\circ$ on large scales \citet{Koumpia2016}. 
This could be due to an intrinsic change of the gas propagation 
direction, or a signature of jet precession. The launching 
direction of the outflow driven by IRAS4A1 is also oriented 
close to north-south direction and faster \citep{Santangelo2015}. 
Part of that SiO outflow (position angle of -10$\degr$) was also 
reported by \citet{Choi2001}.

\noindent{\it Velocity field -}
Detections of inverse P-Cygni profiles in the lines of 
common molecular species such as $^{13}CO$ or CS are 
interpreted as signs of strong infall motions of the outer cloud \citep{DiFrancesco2001,Jorgensen2007}. 
\citet{Belloche2006} detected a centroid velocity gradient of 
about 10 km~s~$^{-1}$~pc$^{-1}$, with a position angle of 38$\degr$, 
while \citet{Ching2016} find a velocity gradient along the axis linking 
NGC~1333 IRAS4A1 and IRAS4A2. Both \citet{Belloche2006} and Gaudel et al (in prep.) 
show however that the interpretation of these velocity gradients 
in terms of rotation is not straightforward in IRAS4A due to the
strong infall motions previously mentioned.

\noindent{\it Our SMA observations -}
As already shown in \citet{Girart2006}, the SMA 0.87mm map 
presents a very compact structure extended in the 
A1-A2 direction. The common envelope extends in the northeast
and southwest direction, in the same direction 
($\sim$30-50$\degr$) {\it i)} that  the velocity gradient found
in \citet{Belloche2006}, {\it ii)} that the outflow takes further 
from the protostar and {\it iii)} that the large-scale B 
orientation observed by SCUBA \citep{Matthews2009}. \\

\noindent {\bf NGC~1333 IRAS4B} 
\vspace{2pt}

\noindent NGC~1333 IRAS4B is also a Class 0 protostar and belongs to the same 
N$_2$H$^+$ arcmin-scale filament as IRAS4A. Submm and mm 
observations reveal a second source located 11\arcsec\ on the 
east (IRAS4B2) \citep{Looney2000}. IRAS4B2 is not detected at 
cm wavelengths, suggesting that it is at a different stage 
of its evolution \citep{Reipurth2002} or just not associated 
at all with IRAS4B. There are still some debate on smaller scales 
on whether the disk of IRAS 4B is observed face- or edge-on 
\citep{Watson2007,Marvel2008}.

\noindent{\it Outflows -} The source is driving an outflow
whose axis is nearly in north-south direction. 
Its short dynamical timescale suggests that this outflow is very young
\citep{Choi2001}.

\noindent{\it Velocity field -}
IRAS4B seems to be collapsing \citep{DiFrancesco2001,Belloche2006} 
but does not show clear signs of rotational motions at 1000 au scales \citep{Yen2013}. 

\noindent{\it Our SMA observations -}
The SMA 0.87mm map of IRAS4B presents a very compact structure 
that seems to be extended in the north-south
direction, namely the direction of the outflow emanating from this 
source. We do not observe the same east-west extension as seen
at 1.3 mm by \citet{Hull2014} and \citet{Yen2013}. \\

\noindent {\bf IRAS16293} 
\vspace{2pt}

\noindent IRAS16293 is a Class 0 protostar system 
located in the $\rho$ Ophiuchi molecular cloud. 
It possesses a bolometric luminosity of $\sim$25$L_{\sun}$ 
and a massive envelope \citep{Correia2004,Crimier2010} surrounding 
two cores separated by 6\arcsec, called A (south) and B (north). 
\citet{Crimier2010} suggests that IRAS16293A/B is a ``separate 
envelope system'' \citep[following the classification proposed by][]{Looney2003}. 
IRAS16293-A is an edge-on system and IRAS16293-B is nearly face-on \citep{Jorgensen2016}.
The first dominates the system luminosity. SMA observations have also 
enabled to separate IRAS16293-A into two components, Aa and Ab, 
aligned along a position angle of 45$\degr$ \citep{Chandler2005}.
No small scale multiplicity has been found for IRAS16293-B.

\noindent{\it Outflows -}
The source shows significant outflow activity.
While IRAS16293-A seems to be driving the main east-west CO 
outflow, the driving source of the northwest-southeast
compact outflow (PA of 145$\degr$) traced with SiO observations is still unclear 
\citep{Yeh2008,Rao2009}. This much younger outflow is probably 
a more robust tracer of the rotation axis, as recent studies revealed 
at various scales an almost edge-on rotational pattern whose axis match 
the SiO outflow orientation \citet{Pineda2012,Girart2014}.
A one-sided bubble-like outflow structure originates from IRAS16293-B 
in the southeast direction \citep{Loinard2013}.

\noindent{\it Velocity field -}
In IRAS16293-A, C$^{34}$S observations reveal a clear velocity 
gradient along PA~40-45$\degr$, i.e. perpendicular to the SiO-traced 
outflow \citep{Girart2014} probably due to infalling motion of the rotating envelope. A disk-like rotation is also detected at 40 -- 60 au 
with ALMA, with the same orientation \citep{Oya2016}. 
In IRAS16293-B, recent ALMA observations reveal inverse P-Cygni profiles 
toward the center of IRAS16293-B in the CH$_3$OCHO-A CH$_3$OCHO-E and 
H$_2$CCO lines, here again indicative of infall motions \citep{Pineda2012}. 
The source is observed face-on: no rotation has been detected.

\noindent{\it Our SMA observations -}
As shown in \citet{Rao2009}, the SMA 0.87mm observations separate 
the two A and B components. IRAS16293-B is brighter than IRAS16293-A. 
The continuum emission of the envelope extends in the north of IRAS16293-A 
and is also quite extended ($\sim$10\arcsec) toward its 
southeast direction, i.e. in the direction of the redshifted 
part of the SiO outflow. \\

\noindent {\bf L1157} 
\vspace{2pt}

\noindent L1157 is a low-mass Class 0 protostar 
and seems to be a single system \citep{Tobin2013}. 

\noindent{\it Outflows -}
It possesses a spectacular bipolar outflow asymmetric on large-scales.
Its prominent CO cavities are likely created by the 
propagation of large bow-shocks  \citep[also traced 
via the NH$_3$ or SiO molecular lines;][]{Tafalla1995,Zhang1995,Gueth1996}. 
This could be the sign that a time-variable 
collimated jet is driving the outflow \citep{Podio2016}. 

\noindent{\it Velocity field -}
L1157 has a flattened and filamentary envelope presenting 
a weak velocity gradient along the filament - thus 
perpendicular to the outflow. Broad linewidths are 
observed in the inner envelope \citep[see][]{Kwon2015}. 

\noindent{\it Our SMA observations -}
Our 0.87mm observations are consistent with those presented in 
\citep{Chen2013}. Neither \citet{Chen2013} nor our observations
detect the elongation structure detected at 1.3 mm with \citet{Stephens2013} 
in the northern (outflow) direction. By contrast, we detect a 
similar eastern extension and western bump. This horizontal elongation 
is also observed at 350\mic\ by \citet{Wu2007}. \\

\noindent {\bf CB230} 
\vspace{2pt}

\noindent CB 230 is an isolated globule hosting a dense core, itself 
hosting two deeply embedded YSOs. These are separated by 
10\arcsec\ and are aligned in the east-west direction 
\citep{Launhardt2001,Launhardt2004}. 
The eastern object, IRS2 is not detected beyond 24 \mic\ 
\citep{Massi2008,Launhardt2013}. 

\noindent{\it Outflows -} 
The north-south CO outflows seem to be driven by the western protostar IRS1. 
More recent 7mm high-resolution observations of IRS1 led to the detection 
of two continuum sources separated by $\sim$100au, 
with an unresolved primary source at the origin of the 
outflow and an extended companion source located in a direction 
perpendicular to the outflow direction \citep{Tobin2013}. 

\noindent{\it Velocity field -}
CB230 presents a velocity gradient $>$10 km~s~$^{-1}$~pc$^{-1}$
increasing from east to west along the axis connecting the two embedded 
YSOs \citep[][]{Chen2007,Tobin2011}. 
The gradient is perpendicular to the outflow axis. 

\noindent{\it Our SMA observations -}
The 0.87mm continuum emission of CB230 is elongated in a boxy-shaped 
like structure in the north-south direction, an elongation also observed at 
1.3mm by \citet{Hull2014}. The southwest elongation could here 
again be linked with emission tracing the edges of the north-south outflow cavity.

\section{Stokes Q and U, polarization intensity and fraction maps of the sample}

\begin{figure*}
\begin{tabular}{p{18.5cm}}
\vspace{10pt}
\includegraphics[width=18.5cm]{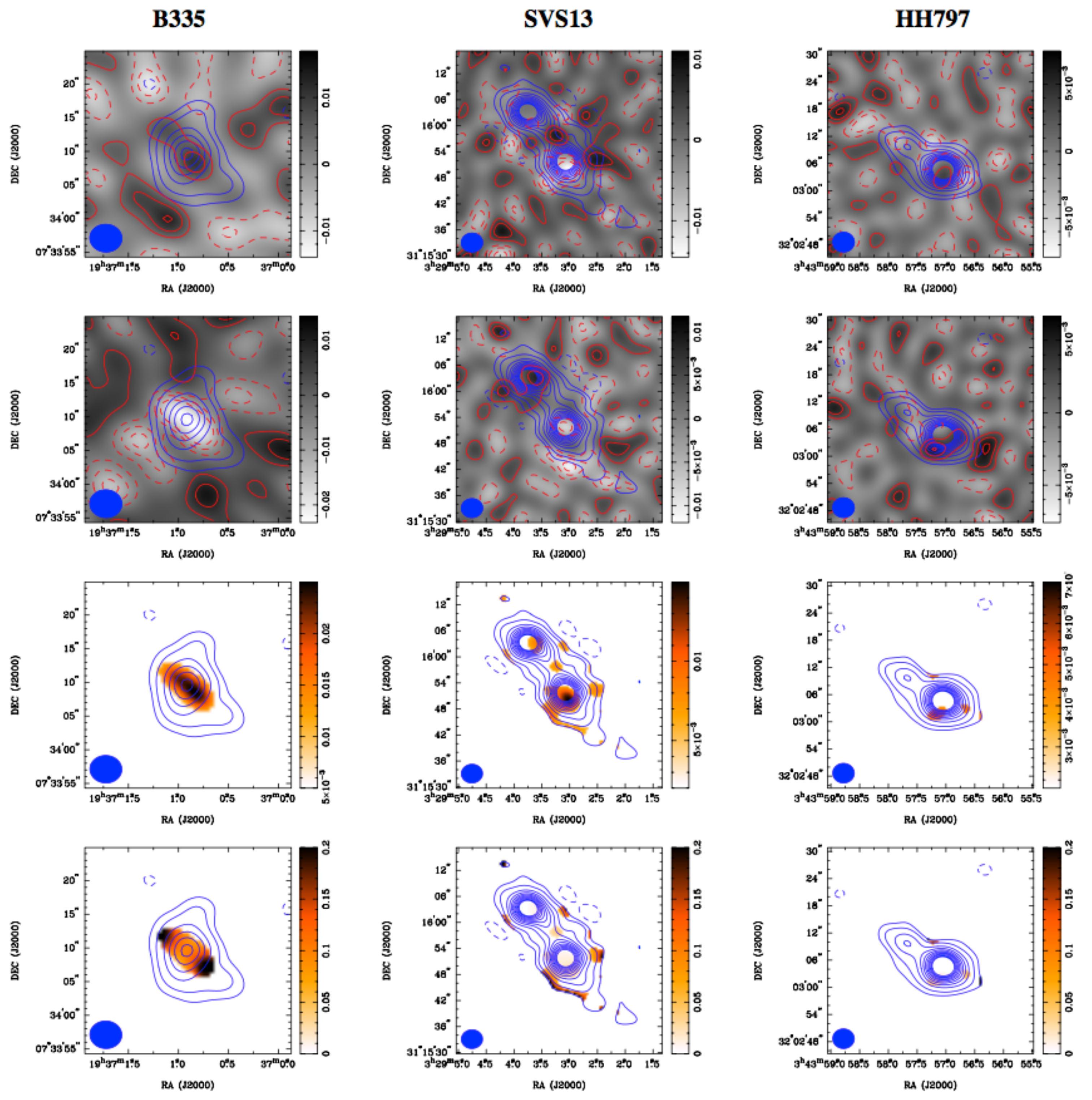} 
\end{tabular}
\caption{Polarization maps of, from left to right, B335, SVS13 and HH797. 
{\it Two top panels:} Stokes Q and U maps. Color scales are in mJy/beam. 
Stokes I contours (same as in Fig.~\ref{StokesI_Borientation}) appear in blue. 
Stokes Q and U contours appear in red and are [-3,-2,-1,1,2,3] $\sigma$.
{\it Third line:} Polarization intensity maps. The color scale is in mJy/beam.
{\it Fourth line:} Polarization fraction map. 
The filled ellipses on the lower left corner indicate the synthesized beam of the SMA maps. }
\label{PolaMaps}
\vspace{10pt}
\end{figure*}
\addtocounter {figure}{-1}
\begin{figure*}
\begin{tabular}{p{18.5cm}}
\vspace{10pt}
\includegraphics[width=18.5cm]{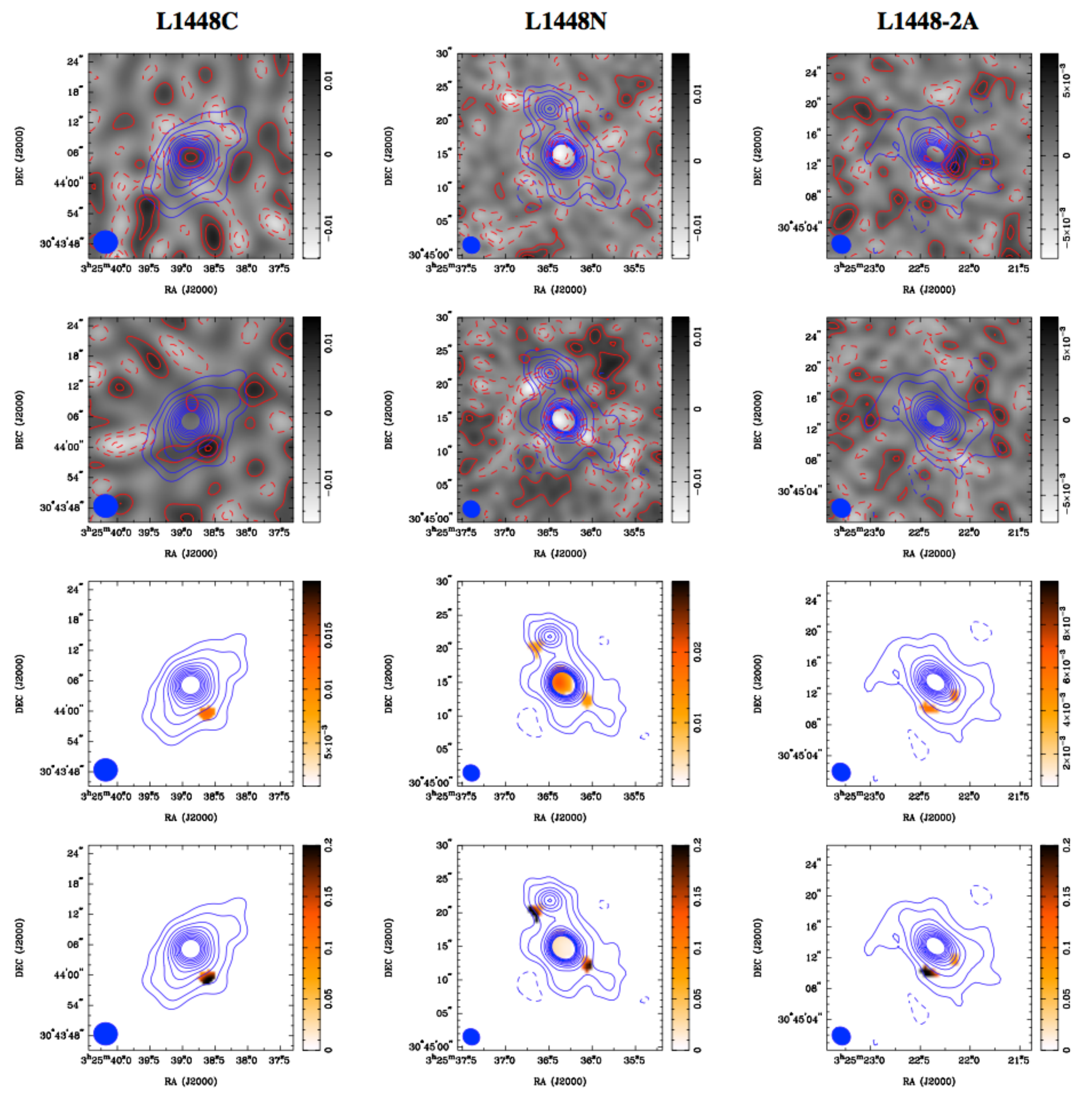} 
\end{tabular}
\caption{Continued. Same for L1448C, N and 2A.}
\vspace{40pt}
\end{figure*}
\addtocounter {figure}{-1}
\newpage
\begin{figure*}
\begin{tabular}{p{18.5cm}}
\vspace{10pt}
\includegraphics[width=18.5cm]{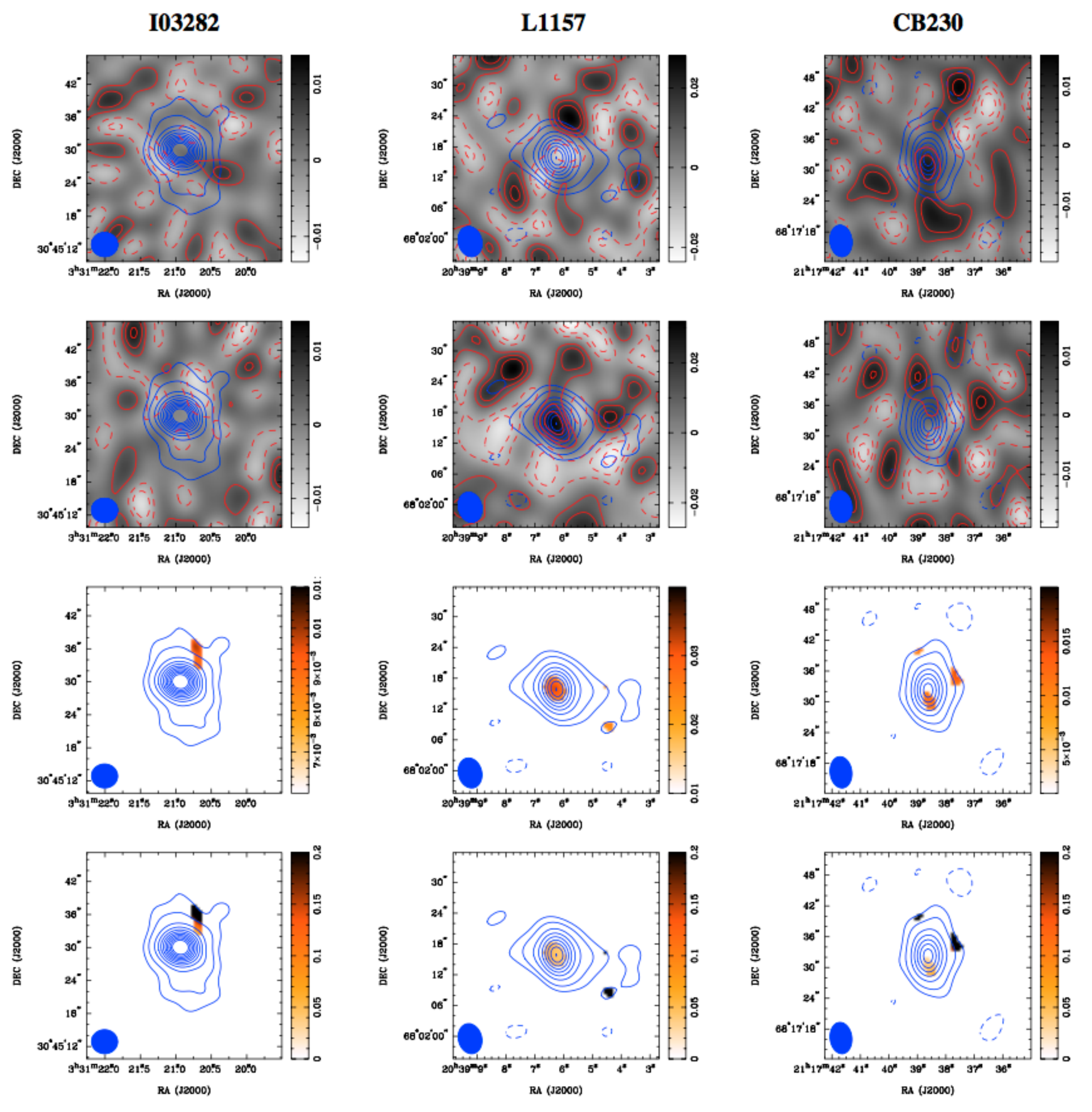} 
\end{tabular}
\caption{Continued. Same for IRAS03282, L1157 and CB230.}
\vspace{40pt}
\end{figure*}
\addtocounter {figure}{-1}
\newpage
\begin{figure*}
\begin{tabular}{p{18.5cm}}
\vspace{10pt}
\includegraphics[width=18.5cm]{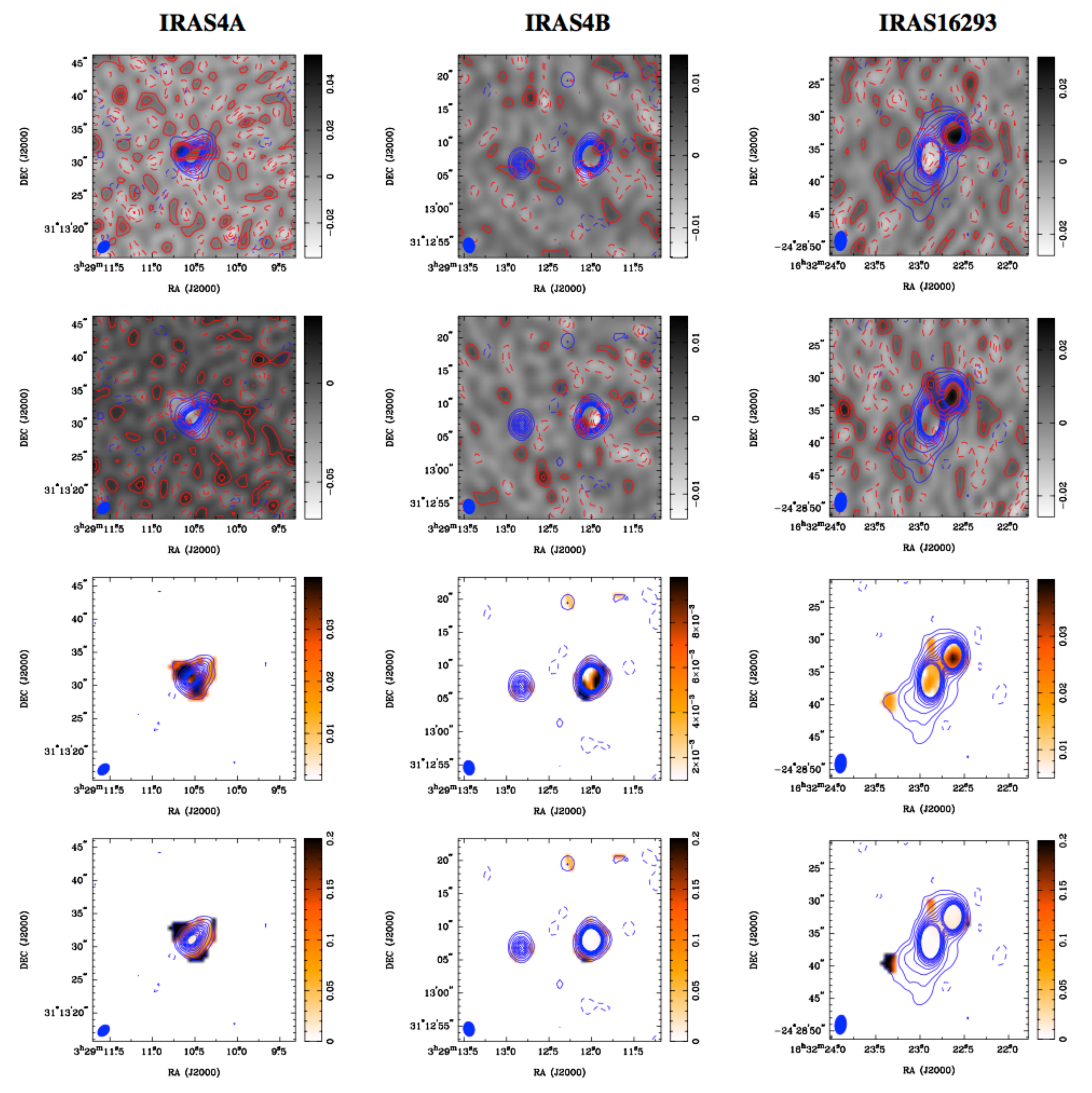} 
\end{tabular}
\caption{Continued. Same for NGC~1333 IRAS4A, 
NGC~1333 IRAS4B and IRAS16293.}
\vspace{40pt}
\end{figure*}

\end{document}